\documentclass{article}
\usepackage{graphicx} 
\usepackage{fvextra}
\usepackage{amsmath}
\usepackage{array}
\usepackage{amssymb}
\usepackage{amsthm}
\usepackage{booktabs}
\usepackage[margin=1in]{geometry}
\usepackage{subcaption}

\usepackage[most]{tcolorbox}

\usepackage{algorithm}

\usepackage{amsfonts}
\usepackage{mathabx}
\usepackage{enumitem}
\usepackage{hyperref}
\usepackage[table]{xcolor}
\usepackage{colortbl}
\newcolumntype{P}[1]{>{\centering\arraybackslash}p{#1}}

\usepackage[numbers]{natbib}

\usepackage{float}
\usepackage{stfloats}

\usepackage{threeparttable}
\usepackage{xspace}
\usepackage{tabularx}

\newcommand{\ai}{\textbf{AI}\xspace}
\newcommand{\self}{\textbf{SELF}\xspace}
\newcommand{\orig}{\textbf{O}\xspace}
\newcommand{\qual}{\textbf{T}\xspace}
\newcommand{\selforig}{\textbf{SELF-O}\xspace}
\newcommand{\selfqual}{\textbf{SELF-T}\xspace}
\newcommand{\aiorig}{\textbf{AI-O}\xspace}
\newcommand{\aiqual}{\textbf{AI-T}\xspace}

\usepackage{authblk}
\definecolor{lightgray}{HTML}{E6E6E6}

\newcommand{\revision}[1]{{\normalfont #1}}

\title{Incentives Shape How Humans Co-Create with Generative AI}
\author[1]{Nathanael Jo\thanks{Corresponding author: nathanjo@mit.edu}}
\author[1]{Manish Raghavan}

\affil[1]{Massachusetts Institute of Technology}
\date{}

\begin{document}

\maketitle

\begin{abstract}

Generative AI is quickly becoming an integral part of people's everyday workflows. Early evidence has shown that while generative AI can increase individual-level productivity, it does so at the cost of collective diversity, potentially narrowing the set of ideas and perspectives produced. Our research stands in contrast to this concern: through \revision{two} pre-registered randomized control trial\revision{s}, we show that \textit{incentives} \revision{can moderate} AI's homogenizing force in \revision{two} creative writing task\revision{s} where participants can use AI interactively. Participants rewarded for originality relative to peers produce collectively more diverse writing than those rewarded for quality alone. \revision{These participants continued to engage with AI, but ultimately incorporated less AI-generated text verbatim.}
Our results reveal that the effects of generative AI depend not only on the technology itself, but also the behavioral strategies and incentive structures surrounding its use. \revision{We release a rich, anonymized dataset capturing fine-grained trajectories of human--AI co-creation across both studies.\footnote{\url{https://huggingface.co/datasets/nathanaeljo/ai-cocreation-incentives}}}

\end{abstract}

\section{Introduction}
As generative AI becomes woven into creative and professional workflows, a
troubling pattern has emerged: AI assistance appears to homogenize outputs,
narrowing the diversity of ideas and expression even when humans retain final
control \cite{santurkar2023whose,wu2024generative, wenger2025we,
moon2025homogenizing, anderson2024homogenization, doshi2024generative,
padmakumar2023doeswriting, agarwal2025ai}. This evidence has led to concerns that widespread AI adoption will compress the creative landscape, as is the case with discussions around ``AI slop'' polluting the Internet \cite{read2024drowning,
wallacewells2024slop}. We argue that this conclusion is premature because existing findings have two shortcomings that limit the extent to which we can extrapolate to the real world.

First, \textbf{incentives matter}. In controlled experiments, participants without proper incentives may simply accept the first AI suggestion because they want to complete the task as fast as possible. This is not necessarily a good proxy for high-stakes, real-world settings. For example, even if prospective students use AI to help write their college admissions essays, we might expect them to expend effort in tailoring their narratives to stand out within a highly competitive applicant pool \cite{isley2026ai}.  With stronger incentives for differentiation, human judgment might mitigate or overcome the homogenizing force of AI \cite{raghavan2024competition}.

Second, much of the evidence on homogenization comes from experimental setups
with limited \textbf{interactivity} between humans and AI. Most studies restrict
participants to a single mode of use: brainstorming
\cite{anderson2024homogenization, doshi2024generative} or text completion
\cite{padmakumar2023doeswriting, agarwal2025ai}\footnote{We note that there exists many qualitative works investigating the impacts of AI on open-ended creative writing, see e.g., \cite{ippolito2022creative, gero2023social, lee2024design}, but these studies are mainly descriptive and have low sample sizes.}. But real-world AI use is
open-ended and iterative. A user may go back and forth with a model to
brainstorm, draft, and refine. Studying only one modality risks overstating the
homogenizing effect, since it prevents users from exercising the kind of
sustained agency that characterizes genuine \textbf{co-creation}. Indeed, we
find that as interaction unfolds, humans progressively diverge from initial AI
suggestions, recovering much (but not all) of the diversity that those
suggestions initially suppress.

\paragraph*{The present work: The impact of incentives on co-creation.}
Our work addresses these shortcomings by embedding a richer model of co-creation
within \revision{two randomized trials. In \textbf{Study 1} ($n=200$), p}articipants write a 250-350 word short
story with or without access to AI (OpenAI's 
GPT5-Mini), interacting freely with
a general-purpose assistant however they see fit. \revision{In \textbf{Study 2} ($n=400$), participants write a 100-125 word product pitch about a pre-specified product, with a similar setup.}
This design is  substantially more interactive than prior work \revision{investigating writing diversity}, and
yields rich longitudinal data on participants' strategies and writing
trajectories throughout the session.  We also vary \textit{incentives}
across participants: half are rewarded for high-quality work, and half for
originality. While neither condition exactly matches real-world conditions, the
\textit{contrast} between them offers a directionally informative window into
how market forces \revision{moderate} human-AI co-creation. The design and analysis plan
were pre-registered on
OSF.\footnote{\url{https://osf.io/3u8tq/overview?view_only=e1a8044dd5044bc59dd0a7d8f3fca3c2}}\footnote{\url{https://osf.io/xv9sa/overview?view_only=e1568bd84959466689af03c2aa362e7f}}

\begin{table*}[t]
\centering
\small

\begin{tabularx}{\textwidth}{lXX}
\toprule
& \textbf{Study 1: Short-story writing}
& \textbf{Study 2: Product pitch} \\

\textbf{Task}
& Write a 250--350 word short story
& Write a 100--125 word product pitch for a digital camera \\
\midrule
\textbf{Task duration}
& 5-min brainstorming phase, then a 20-min writing phase
& Single writing task \\

\textbf{AI access}
& No AI during brainstorming; AI access randomized during the writing phase
& AI access randomized during the writing task \\

\textbf{Task constraints}
& Soft creative constraints (e.g., writing in the first person or having the events take place within a single day)
& Product specifications provided to participants, such as 5$\times$ optical zoom and available colors \\

\textbf{Technical quality}
& Organization and writing technique, see Appendix~\ref{appsec:prompt_phase}
& Information and communication quality \\

\bottomrule
\end{tabularx}
\caption{\revision{Comparison of the two experimental studies.}}
\label{tab:study_comparison}
\end{table*}

\paragraph*{Results.} Across both studies, we find the following:

\begin{itemize}
    \item \textbf{Human editing/co-creation substantially increases diversity.} When AI is used to generate full drafts, the resulting texts are strikingly similar to one another, often sharing stylistic patterns (e.g., short stories starting with ``I woke up\dots''). However, participants rarely accept these drafts as-is. Instead, they edit, revise, and re-prompt the AI with more specific preferences, producing final submissions that are noticeably more diverse and much closer to the distribution of human-only writing. This suggests that human involvement produces a natural force away from the generic ``AI voice''. Even so, AI-assisted submissions remain more homogeneous than human-only submissions on most embedding-based metrics.
    \item \textbf{Incentives increase diversity with AI usage.} When participants using AI are rewarded for originality, they produce more diverse writing than those rewarded for technical quality. This effect is strongest in Study 2, and directionally consistent (but not statistically significant) in the smaller Study 1. However, they do not achieve the same amount of diversity as humans writing alone. We interpret this finding as \textit{directional}: incentives increase diversity, but the size of the effect likely scales with stronger (or weaker) incentives.
    \item \textbf{Increased diversity through incentives is associated with lower
      adoption of AI text, not disengagement from AI.} Participants rewarded for originality incorporate less AI-generated text verbatim into their final submissions, yet continue to engage with the AI tool. Additional prompting effort does not translate proportionally into diversity: in Study 1, these participants spent up to 50\% more time crafting prompts without a commensurate gain in diversity, while in Study 2 they exerted little additional effort yet achieved markedly higher diversity. Much of the recovered diversity thus appears to come from participants' own edits.
    \item \textbf{Greater diversity comes with a modest trade-off in technical quality.} Submissions under the originality incentive score somewhat lower on technical quality, but within-quality comparisons indicate that the diversity gains among AI-assisted participants cannot be attributed solely to weaker writing.
    \item \textbf{Participants with greater AI experience tend to rely more
      heavily on AI suggestions}, resulting in more homogeneous outputs. Despite
      their familiarity with these tools, they do not appear to adopt more
      effective or novel strategies for achieving originality.
\end{itemize}

More broadly, these findings suggest that human-AI interaction is best viewed as
a strategic, adaptive process, where outcomes reflect not just model
capabilities, but how users allocate effort, attention, and trust in response to
the task environment. Designing AI systems---and incentives around them---thus plays a central role in determining whether co-created artifacts reflect the values we care about.

\revision{We also release a rich, anonymized dataset from both studies, providing a fine-grained record of human--AI co-creation through  interaction transcripts and 5-second snapshots of participants' evolving work.\footnote{\url{https://huggingface.co/datasets/nathanaeljo/ai-cocreation-incentives}} This dataset enables researchers to study not only final outcomes, but also how people interact with, adopt, and revise AI-generated content throughout a session.}
\section{Related Work}
\paragraph{Co-creation and Diversity.}
Our work is closely related to a growing literature on homogenization in generative AI systems. Prior work has shown that AI models, when prompted independently, tend to produce highly similar outputs \cite{wu2024generative, wenger2025we}. Beyond studying AI-generated text in isolation, an increasing body of research examines settings in which humans and AI co-create to varying degrees. For example, \citet{doshi2024generative} show that using AI for creative ideation leads to more homogeneous writing, while \citet{agarwal2025ai} and \citet{padmakumar2023doeswriting} find similar effects when AI is used for text completion. Together, these results suggest that homogenization arises not only from fully AI-generated content, but also from human exposure to AI suggestions. Our work builds on this line of research by studying homogenization in a more natural co-creation setting that allows for iterative prompting, editing, and selective adoption of AI outputs. We also explicitly manipulate incentives for diversity, allowing us to examine how users strategically adjust their interaction with AI in response to the task environment. Most closely related is \citet{kosmyna2025your}, which also studies a relatively unrestricted co-creation setting. However, they focus on studying brain function and cognitive debt rather than, as we do, on how incentives shape user strategies and the diversity of outputs.


\paragraph{AI assistance, creativity, and reliance.}
A large body of work studies how AI assistance shapes human creativity. Much of
the cognitive science and human-computer interaction literature advocates using
generative AI to \emph{enhance} creative processes rather than to replace human
effort \cite{wu2021ai, haase2024human, chompunuch2025ai}. Creativity in these
settings is often operationalized through divergent thinking tasks, where
success is measured by the ability to generate a wide range of novel ideas
\cite{runco2012divergent}, see e.g., Torrance tests \cite{torrance1966torrance}.
Empirically, recent studies paint a mixed picture. \citet{sun2025and} conduct a
field experiment and find that AI assistance is most beneficial for workers who
already engage deeply with their tasks. Other work suggests that AI alone or
human-AI teams can yield short-term gains in novelty or productivity, but may undermine independent creative performance once assistance is removed \cite{kumar2025human, huang2025unlocking, medeiros2025human, sternberg2024not}. Our work complements these findings by adopting a broader notion of creativity that extends beyond ideation to include writing style, narrative structure, and thematic exploration in creative writing.

A recurring concern in this literature is the risk of overreliance on AI to achieve tasks, whether that be creative or otherwise. While human-AI collaboration can improve immediate performance, recent evidence suggests it may reduce intrinsic motivation or engagement when the tool is no longer available \cite{wu2025human}. Related work also documents forms of cognitive offloading, in which users defer critical thinking or decision-making to AI systems \cite{kosmyna2025your, wang2025genai, hou2025role}. Although our study does not examine long-term effects of AI use, we observe clear short-term costs of overreliance: participants who heavily anchor on AI-generated drafts produce less diverse stories.

\paragraph{Incentives in human--AI collaboration.}
Recent work has increasingly emphasized the role of incentives in human--AI decision-making. A central critique is that many studies overlook how strongly incentive structures shape participant behavior and, consequently, study outcomes \cite{kaur2026incentive}. Most related to our work is \citet{holstein2025thinking}, which shows that altering incentives can substantially reduce overreliance on AI in decision-making tasks. To our knowledge, our study is the first to explicitly manipulate incentive structures in a human--AI \emph{co-creation} setting involving generative AI tools.

\section{Experiment Design and Data}

We run a pre-registered, between-subjects experiment in which participants are assigned via block randomization to one of four conditions in a $2\times 2$ factorial design, which we elaborate below. Participants complete either a short story writing task (Study 1) or a product pitch writing task (Study 2). We collect both the final outputs and detailed interaction logs (snapshots of text every 5 seconds as well as the complete transcripts of their AI conversations). The design and analysis plan were pre-registered on OSF.\footnote{\url{https://osf.io/3u8tq/overview?view_only=e1a8044dd5044bc59dd0a7d8f3fca3c2}}\footnote{\url{https://osf.io/xv9sa/overview?view_only=e1568bd84959466689af03c2aa362e7f}} Note that some of the analyses in this paper are exploratory, and we will flag them when relevant.

\subsection{Task and Recruitment}

Participants were instructed to write a short story of 250--350 words (Study 1) or a product pitch of 100--125 words (Study 2) using a browser-based rich-text editor. After revealing the task and instructions (Figures~\ref{fig:instruction_phase1} and~\ref{fig:prompt_phase}), participants were given at most 5 minutes to brainstorm and outline their short story (see Figure~\ref{fig:brainstorm}; only Study 1). This stage gave participants time to scaffold their ideas so that they could begin writing with a concrete plan, which was important for enabling them to complete their stories within the limited time available. After the brainstorming phase, they were taken to a writing window, where they had at most 20 minutes (9 minutes) to write their story for Study 1 (Study 2) (see Figures~\ref{fig:writing_phase} and~\ref{fig:writing_phase2}).
If a participant's story was not within the desired word range after the time limit, they were given an additional 5 minutes (2 minutes) to adhere to the word limit, after which their story was automatically submitted (and accepted). We also imposed minimum times to ensure participants were faithfully exerting effort into the task: 1 minute for brainstorming and 5 minutes for writing (same for Study 2).

We recruited a total of $n=200$ ($n=400$) participants from Prolific for Study 1 (Study 2). Participants received \$5 (\$4) for completing the study, which took on average 25 minutes (12 minutes). This is equivalent to on average \$12 (\$20) per hour of work, though we also include bonuses as part of the incentive structure, see next section.

\textbf{Ensuring attention in task.} We incorporated extensive checks to ensure the fidelity of the online experiment. For example, the interface can only be done in full screen, without the ability to switch tabs or copy paste from outside the current window. Additionally, we implemented inactivity checks by tracking mouse and keyboard movements; after six 30-second periods of inactivity as well as numerous warnings, we discarded the participant's work. We also conducted post-hoc analyses of participants' writing patterns and filtered out abnormally confident sessions with very little backtracking and pauses. Details on other mechanisms to ensure attention in this experiment can be found in Appendix~\ref{appsec:attention}.

\subsection{Treatment Groups}
\revision{In each study,} participants were block-randomized \revision{based on enrollment order} into one of four groups in a $2\times2$
design: (\ai, \self) $\times$ (\orig, \qual), where \ai denotes access to an AI
assistant, \self is the human-only baseline, \orig rewards originality, and
\qual rewards technical quality. Therefore, each group consists
of $n=50$ ($n=100$) participants for Study 1 (Study 2).

\subsubsection{Factor 1: AI Usage}
In the \self conditions, participants wrote entirely on their own. In the \ai
conditions, participants had access to an integrated writing assistant designed
to mimic general-purpose LLMs (see Figure~\ref{fig:writing_phase}). Participants
were required to use the AI tool at least once, though they were not obligated to
use any of the output nor interact with the tool further. However, we emphasized
that participants are ultimately responsible for the text they submit. Note that we did not provide
the AI tool during the brainstorming phase because we wanted to isolate the
effect of AI assistance on writing style and execution more broadly; there
already exists numerous studies investigating the impact of AI on creative
ideation \cite{doshi2024generative, kumar2025human, medeiros2025human}. 

In Study 1, those in
the \ai condition on average spent 18.6 minutes writing, compared to those in \self who took on average 20.6 minutes. We similarly observe that \ai participants spent on average less time than \self for Study 2. Both groups spent the same amount of time brainstorming in Study 1. While participants with AI spent slightly less time writing on average, the difference is modest. One likely reason is that participants were incentivized to produce high-quality or highly original writing, which may have reduced the extent to which they could rely on AI to save time. In addition, using AI required participants to read, evaluate, and revise model outputs, which introduces its own cognitive overhead under a fixed time limit.

\subsubsection{Factor 2: Task Incentives}

Apart from randomizing AI assistance, we also randomized participants into two conditions: a condition incentivizing \textbf{originality (\orig)} and another rewarding \textbf{technical quality (\qual)}. Table~\ref{tab:div_conv_instructions} compares the instructions given in both conditions for Study 1. In \orig, participants \revision{were asked to imagine being in a competitive environment, and} that their bonus is a function of originality and uniqueness. \revision{Specifically, the bonus structure was set up to be \textit{scarce}: the top 2-25\% most original submissions receive substantial tiered bonuses. Meanwhile, }in \qual, participants' bonuses were based \revision{on a rubric that assessed }organization \& technique \revision{for Study 1, and information \& communication quality for Study 2; there was no limit to the number of bonuses given so long as they passed a certain quality threshold. See Appendix~\ref{appsec:prompt_phase} for details on the incentive structure. Thus, the difference between \orig and \qual should be interpreted as a comparison between two incentive regimes: a competitive, rank-based setting in which scarce rewards make differentiation valuable, and a noncompetitive, criterion-based setting in which rewards depend on technical quality.} We refer to each group using hyphen notation, e.g.,
\aiorig.


\begin{table}[H]
    \centering
    \begin{tabular}{>{\centering\arraybackslash}p{8cm}|>{\centering\arraybackslash}p{8cm}}
    \toprule
      Originality (\orig)   & Technical Quality (\qual) \\
      \midrule
       ``You are participating in a short story competition. There are thousands of submissions, so \textbf{your goal is to stand out} as much as possible. Find your voice and be as creative as possible! The short story should be around 250--350 words.'' 
       
    \textit{Meta instruction:} ``Your bonus will be determined based on originality and uniqueness.''  & ``You are starting an Intro to Writing class. Your first assignment is to create a 250--350 word short story. \textbf{Your goal is to get an A by submitting a high-quality piece of work}.''  
    
    \textit{Meta instruction:} ``Your bonus will be determined based on the grade you receive.''

    \textit{Grading is based on a rubric, see Figure~\ref{fig:prompt_phase}.}
    \\
    \bottomrule
    \end{tabular}
    \caption{Instructions given to the originality (\orig) and technical quality (\qual) conditions.}
    \label{tab:div_conv_instructions}
\end{table}

\subsection{Datasets and Preprocessing}

We took snapshots of the plain text editor every 5 seconds of activity, resulting in an incredibly rich temporal dataset of each participant's progress throughout the task. We also saved the entire conversation transcript for participants who had access to an AI tool. This dataset forms the basis for the bulk of our analyses.

In addition to the temporal dataset, we also collected information about participants' characteristics. After the writing task, participants completed a short survey, which asked questions about their general experience with generative AI tools, and their experience with these tools for writing tasks specifically (both on a 5-point Likert scale from ``Not at all familiar'' to ``Extremely familiar''). We also obtained non-identifying demographic information of participants, provided by Prolific. This includes self-reported ethnicity, age, sex, education level, and employment status.

\paragraph{AI Usage.} In Study 1, those with access to an AI tool ($n=100$) on average took 3.57 conversation turns (3.23 standard deviation), resulting in a total of 357 queries. Many of them were \textbf{editing} requests (38\%; condensing, restructuring, partial edits), then \textbf{drafting} (30\%; asking for full stories), \textbf{creative consulting} (14\%; asking opinions or help with story), \textbf{mechanical assistance} (3\%; spelling or grammar checks), and others (15\%). \revision{In Study 2, the requests followed a similar distribution, except that editing requests were relatively rare; we believe this is due to the constrained nature of Study 2, which does not leave much space for changes in phrasing. On the other hand, Study 1 requires users to generate short stories where phrasing sentences may have been more difficult.} Overall, 91\% \revision{(83\%)} of participants requested at least one complete draft \revision{for Study 1 (Study 2)}. This does not imply that users adopted these drafts verbatim, but it shows that AI-generated drafts were a central part of the writing process. Motivated by this pattern, we manually annotated draft suggestions and later analyze how participants revise, edit, and incorporate them (Section~\ref{sec:strategies}). See Appendix~\ref{appsec:addl_results2} for a more detailed breakdown of AI usage in Study 2.




\section{Diversity of Stories}

A central question that we ask is how stories become more or less diverse when participants are given access to an AI tool, under various incentive structures. All of the analyses in this section were pre-registered except for the results in Figure~\ref{fig:effect_heatmaps} (right). 

\subsection{Methods}
\subsubsection{Diversity/Similarity Metrics}
Diversity of natural language text is, in general, hard to measure since natural language is incredibly high-dimensional. Historically, the NLP literature has focused on measuring \textit{lexical diversity} through n-gram or bit-level analyses \cite{lin2004rouge,shaib2024standardizing}. Recent works have started to use more advanced techniques, such as passing through some embedding models and calculating the similarity of these vectors \cite{doshi2024generative, wu2024generative, guo2025benchmarking}. Unfortunately, embedding models do not provide interpretable dimensions of natural language, and there are millions of embedding models to choose from -- leading to a selection problem. One can generally view embeddings as capturing a combination of \textit{semantic meaning} and \textit{style}. There is currently no standard and there are many perspectives from the respective works we cited. Our view is that diversity in natural language is a latent construct; instead of taking a particular view of diversity, we pre-specify a suite of diversity metrics, each providing correlated aspects of text diversity, see Table~\ref{tab:similarity_metrics}.

\begin{table}[ht]
    \centering
    \begin{threeparttable}
    \begin{tabularx}{\linewidth}{l>{\raggedright\arraybackslash}X}
    \toprule
     Name &  Description \\
     \midrule
     \rowcolor{lightgray}\multicolumn{2}{l}{\textbf{Embeddings} (using cosine similarity)} \\
    all-MiniLM-L12-v2\tnote{a} & One of the most popular lightweight embedding models \\
    all-mpnet-base-v2\tnote{b} & One of the most popular lightweight embedding models \\
    Qwen3-Embedding-0.6B \cite{zhang2025qwen3} & Lightweight ($<$1B) but high-performing on embedding benchmarks\tnote{c} \\
    embeddinggemma-300m \cite{vera2025embeddinggemma} & Lightweight ($<$1B) but high-performing on embedding benchmarks\tnote{c}\\
    Style-Embedding \cite{wegmann2022author} & Intended to capture style more than content (for author attribution) \\
    STAR \cite{huertas2024understanding} & Intended to capture style in social media text \\
    byt5 \cite{xue2022byt5} & Breaks into byte-level instead of token-level data (might capture style) \\
    \midrule
    \rowcolor{lightgray}\multicolumn{2}{l}{\textbf{Lexical-based}} \\
    Compression ratio \cite{shaib2024standardizing} &  Size of original file $\div$ size of compressed file (e.g., via gZip)\\
    BLEU \cite{papineni2002bleu} & Geometric mean of $n$-gram precision between two texts \\
    Rouge-L \cite{lin2004rouge} & F-measure of precision \& recall based on the longest common subsequence between two texts \\
    $n$-gram diversity score$^\star$ \cite{li2023contrastive, padmakumar2023doeswriting} & $\sum_{n=1}^N ($\# unique $n$-grams $\div$ \# all $n$-grams$)$\\
    \bottomrule
    \end{tabularx}
    
    \begin{tablenotes}
    \footnotesize
    \item[a] \url{https://huggingface.co/sentence-transformers/all-MiniLM-L12-v2}
    \item[b] \url{https://huggingface.co/sentence-transformers/all-mpnet-base-v2}
    \item[c] \url{https://huggingface.co/spaces/mteb/leaderboard}
    \end{tablenotes}
    
    \end{threeparttable}
    \caption{Table of similarity metrics used, in that higher similarity means reduced diversity. $^\star$ $n$-gram diversity score is the only diversity measure, where a higher score means more diversity.}
    \label{tab:similarity_metrics}
\end{table}

For each participant $i$ in cell $g\in\{\ai,\self\}\times\{\orig,\qual\}$, let $X_i$ be their text.
Let $S(X_i,X_j)\in\mathbb{R}_+$ be a similarity function.
Define the participant-level outcome as the
leave-one-out average similarity within the same cell $g$:
\[
Y_i^{(g)}\;:=\;\frac{1}{\,n_g-1\,}\sum_{\substack{j\in g\\ j\neq i}} S(X_i,X_j).
\]

All four $n$-gram/token based measures are similarity functions (in that higher similarity means reduced diversity), except for $n$-gram diversity score (NGDS). For the embedding models, we will use the cosine similarity of the embedding vector as $S$. Formally, if $f$ is an embedding model, then $S(X_i, X_j) := \cos (f(X_i), f(X_j))$.

\subsubsection{Hypothesis Tests}
Fix a similarity metric $D$. We construct the following hypotheses:

\paragraph{Hypothesis 1 (Less diversity under AI).}
\[
H_0:\ D_{\mathrm{AI}, j} - D_{\mathrm{SELF}, j} \leq 0
\qquad\text{vs}\qquad
H_1:\ D_{\mathrm{AI}, j} - D_{\mathrm{SELF}, j} > 0, \quad \forall j \in \{\orig, \qual\}
\]

\paragraph{Hypothesis 2 (More diversity when originality is rewarded).}
\[
H_0:\ D_{j, \mathrm{T}} - D_{j, \mathrm{O}} \leq 0
\qquad\text{vs}\qquad
H_1:\ D_{j, \mathrm{T}} - D_{j, \mathrm{O}} > 0, \quad \forall j \in \{\ai,\self\}
\]

We construct the reverse for diversity metrics, as is the case for NGDS. From the data, we get an estimate of $D_g$ by taking $\bar{Y}^{(g)} = \frac{1}{n_g} \sum_{i \in g} Y_i^{(g)}$. We then conduct a two-sample difference in means test for each metric in Table~\ref{tab:similarity_metrics}. \revision{Because each participant-level similarity score is computed against all other participants in the same group, these scores are not independent and standard two-sample $t$-tests would underestimate uncertainty. We thus treat the group mean as a second-order U-statistic and estimate the variance of the difference between groups using a leave-one-participant-out jackknife. We then form a normal-approximation test statistic using the resulting jackknife standard error, see \cite{arvesen1969jackknifing}.}

Since these metrics are statistically correlated reflections of the same hypothesis, we do not adjust for multiple hypothesis testing. Instead, we will interpret the results as evidence about the same latent phenomenon (writing diversity), emphasizing the overall pattern across metrics rather than the significance of any single $p$-value. However, even with Benjamini-Hochberg adjustment for multiple hypothesis testing \cite{benjamini1995controlling}, the results remain significant.

\begin{figure*}
    \centering
    \includegraphics[width=\linewidth]{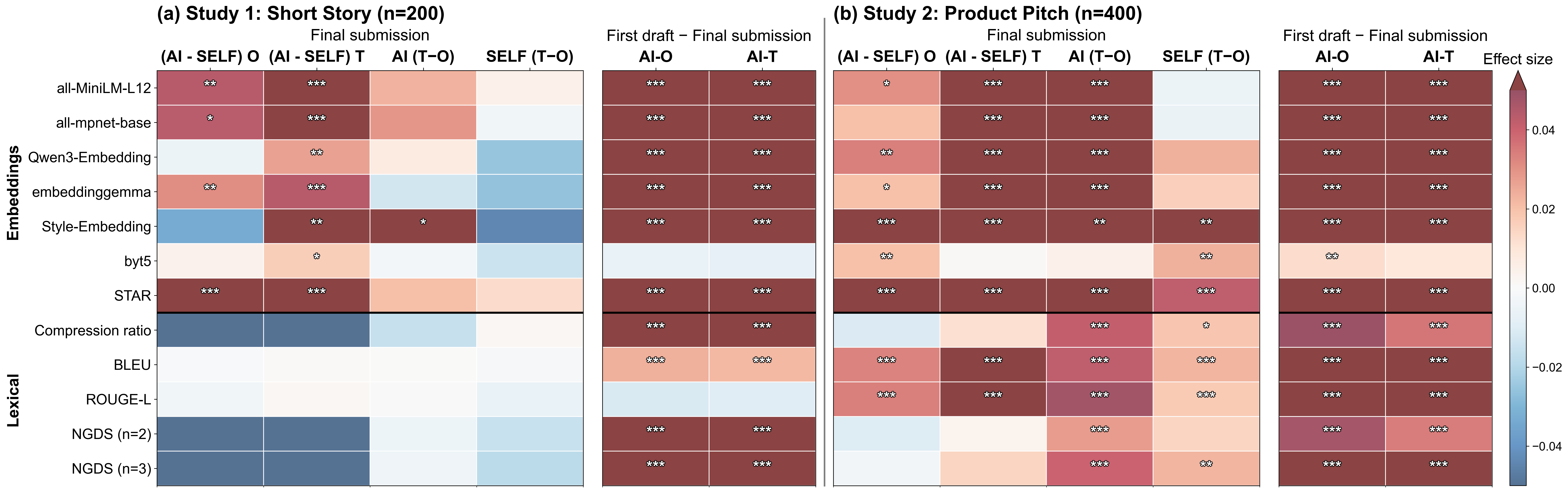}
\caption{\revision{Differences in U-statistic estimates of average pairwise similarity, with uncertainty and one-sided significance tests computed using a leave-one-participant-out jackknife normal approximation:} $^{*}$ $p<0.1$, $^{**}$ $p<0.05$, $^{***}$ $p<0.01$. \revision{(a) and (b) show results for Study 1 and Study 2, respectively. For each subfigure,} results are shown for comparisons between final submissions (left) and between the first valid AI-generated draft and the corresponding final submission (right). A higher positive value means that the first
group in the difference-in-means test is more homogeneous. All rows are
similarity metrics, except for $n$-gram diversity score (NGDS), which is a
diversity metric. As such, we flip the comparison for NGDS so that the effect
sizes are directionally consistent. See Appendix
Tables~\ref{tab:effects_combined} and~\ref{tab:effects_combined2} for specific effect sizes and multiple
hypothesis correction, which \revision{mostly} remain statistically significant.}
    \label{fig:effect_heatmaps}
\end{figure*}

\subsection{Results}
Our first analysis is the pre-registered, difference-in-means test that quantifies the effect of AI and incentives for originality on participants' final submissions (Figure~\ref{fig:effect_heatmaps}, left). In Figure~\ref{fig:effect_heatmaps}
(right), we additionally display the difference in means between the first full draft
that the AI produces versus the final submission. This difference gives a
glimpse on how participants change and edit substantially before being satisfied
with their endproduct.

\paragraph{On lexical-based measures.} First, we note that the \revision{lexical-based} metrics start off as having large effects at the first-draft stage, but this effect virtually disappears at final submission \revision{in Study 1}. In contrast, embedding-based \revision{effect sizes} remain statistically significant. This pattern suggests that when participants edit \revision{their short stories}, they manage to diversify wording and phrasing, but deeper narrative structure and style remain more homogeneous when AI assistance is involved. \revision{This effect is much less pronounced in Study 2, where we do observe statistically significant differences upon final submission for lexical-based metrics. We believe this difference stems from the more constrained nature of Study 2; participants only had at most 125 words to write a product pitch with pre-specified features, leaving less room to diversify at the token level compared to Study 1, which was a free-for-all short story task. Additionally, in Study 2, \aiqual were partially assessed by their ability to communicate product features, which likely nudged participants to use the product features verbatim in their pitch.} We are generally more interested in characterizing higher-order features of style and narrative, and thus we will restrict our attention to the embedding-based metrics.\footnote{Note that because we conduct one-sided hypothesis tests, we cannot rule out the possibility that some metrics exhibit a statistically significant effect in the opposite direction when the estimand is negative.}

\paragraph{Hypothesis 1.} For the first hypothesis\revision{---AI homogenizes outputs---}we find \textbf{strong evidence against the null hypothesis in both incentive conditions} (\orig and \qual). Across most embedding-based metrics \revision{(19 of 28 tests across both Studies, or 67.9\%)}, AI-assisted outputs are significantly \revision{($p < 0.05$)} more similar than human-only outputs at final submission. We also find that the first drafts provided by AI are much more similar than the AI-assisted final submissions, on average being \revision{43\% (36\%) more similar across embedding (lexical) metrics}. This result indicates that participants often do not like the first AI draft and exert significant effort to express their own voice. Despite this additional effort, however, the AI suggestions still have a lingering homogenizing effect at final submission relative to the human-only condition. \revision{This finding is consistent with prior work \cite{anderson2024homogenization, doshi2024generative,padmakumar2023doeswriting, agarwal2025ai}, but we additionally find that the opportunity to interact in free-form allows participants to break from the AI mold to a limit.}

\paragraph{Hypothesis 2.} For the second hypothesis, \revision{the evidence is strongest among AI-assisted participants. In Study 2, \aiorig submissions are significantly more diverse than \aiqual submissions on 6 of 7 (85.7\%) embedding metrics. In Study 1, estimates are directionally correct for most metrics but not statistically significant, likely due to the smaller sample sizes. Overall, this finding} indicates that the incentive to be original
was strong enough to induce those with AI access to make attempts to differentiate themselves. We will unpack the strategies participants take to
achieve diversity in Section~\ref{sec:strategies}. \revision{Among \self participants, results are mixed: there is little difference between \selforig and \selfqual in Study 1, while Study 2 shows significant effects on 3 of 7 (42.9\%) embedding metrics. We suspect that this difference across studies stems from the nature of the tasks: Study 1 involves a longer, more open-ended writing task that naturally allows participants to produce diverse outputs even without explicit incentives, whereas the more constrained task in Study 2 leaves greater room for incentives to influence diversity.}

\revision{Given these qualitative differences between \self and \ai, we test directly whether the incentive effect differs with AI access through an interaction test (Appendix Table~\ref{tab:interaction}), though we emphasize that this analysis is purely exploratory. In Study 2, the interaction is positive for 6 of 7 embedding metrics (85.7\%) and significant for 3 (42.9\%, $p< 0.05$), indicating a larger incentive effect among AI-assisted participants. In contrast, Study 1 estimates are directionally consistent but not significant, again likely due to the smaller sample size. We also report a scale-free measure of the difference in means tests, see Appendix~\ref{app:effect_size_interpretability}.}

\section{Characterizing Writing Trajectories}

In the previous section, we confirmed our hypotheses regarding the effects of AI assistance and originality incentives. We also provided an initial look at how stories evolve from the first AI draft to the final submission. In this section, we examine these trajectories in greater depth, asking \emph{when} and \emph{in what ways} stories shift throughout the session. From this point onward, our analyses are exploratory.

\subsection{Methods}\label{sec:characterizing_stories}

\paragraph{Diversity through time.} 
We apply the same similarity metrics from Table~\ref{tab:similarity_metrics},
now evaluated at fixed time intervals. Taking $t=0$ as the start of the writing
session, we use snapshots of each participant’s text at 30-second intervals until the session concludes. While we have data for every 5 seconds, we believe 30 seconds is sufficient granularity to report.
If a participant finishes before some time $t=T$, we use
their final submission for all subsequent time points. To ensure that embeddings
meaningfully capture story content, we restrict attention to text containing at
least 200 (75) words for \textbf{Study 1} (\textbf{Study 2}). Shorter (incomplete) drafts may yield unstable or biased
embedding representations.


\paragraph{Qualitative structure of stories.} 
Beyond measuring diversity through time, we also ask whether different groups
produce qualitatively distinct types of stories. For example, even if
AI-assisted stories are less diverse, they may still occupy the same overall
semantic space as \self stories---or they may form a distinct cluster. To investigate this, we embed all stories and perform principal component analysis (PCA) on the embedding vectors. PCA extracts the principal axes of variation, allowing us to visualize whether stories cluster differently across conditions and through time.


\begin{figure}
    \centering
    \includegraphics[width=0.95\linewidth]{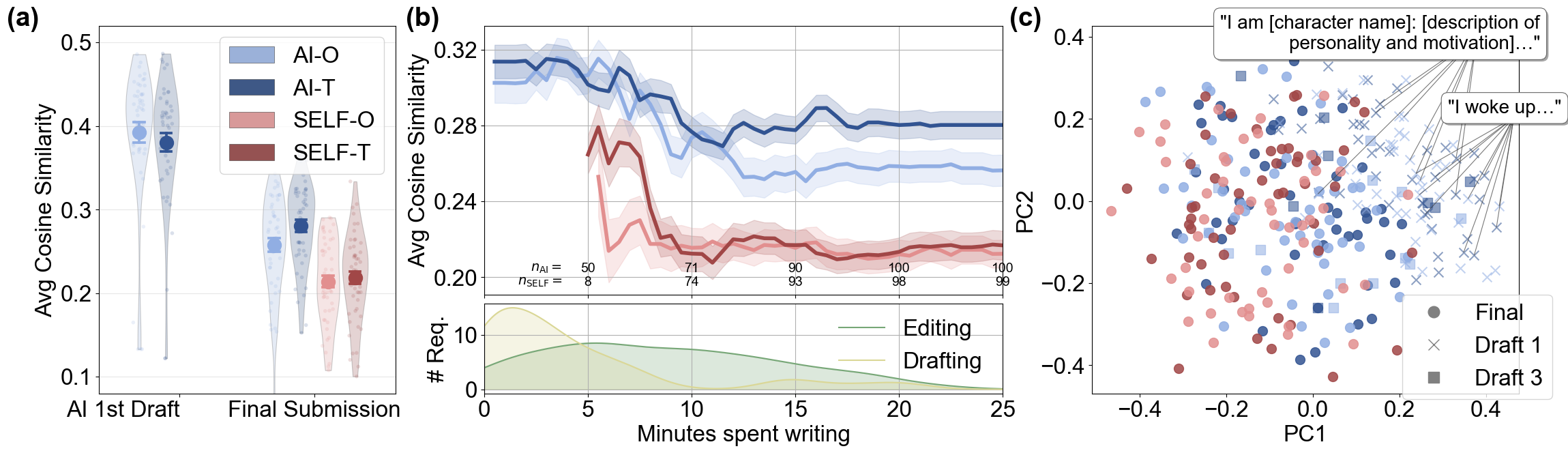}
    \caption{(a) Average cosine similarity of embeddings across randomized groups, for both the final submission and the first full draft the AI produces (when available). (b) [Top] Average cosine similarity of embeddings across randomized groups through time spent in the session, along with sample sizes at 5-minute intervals. One sample was not included in this analysis because it failed to reach the 200-word threshold. We only include text that has at least 200 words for every timestep to ensure that the embeddings capture enough information about the story. [Bottom] Number of requests over time spent in the session. (c) First and second principal component of the embeddings across groups, for both the drafts that AI produces (when available) and the final submissions. All plots use the embedding model \texttt{all-MiniLM-L12-v2}. See Appendix~\ref{appsec:trajectories} for other metrics as robustness checks. 
    }
    \label{fig:outcome_level}
\end{figure}

\subsection{Results}
We now dive deeper into participants' writing trajectories, analyzing \emph{when} and \emph{in what ways} stories shift throughout the session. For clarity, we focus on results using the \texttt{all-MiniLM-L12-v2} embedding model for Study 1, though the qualitative patterns hold across alternative similarity measures (see Appendix~\ref{appsec:trajectories}).

\paragraph{Diversity through time.} We find that diversity increases steadily throughout the session for AI-assisted
participants (Figure~\ref{fig:outcome_level}(b)), despite most beginning from
highly homogeneous AI drafts. \revision{This recovery may reflect participants increasingly expressing their own voice through editing and iteration. However, estimates at earlier time points may also be affected by sample selection: because we restrict the analysis to stories with at least 200 words at each time point, participants who reach this threshold earlier may differ systematically from those who reach it later. By the 10-minute mark, however, most AI-assisted participants ($n=71$) meet this threshold, substantially reducing concerns that subsequent trends are driven by changes in sample composition. Importantly, \aiorig and \aiqual begin to diverge around 12 minutes into the session,} at a point where enough
participants shift from asking for drafting suggestions to actively editing with
or without AI. The timing suggests that \revision{our incentive structure may have shaped} how
participants modify and incorporate AI outputs, rather than affecting the
content of the initial AI suggestions themselves. We examine these behavioral
strategies in greater detail in Section~\ref{sec:strategies}.

\paragraph{Qualitative structure of stories.} Apart from reduced diversity, we also find that the AI-generated stories occupy a markedly different distribution to human stories\revision{, which adds to the evidence found in \cite{abdulhai2026llms}}. Figure~\ref{fig:outcome_level}(c) visualizes the top two principal components of story embeddings across all stories. AI-generated drafts form a distinct cluster, separated from both AI-edited final submissions and human submissions. Qualitatively, many AI drafts share similar structures; for example, starting the story with ``I woke up in\dots'', or introducing the main character with ``I am [name], [description of character's disposition and motivations]''. A subset of participants repeatedly prompted the AI for full drafts ($n=23$ requested at least three drafts). Across successive drafts, \revision{the AI-assisted writing} moves closer to the distribution of final human submissions. This shift \revision{is achieved through a combination of prompting AI and editing the text themselves; we will partially disentangle these mechanisms in the next section.}

\section{Differential Strategies in Using AI}\label{sec:strategies}
Apart from characterizing the trajectories of participants' stories, our data is uniquely positioned to answer the question of \textit{how} participants used AI to produce their short stories. In other words, how do participants' strategies in using AI differ so that they achieve greater diversity in the \aiorig condition? Again, we note that the following analyses are exploratory and were not pre-registered. In particular, adoption (defined below) is a post-treatment measure, and thus this analysis is purely descriptive.

\subsection{Methods}

\begin{figure}
    \centering
    \includegraphics[width=0.75\linewidth]{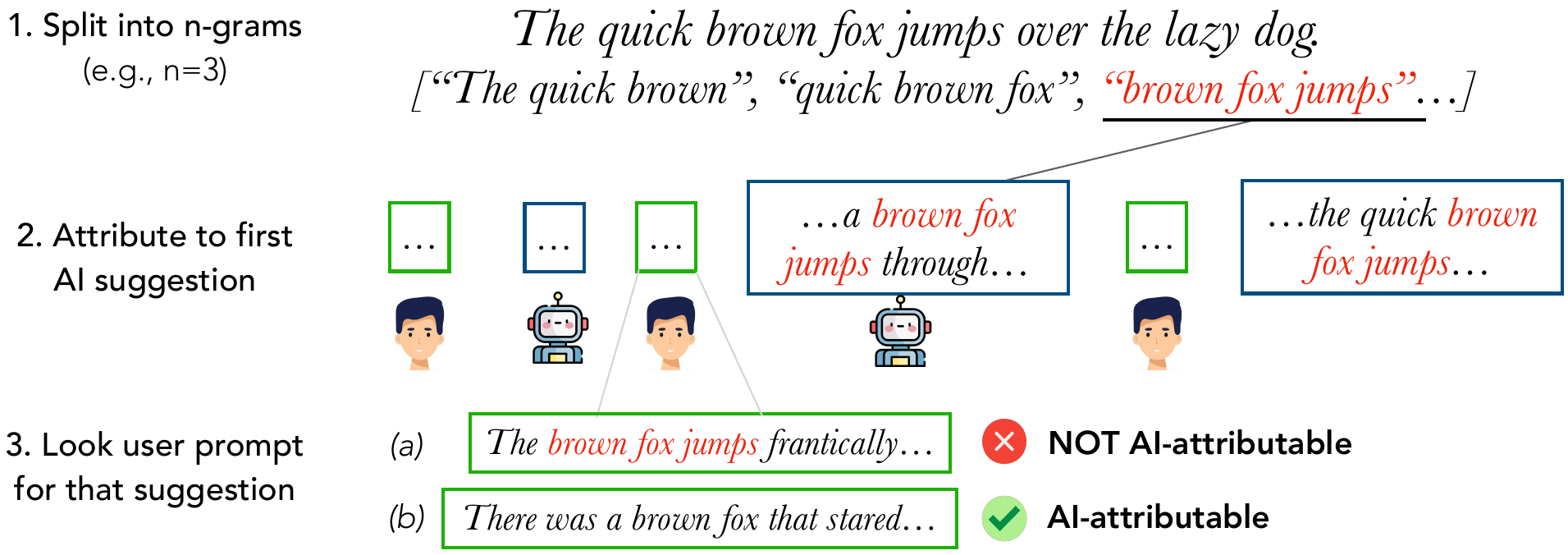}
    \caption{Example of the AI attribution procedure to construct the adoption metric.}
    \label{fig:example_attribution}
\end{figure}

\paragraph{Measuring adoption of AI suggestions.} 

To summarize the complex and heterogeneous trajectories participants take throughout the session, we construct a single metric that captures adoption of AI suggestions, denoted $A$. As we show later, this measure is highly informative of participants’ overall strategies.

Our approach is based on lexical attribution. First, we extract $n$-grams\footnote{A sequence of $n$ consecutive
natural language tokens, for some fixed $n$.} for each final submission. For each $n$-gram, we check whether it appears in any AI-generated output within the interaction transcript. If a match is not found, the $n$-gram is \textit{not AI-attributable}. If a match is found, we find the earliest AI suggestion in which the $n$-gram appears and check the corresponding user prompt. If the $n$-gram also appears in the corresponding prompt, then the $n$-gram is \textit{not AI-attributable} because it was originally authored by the user. Otherwise, it is AI-attributable. See Figure~\ref{fig:example_attribution} for a
representative illustration of AI attribution. Adoption $A$ is then defined as:
$$A = \frac{\text{\# AI-attributable $n$-grams}}{\text{\# total $n$-grams}}.$$

This metric is particularly useful because it accommodates the wide range of strategies participants use when interacting with AI---whether drafting entire stories, requesting partial edits, or asking for specific phrases. In Appendix~\ref{appsec:adoption_details}, we conduct multiple robustness checks: \textit{(1)} we test different values of $n$; \textit{(2)} we also construct an embedding-based attribution measure that better accounts for semantic similarity and paraphrasing, whereas the $n$-gram approach cannot. \revision{In Study 1, we use $n=10$, and} all adoption metrics are highly correlated ($\rho \geq 0.94$; Appendix Figure~\ref{fig:adoption_robustness}), suggesting that our conclusions are not sensitive to the specific measure of adoption. \revision{In Study 2, we use $n=3$ rather than $n=10$ because of the much shorter text length. Using $n=10$ on 100-125 words would span nearly 10\% of the text, meaning that minor changes would cause substantially reused AI text to go undetected. A 3-gram covers a comparable share of a Study 2 pitch (about 3\%) as a 10-gram does of a Study 1 story. Adoption measures remain strongly correlated across choices of $n$ and with the embedding-based measure in Study 2 ($\rho \geq 0.89$), and our main conclusions hold under $n=10$ (Appendix Figure~\ref{fig:reliance_time_requests2}).}

\subsection{Results}

We first examine how participants differed in their adoption of AI-generated text, and then show how these strategic differences account for the diversity patterns observed in the previous section.


\begin{figure*}
    \centering
    \includegraphics[width=0.9\linewidth]{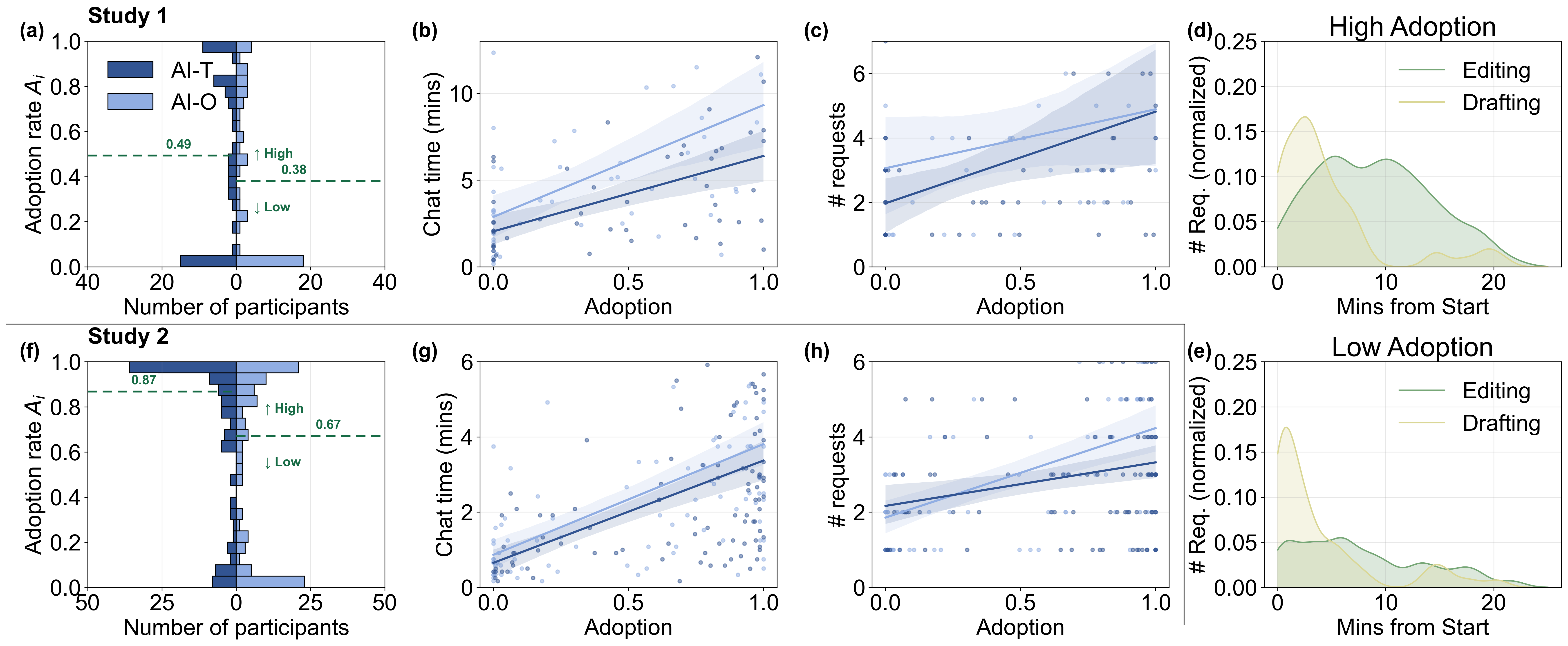}
    \caption{\revision{(a) Histogram of adoption score $A_i$ for both the \aiqual
    and \aiorig groups in Study 1, with median indicated by dashed green line. (b) Time spent writing AI prompts by adoption
  and group. (c) Number of AI requests by adoption and group. (d) and (e) Distribution of editing and drafting requests normalized between high and low adopters. (f, g, h) are similar plots to (a, b, c) for Study 2. }}
    \label{fig:reliance}
\end{figure*}

\paragraph{Adoption as a driver for diversity.} We find that adoption captures the behavioral differences associated with the diversity patterns from the previous section. First, adoption is highly bimodal \revision{for both studies}: most participants either anchor heavily on AI suggestions or make minimal use of them (Figure~\ref{fig:reliance}(a\revision{, f)}). However, relative to \aiqual, the \aiorig condition exhibits a clear shift toward lower adoption. More \aiorig participants fall in the \textit{low-adoption} range ($A < \text{median}(A)$), and fewer in the \textit{high-adoption} range ($A \geq \text{median}(A)$). This pattern suggests that incentives to diversify encourage participants to distance their final submissions from AI-generated text, possibly because they have priors that AI outputs are generic. 

\begin{figure*}
    \centering
    \includegraphics[width=\linewidth]{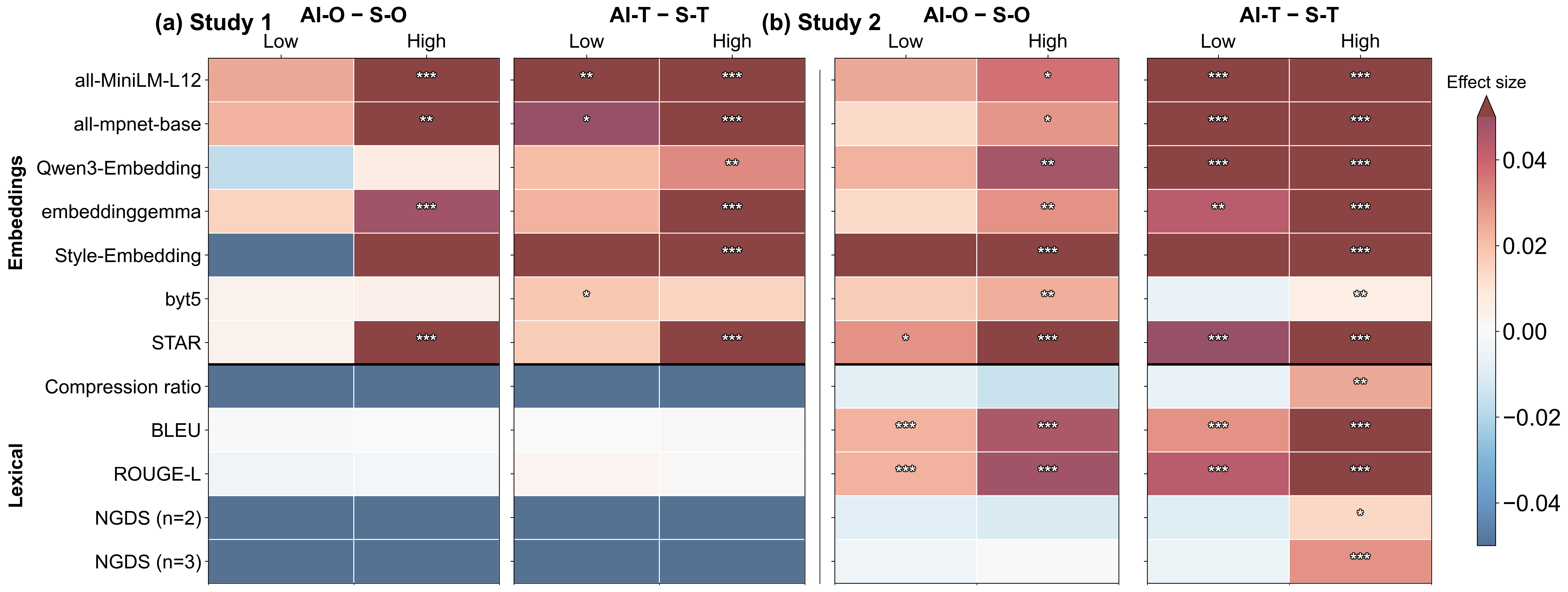}
    \caption{\revision{Differences in U-statistic estimates of average pairwise similarity, with uncertainty and one-sided significance tests computed using a leave-one-participant-out jackknife normal approximation: $^{*}$ $p<0.1$, $^{**}$ $p<0.05$, $^{***}$ $p<0.01$. Results are shown for Study 1 (a) and Study 2 (b), over the final submissions stratified by low and high adoption of AI (lower than median and higher than median, respectively)
among the AI-assisted group. See Appendix Tables~\ref{tab:effects_low_high} and~\ref{tab:effects_low_high_part2} for
specific effect sizes.}}
    \label{fig:effects_low_high}
\end{figure*}

This modest shift in adoption \revision{is consistent with} the gap in diversity between \aiorig and \aiqual. In Figure~\ref{fig:effects_low_high}, we re-estimate the difference-in-means tests after stratifying AI-assisted participants into low- and high-adoption groups. Among low-adoption participants, final submissions are \revision{generally} indistinguishable in diversity from those in the \self condition, perhaps because these participants largely author their stories independently. In contrast, high-adoption participants produce substantially more similar outputs compared to \self. \revision{We note that this pattern is less pronounced in Study 2 (Figure~\ref{fig:effects_low_high}(b)), where low-adoption \aiqual participants remain more homogeneous than \selfqual. This partly reflects our median-based definition of adoption groups: because adoption is much higher overall in Study 2 (median $A$ of 0.67 in \aiorig and 0.87 in \aiqual), many participants classified as ``low’’ adopters still incorporated substantial AI-generated text.}

\paragraph{Low-adoption users still engage with AI.} We find that low adoption does not imply disengagement from the AI tool. In
fact, participants in the low-adoption group still interacted with the AI
multiple times, averaging just under three prompts per session
(Figure~\ref{fig:reliance}(d)) and spending roughly three minutes composing
these prompts (Figure~\ref{fig:reliance}(b)) \revision{in Study 1}. These users are more likely to
request full drafts than high-adoption users, but make far fewer subsequent
attempts to engage with the AI (Figure~\ref{fig:reliance}(d, e)). \revision{We see similar patterns for Study 2 (Figure~\ref{fig:reliance}(g, h).} This pattern
suggests that low-adoption participants made genuine attempts to use the AI, but
ultimately chose not to incorporate its suggestions---perhaps due to strong
preferences over narrative direction or voice, or skepticism that the AI could
further improve their story.

\paragraph{Participants exert more effort in using AI when incentivized to be original, with mixed results.} \revision{Additional effort in prompting may not have translated to any meaningful change in diversity by our measure. In Study 2, participants in \aiorig barely exerted more effort compared to \aiqual (Figure~\ref{fig:reliance}(g, h)), yet \aiorig achieved markedly higher diversity upon submission (Figure~\ref{fig:effect_heatmaps}(b)). This pattern indicates that much of the diversity recovered comes from participants' edits beyond their interactions with AI. In Study 1, }participants in \aiorig exerted more effort in prompting AI: in particular, they spent up to 50\% more time crafting prompts than participants in \aiqual (Figure~\ref{fig:reliance}(b)). \revision{Yet this additional deliberation does not result in a proportional increase in diversity upon submission (Figure~\ref{fig:effect_heatmaps}(a)). While crafting prompts may have helped in the margin, it is likely that the largest gains come from participants' own writing and editing, as we have seen in Study 2.}

\paragraph{Users with more AI experience tend to anchor more on AI suggestions.}  Finally, we find that users with more AI experience tend to anchor more on AI suggestions. Participants in the high-adoption group report substantially greater AI experience than those in the low-adoption group (e.g., 1.84 vs.\ 1.26 for AI writing experience in Study 1\revision{, and similarly for Study 2)}. See Appendix Tables~\ref{tab:summary_2col} and~\ref{tab:summary_2col_part2} for a detailed comparison. Unsurprisingly, participants with more AI experience tend to prompt more, spend more time interacting with the tool, and adopt similar prompting strategies to those in the high-adoption group. In particular, they are more likely to iterate with the AI by requesting partial edits throughout the session (Figure~\ref{fig:reliance}(d)). This pattern suggests that greater familiarity with AI is associated with increased trust in, and reliance on, AI tools. In the context of our task,  this tendency \revision{contributes to} lower diversity. Note, however, that we cannot rule out the possibility that these more experienced users are using AI for some other goal that our metrics do not capture. A qualitative review of the AI prompts reveals no substantial differences in sophistication across experience levels.

\revision{
\section{Quality of Stories, and the Diversity-Quality Trade-Off}

Thus far, we have focused on how the originality incentive \orig affects the diversity of writing. Greater differentiation, however, may come at the expense of \textit{technical quality}---the dimension explicitly emphasized by the \qual condition. We now investigate this trade-off that might have arisen.

\subsection{Methods}\label{sec:quality_methods}
\paragraph{Measuring technical quality.} To measure the technical quality of each final submission in both studies, we use GPT-6-Luna as an LLM judge. We evaluate stories along the same two dimensions presented to participants in the \qual condition: \textit{organization} and \textit{technique} in Study 1, and \textit{communication quality} and \textit{information quality} in Study 2. Each dimension is scored from 1--4 using the original rubric, and we define overall technical quality as the sum of the two scores.

In order to assess the internal validity of this procedure, for Study 1, we independently evaluate each story three times and take the median score across runs for each dimension. The ratings exhibit good within-LLM reliability across runs (ICC(2,1) $=0.75$ for organization, $0.83$ for technique, and $0.85$ for total quality), with all three runs assigning identical organization and technique scores to approximately three-quarters of submissions. As an external validity check, an independent human annotator applies the same rubric to a subset of 40 stories in Study 1, and we compare the resulting human and LLM ratings. The human annotator was only exposed to the raw text. Appendix Table~\ref{tab:quality_validation} reports the full reliability and validation results, and Appendix~\ref{appsec:quality_details} provides additional details on the LLM-as-judge implementation.

\subsection{Results}
Figures~\ref{fig:quality}(a, c) show the distribution of technical-quality scores obtained using our LLM-as-judge procedure for Studies 1 and 2, respectively. See Section~\ref{sec:quality_methods} and Appendices~\ref{appsec:quality_details} and~\ref{appsec:quality_analysis} for implementation details and validation. We emphasize, however, that this quality measure is affected by the incentive manipulation, and thus this analysis is purely descriptive.

Across both studies, technical quality is higher under \qual than under \orig, indicating an aggregate trade-off between technical quality and differentiation. The magnitude of this difference is modest, however, and submissions under both \aiorig and \selforig remain relatively high in technical quality. We also find that AI assistance increases technical quality across incentive conditions, consistent with AI being particularly effective at producing conventionally organized and polished writing.

The lower average technical quality under \orig, however, does not by itself explain the greater diversity observed. To examine this alternative explanation, we conduct a within-quality analysis by recomputing the mean pairwise similarity within each experimental group \textbf{and} technical quality score. Each point then compares the diversity of stories that received the same quality score, rather than differences arising from the groups' \textit{overall} quality distributions. Figure~\ref{fig:quality}(b) and (d) report these conditional similarity estimates. 

In Study 1, among AI-assisted participants, \aiorig submissions remain more diverse than \aiqual submissions conditional on scores 5 and 7. These strata also happen to be the most highly represented, i.e., the largest groups. We observe even stronger evidence in Study 2, where all scores from 5-8 have significant differences in similarity between the two condition. This suggests that the modest increase in diversity under \aiorig cannot be attributed solely to the production of less organized or technically weaker writing. Note that we do not display results for the \self condition  for brevity, and since there is little aggregate incentive effect to begin with, see Figure~\ref{fig:effect_heatmaps}.

\begin{figure*}
    \centering
    \includegraphics[width=\linewidth]{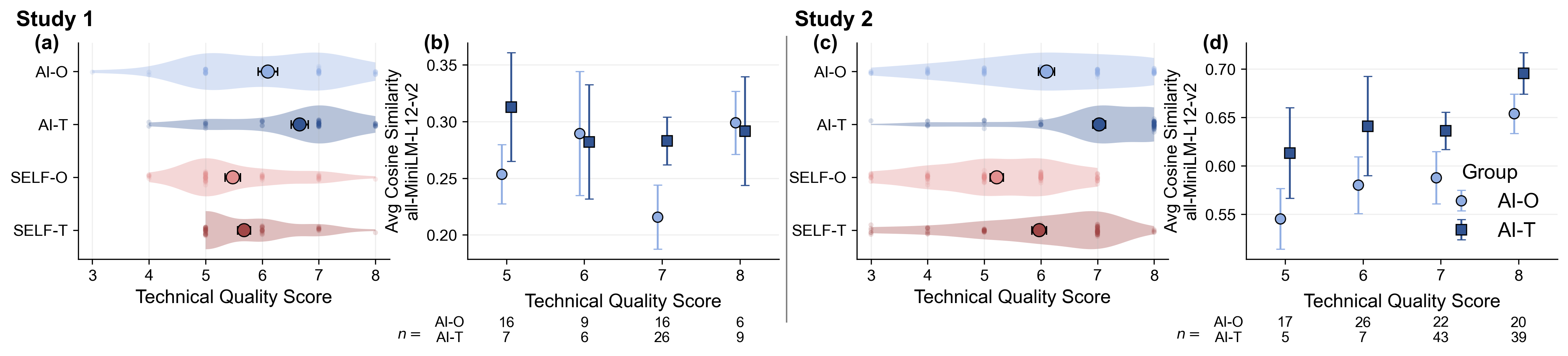}
    \caption{\revision{(a) Distribution of technical quality score across the four conditions for Study 1, using an LLM-as-judge procedure. See Section~\ref{sec:quality_methods} and Appendix~\ref{appsec:quality_details} for implementation details. (b) Mean and standard error of average cosine similarity (using \texttt{all-MiniLM-L12-v2}) for \aiorig and \aiqual, disaggregated by quality score, using a U-statistic and jackknife normal approximation. Similarities are re-calculated relative to both the incentive condition \textbf{and} the quality score. (c) and (d) are similar plots for Study 2.}}
\label{fig:quality}
\end{figure*}
}

\section{User Attitudes About AI Use}\label{sec:attitudes_ai}

We now describe users' self-reported strategies in using AI to complete their tasks.

\paragraph{Participants felt agency and used AI only as a tool.} Most participants treated the AI as a tool to generate an initial draft or structure, while viewing themselves as the agent who refines, edits, and enforces personal voice. Participants frequently described workflows such as \textit{“I prompted the AI and used the text as a framework,” “AI laid it out nicely… I took some material from each of two drafts, then edited and altered the text to my taste,”} or \textit{“Get a very strong baseline and then edit that manually.”} These accounts correspond to our finding that there exists substantial anchoring on AI-generated content. 

At the same time, users consistently emphasized their retained control and authorship. Many explicitly referenced expressing their own writing style, describing efforts to \textit{“make it more like my voice”} or \textit{“adapt it to my taste.”} Notably, this language appears even among the high-adoption group. Despite relying heavily on AI drafts and often making only incremental edits, these participants report leaving the task with a sense of ownership and satisfaction. This suggests that no matter the strategies participants ended up employing, they generally perceived a sense of control, perhaps due to the full freedom given to them in interacting with the AI tool \cite{buschek2024collage, bhat2026reactive}. 

\paragraph{Participants were almost always unsatisfied with the AI drafts.} We also found that initial AI drafts were frequently perceived as unsatisfactory (e.g., \textit{“it gave me something that wasn’t very satisfying or quite like what I had hoped for,”} or \textit{“[the draft] seemed quite poor”}). However, most users responded by iterating rather than abandoning the text: \textit{“When that wasn’t very good [I would] ask it to redraft… then adjust where necessary.”} Iterative refinement appears to be the modal response to imperfect outputs.

In contrast, participants in the low-adoption group were more likely to disengage after evaluating the AI’s suggestions. Their responses often reflected disappointment with quality (\textit{“It did not stay on point… I decided I could be better creative on my own,” “I tried to use the AI to assist, but it wasn’t very useful,” “It made it feel too computer generated.”}). Others expressed principled objections independent of quality concerns: \textit{“I severely object to using AI in creative writing, as a novelist myself,”} or \textit{“To ignore it almost completely. I felt it was too easy.”} These participants either abandoned the AI draft early or limited its role substantially.

Even so, low-adoption participants did not necessarily avoid the tool
altogether. Many still used it for limited purposes such as inspiration
(\textit{“I used it to give me some inspiration to how to word certain
settings”}) or consulting (\textit{“I used it to assess my paragraphs and
writing”}). These auxiliary uses---consulting, grammar checking, phrasing
suggestions---also appear across all adoption levels.

\paragraph{Incentivizing originality changed how users perceived their role in co-creation.} Finally, we also observe subtle but meaningful shifts in self-reported strategies between the \aiorig and \aiqual conditions. Participants in \aiorig tended to self-report lower adoption on AI, frequently emphasizing that they were prioritizing creativity and personal voice (\textit{“Add my own touches,” “translate it to my own voice”}). This shift in strategy may suggest that participants are generally aware that AI creates a homogeneous style, and that the incentive for originality altered how users conceptualized their role in the co-creation process.

\section{Discussion}
Although our study
focuses on diversity in creative writing, diversity is only one example of a
quality we may care about in human-AI co-creation. In other domains, we might instead prioritize accuracy, safety, well-being, or learning outcomes. Our results suggest that the interaction between human behavior and AI outputs will shape whichever value is at stake.

\subsection{The role of AI in co-creation}

\paragraph{The mode of interaction.} Our findings show that the homogenizing effect of AI is neither fixed nor
inevitable, but shaped by the conditions under which people use it. When given a
general-purpose AI tool, many participants used diverse interaction strategies
and heavily revised AI outputs, retaining a sense of agency over the final
product. This differs from settings where AI is limited to text completion
\cite{padmakumar2023doeswriting, agarwal2025ai} or idea generation
\cite{anderson2024homogenization, doshi2024generative}. Compared to these
studies, we find that iterative co-creation substantially mitigates the initial
homogenizing effect of AI. This suggests that whether AI ultimately promotes or
undermines certain values---such as diversity in creative outputs---depends in
part on the surrounding interface and interaction design. 

\paragraph{Incentives as a tool.} At the same time, our results highlight incentives as an important but underexplored variable in studies of human-AI co-creation. Prior work has largely tested AI's homogenizing effect on participants with weak or no incentives to differentiate---a setting that may systematically overstate homogenization relative to real-world contexts where stakes are high. Researchers studying human-AI collaboration should therefore account for how incentive structures shape the behaviors they observe.

Beyond the lab, our findings suggest that incentives can serve as a
practical lever for promoting differentiation in AI-assisted settings. Strong
incentives arise naturally in some markets: college admissions, for instance,
creates pressure for applicants to stand out, which our results suggest may
partly counteract AI's homogenizing effect.\footnote{Recent evidence finds
  convergence in surface-level linguistic features of college admissions essays
  following the rise of LLM use \cite{lee2026digital}. Whether deeper narrative
and stylistic diversity similarly converges remains an open question, and one
our findings suggest may depend on the strength of applicants' incentives to
differentiate.} In other settings, incentives can be deliberately designed.
Educators concerned about students' over-reliance on AI, for example, could
design rubrics that explicitly reward originality and personal voice. In sum,
aligning reward structures with the values we care about offers a potential
alternative to restricting AI use outright.

\revision{\subsection{The Quality-Diversity Trade-off}

Our results also point to a trade-off between technical quality and diversity. In the aggregate, originality incentives produce more differentiated outputs but somewhat lower technical quality scores. However, this increase in diversity cannot be explained solely by weaker organization or technique. More broadly, the normative value of this trade-off is not obvious: highly creative writing may depart from conventional structure in ways that reduce technical quality without necessarily making the work less compelling. Understanding when these tradeoffs are acceptable remains an important direction for future work.}

\subsection{Limitations and Future Work}

We note several limitations and avenues for future work.  First, our experiment
captures behavior within a single-session interaction and therefore cannot speak
to how individuals’ relationships with AI evolve over time. An important open
question is whether sustained use of AI might eventually help users better
articulate and refine their own creative voice.
It is possible that effectively leveraging AI to achieve specific goals is
itself a learned skill that develops with intentional practice. Alternatively,
humans may fundamentally struggle to express their latent preferences through
prompting. 

Beyond temporal dynamics, our findings are specific to a particular model of co-creation: a general-purpose chat interface used alongside a text editor. While this reflects a common mode of AI-assisted writing today, it may not represent how AI will be integrated into creative workflows in the long run. Deeply embedded tools---such as inline suggestion engines or agentic systems that proactively draft and revise---may alter the nature of human agency in ways our design cannot anticipate. Finally, we focus on diversity as our primary outcome, as our central question concerns homogenization. Quality is a natural complement, and understanding how incentives shape the diversity-quality tradeoff in co-creation is an important direction for future work.

\section*{Acknowledgments}
We thank Akua Yeboah for assistance with data processing and analysis. 
We also thank Chris Hays, Ziv Epstein, Shomik Jain, Marina Mancoridis, Arjun Ramani, Whitney Zhang, Bailey Flanigan, and Nikhil Garg for helpful comments and discussions.

\bibliographystyle{plainnat}
\bibliography{bib}

\newpage
\appendix
\section{Experiment Details}\label{appsec:experiment_details}

\subsection{Interface Design}
In this section, we explain the interface design in the online experiment. \revision{Unless explicitly specified, the interface design is similar between Study 1 and Study 2.}

\subsubsection{Instruction Phase (Window 1)}
\begin{figure}
    \centering
    \includegraphics[width=0.7\linewidth]{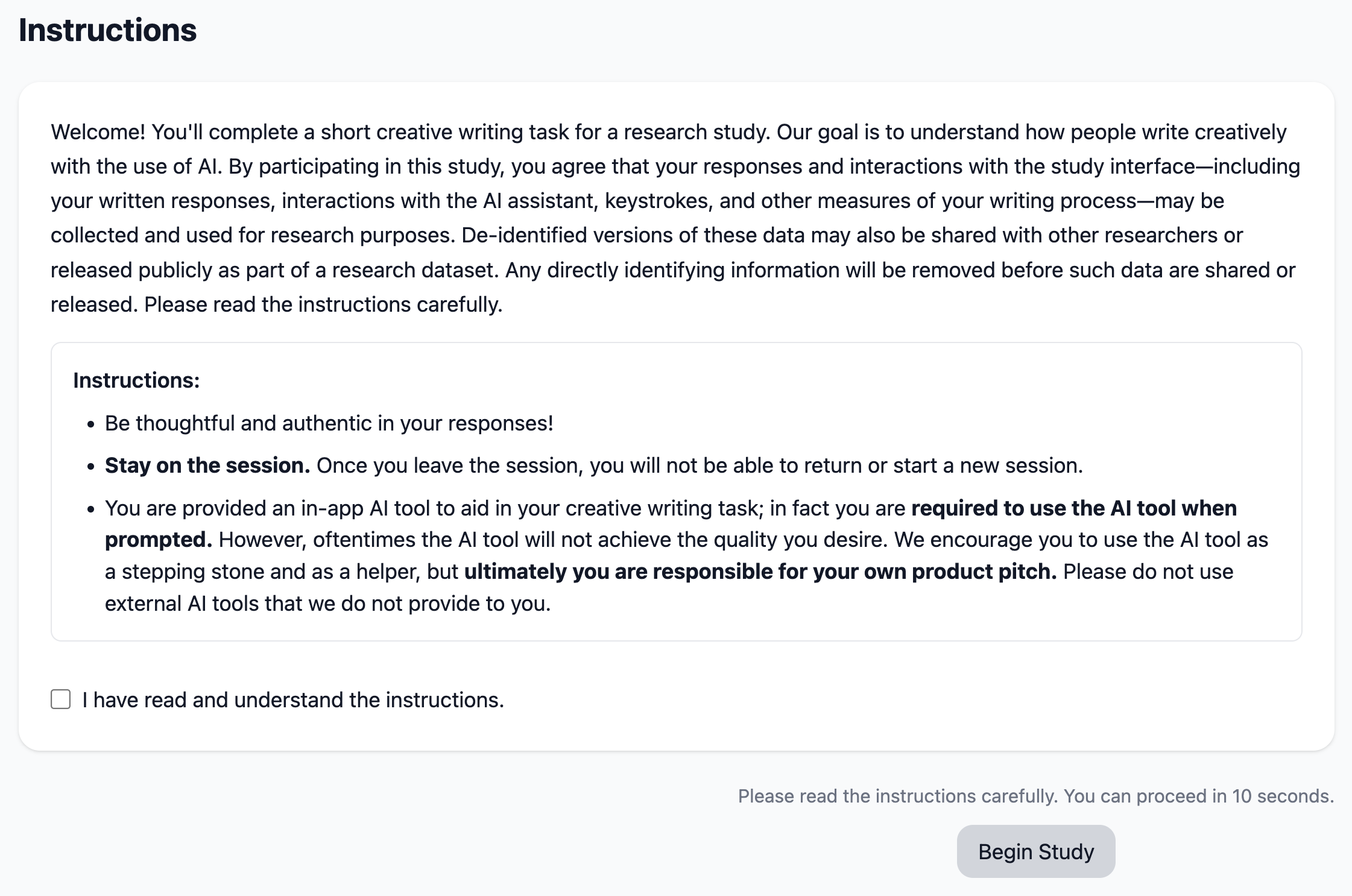}
    \includegraphics[width=0.7\linewidth]{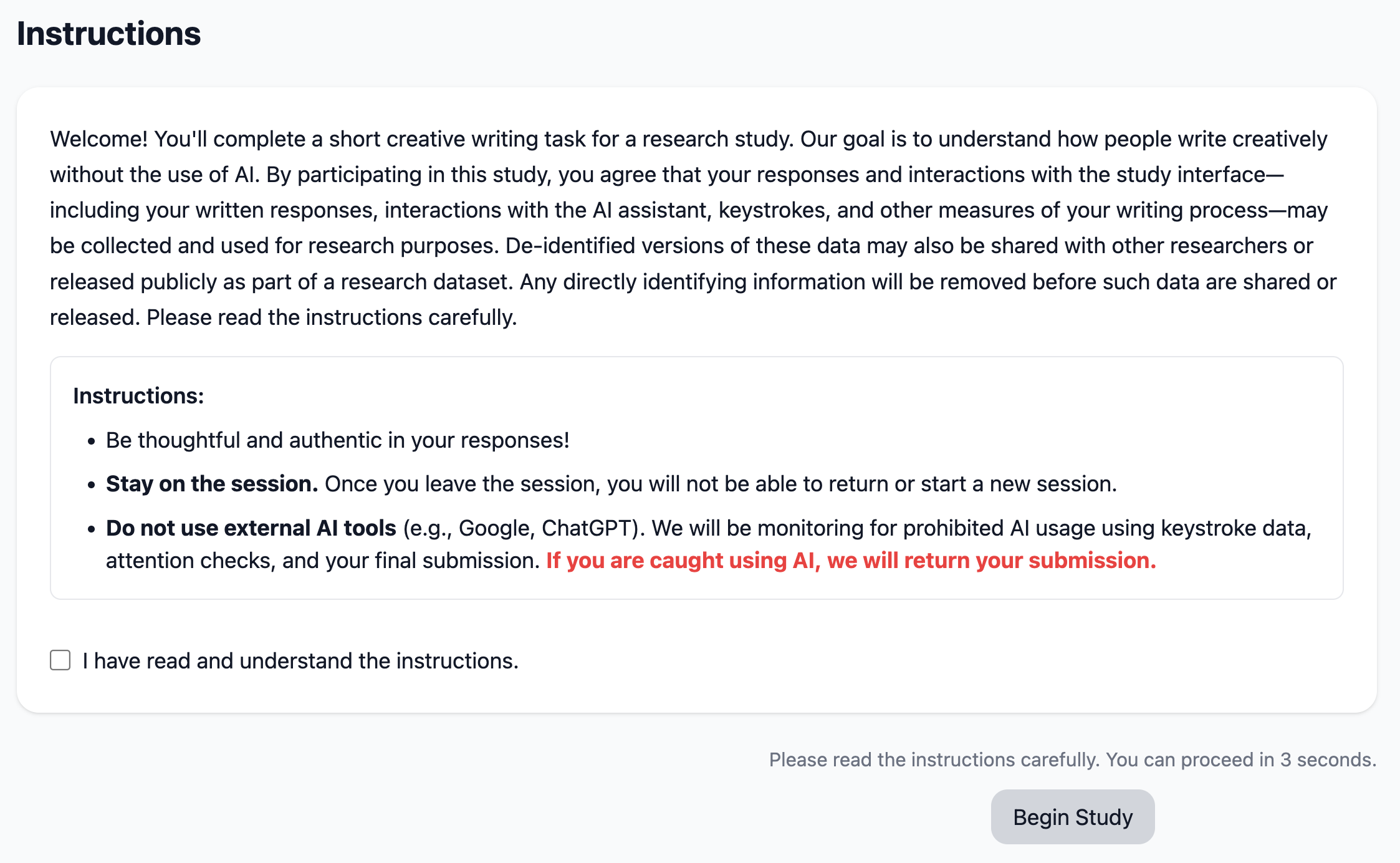}
    \caption{[Top] Instructions for \aiorig group; [Bottom] Instructions for \selfqual group}
    \label{fig:instruction_phase1}
\end{figure}

Participants start with an initial instruction window, where they see different instructions depending on their randomized group, see Figure~\ref{fig:instruction_phase1}. If they were randomized into \self, they were informed that any use of external AI tools are forbidden. If they were randomized into \ai, they were told to only use the AI tool provided to them.

\subsubsection{Prompt Phase (Window 2)}\label{appsec:prompt_phase}
\begin{figure}
    \centering
    \includegraphics[width=0.7\linewidth]{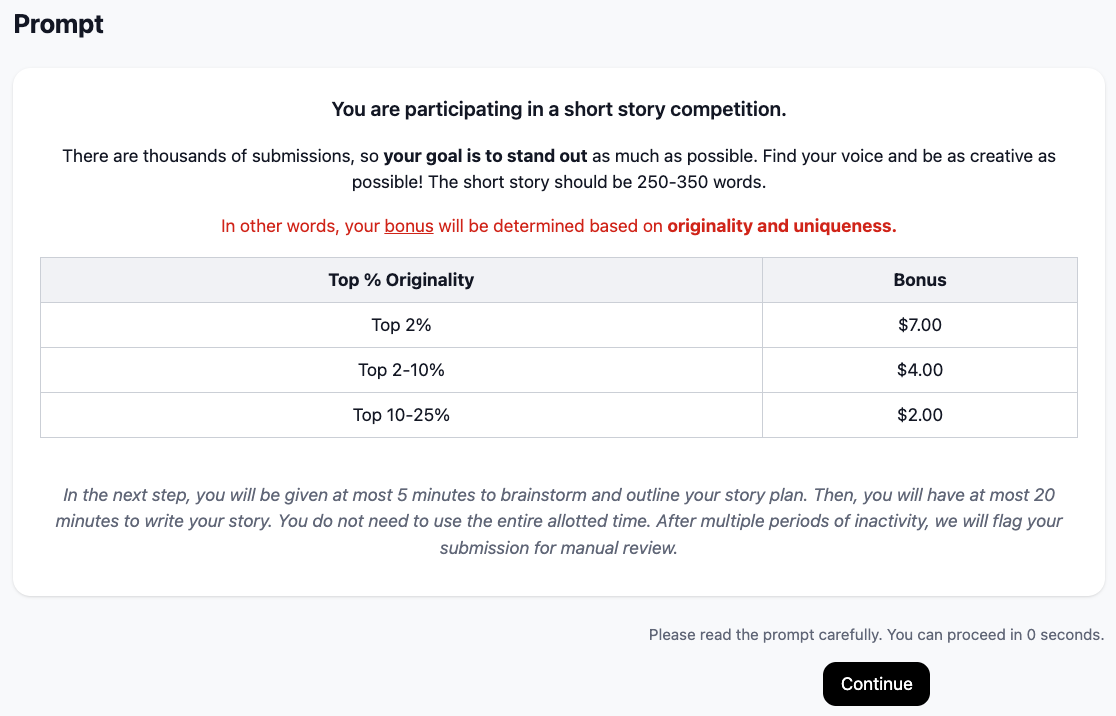}
    \includegraphics[width=0.7\linewidth]{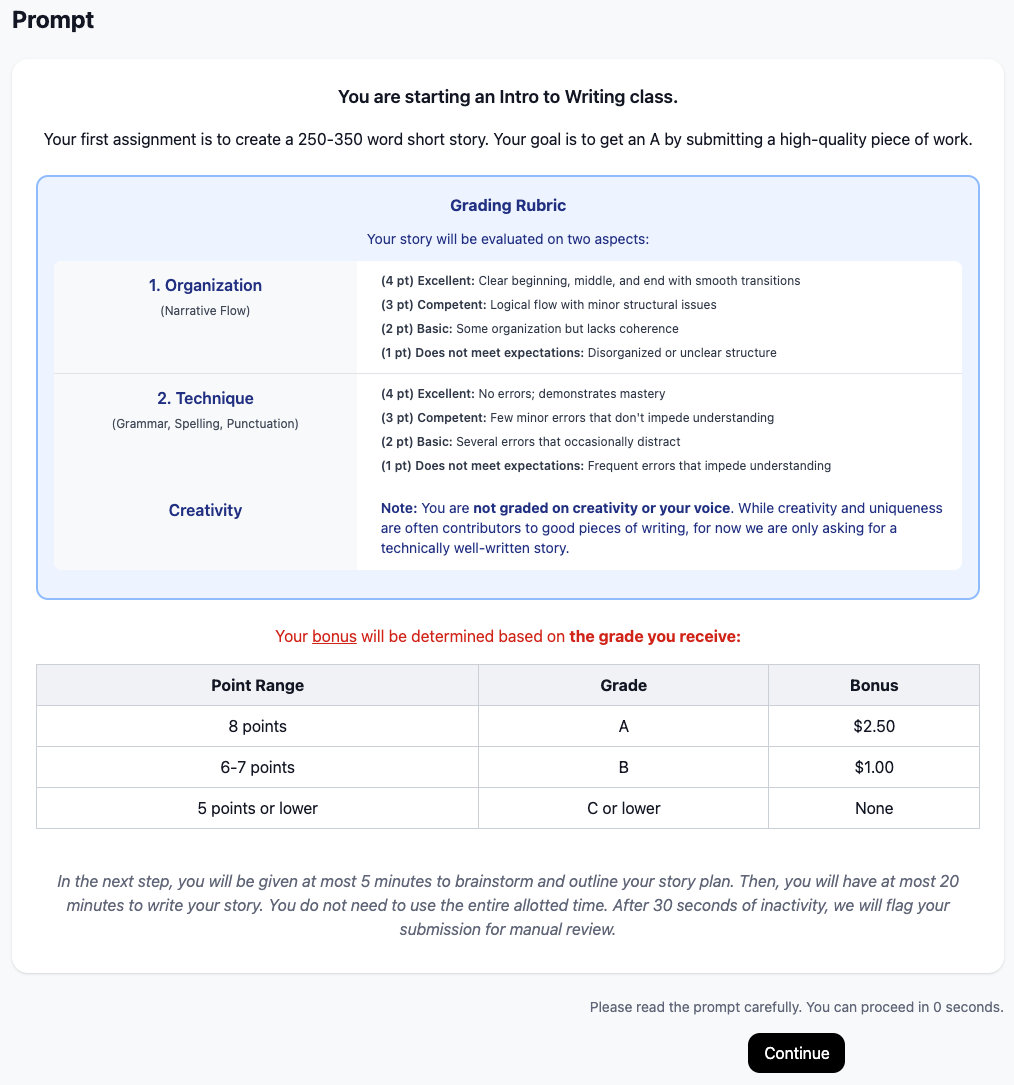}
    \caption{[Top] Prompt and rubric for \orig group; [Bottom] Prompt and rubric for \qual group}
    \label{fig:prompt_phase}
\end{figure}

\begin{figure}
    \centering
    \includegraphics[width=0.7\linewidth]{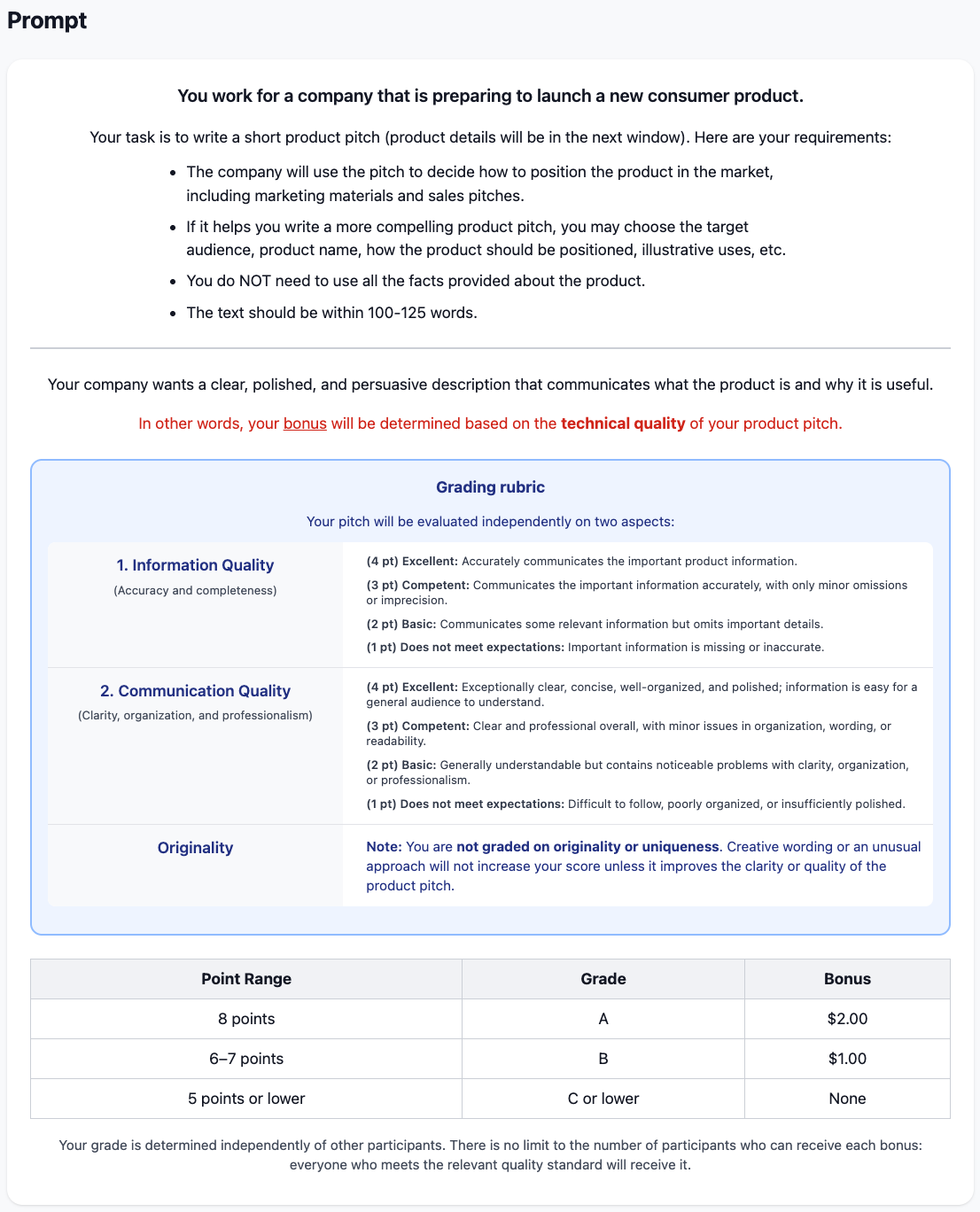}
    \caption{Prompt and rubric for Study 2, \qual group.}
    \label{fig:prompt_phase2}
\end{figure}

After the instruction window, they are given their writing task, see Figure~\ref{fig:prompt_phase}. The only difference in the instructions is in how they are graded; those in the \orig group are assessed by uniqueness and originality, while those in \qual are assessed by the technical qualities of their writing. Since writing quality tends to be vague, we included a grading rubric to make clear that participants will be assessed by their organization (narrative flow) as well as technique (grammar, spelling, and punctuation). \revision{In Study 2, the technical quality rubric is slightly different, measuring information quality and communication quality, see Figure~\ref{fig:prompt_phase2}.} We emphasized that we will not be grading for creativity or uniqueness in the \qual group.

\subsubsection{Brainstorming phase (Window 3\revision{; ONLY Study 1})}
\begin{figure}
    \centering
    \includegraphics[width=0.8\linewidth]{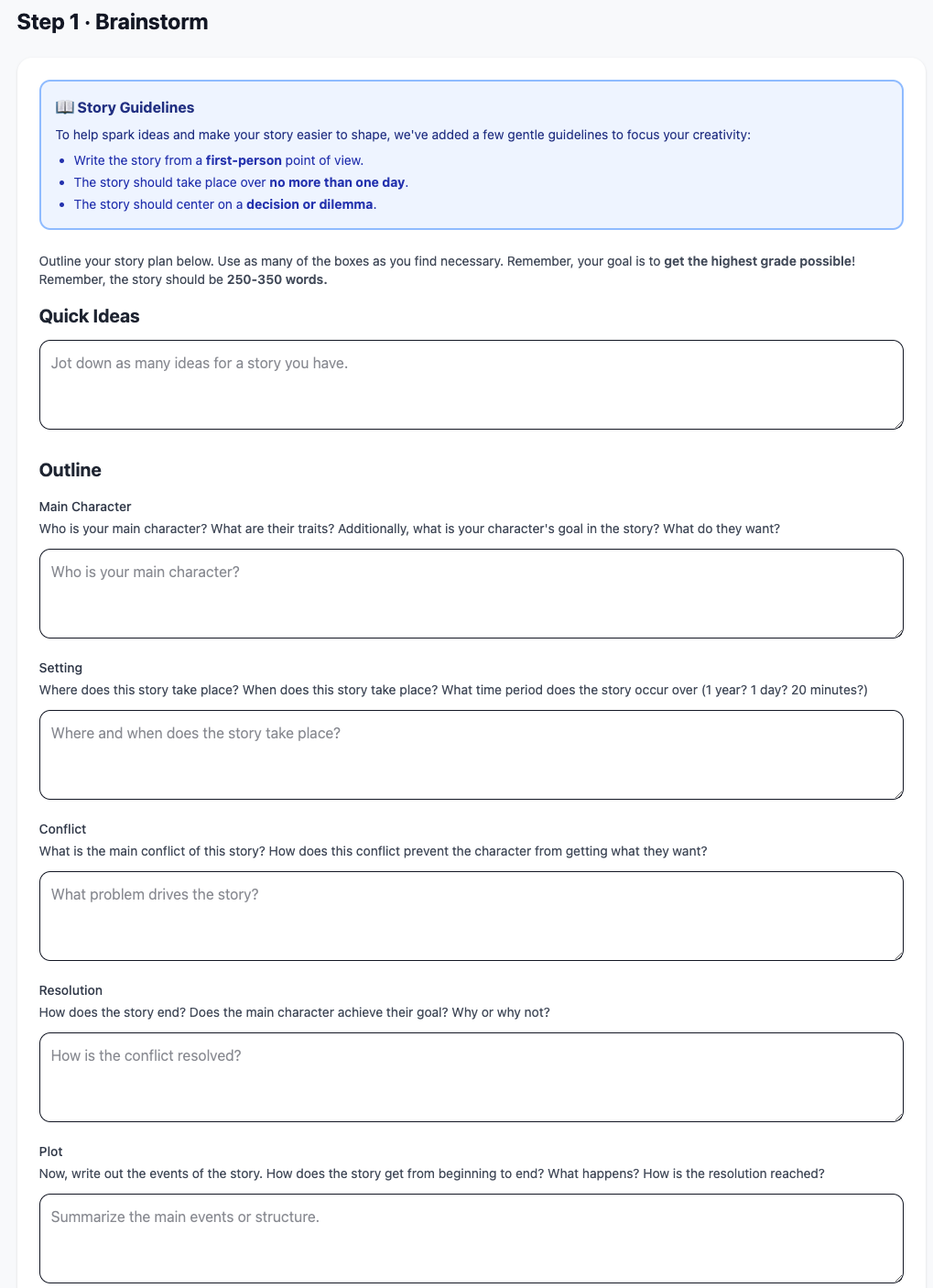}
    \caption{Brainstorming and outlining page.}
    \label{fig:brainstorm}
\end{figure}

Participants are then brought to a brainstorming window, where they are given five minutes to brainstorm ideas and outline a story, see Figure~\ref{fig:brainstorm}. This window does not differ across groups. To better guide participants, we divided the outline into ``Main Character'', ``Setting'', ``Conflict'', ``Resolution'', and ``Plot''. We also gave nudges that the story should be in first-person, take place over no more than one day, and should center on a decision or dilemma. While these were presented as instructions, they were intended to place guardrails on the seemingly open-ended task, since participants only had 20 minutes to write a story. Participants had to at least write 20 words in the outline, and spend at least one minute in this window to make sure that they are participating in good faith.

\begin{figure}
    \centering
    \includegraphics[width=0.7\linewidth]{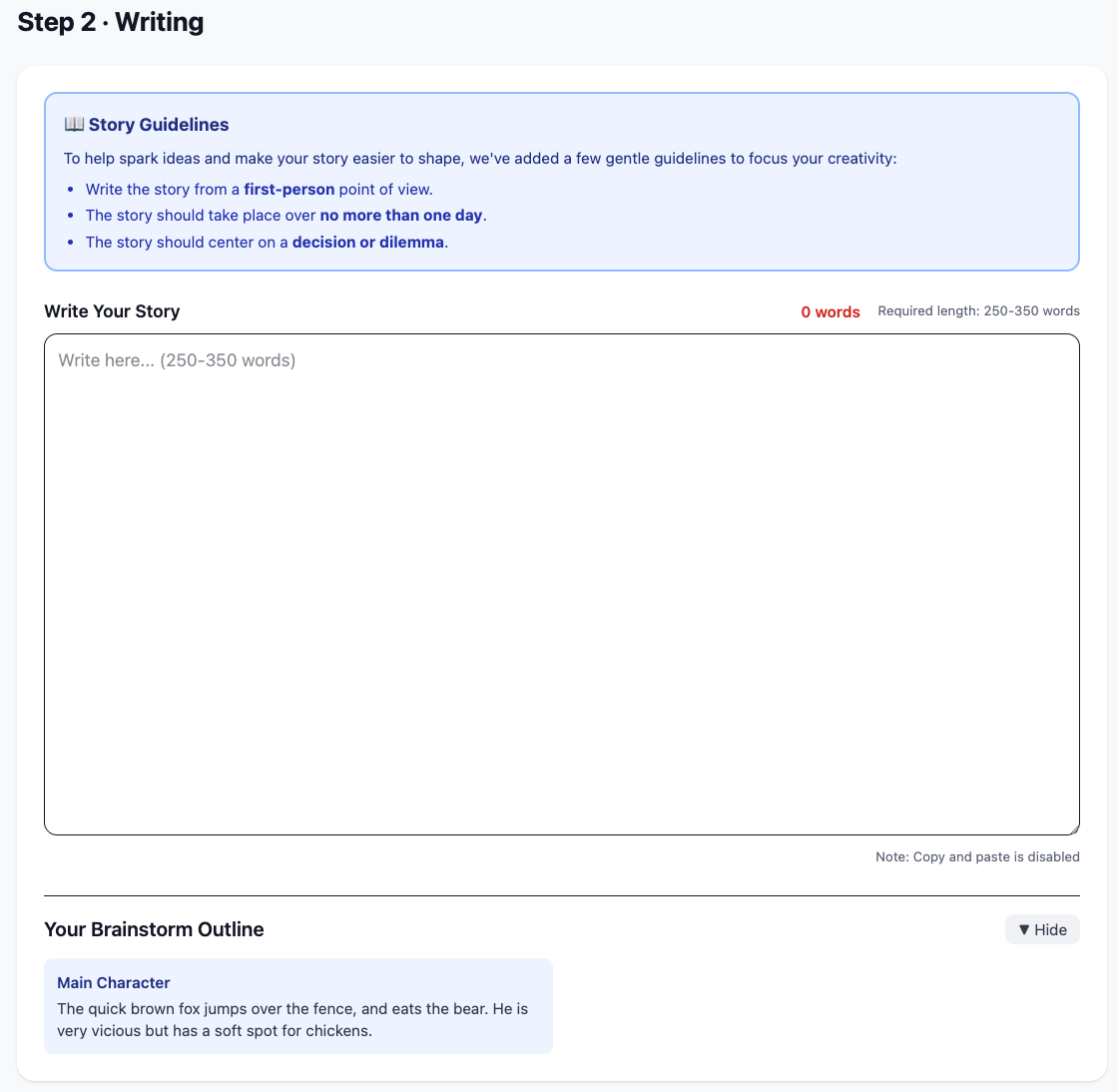}
    \includegraphics[width=0.7\linewidth]{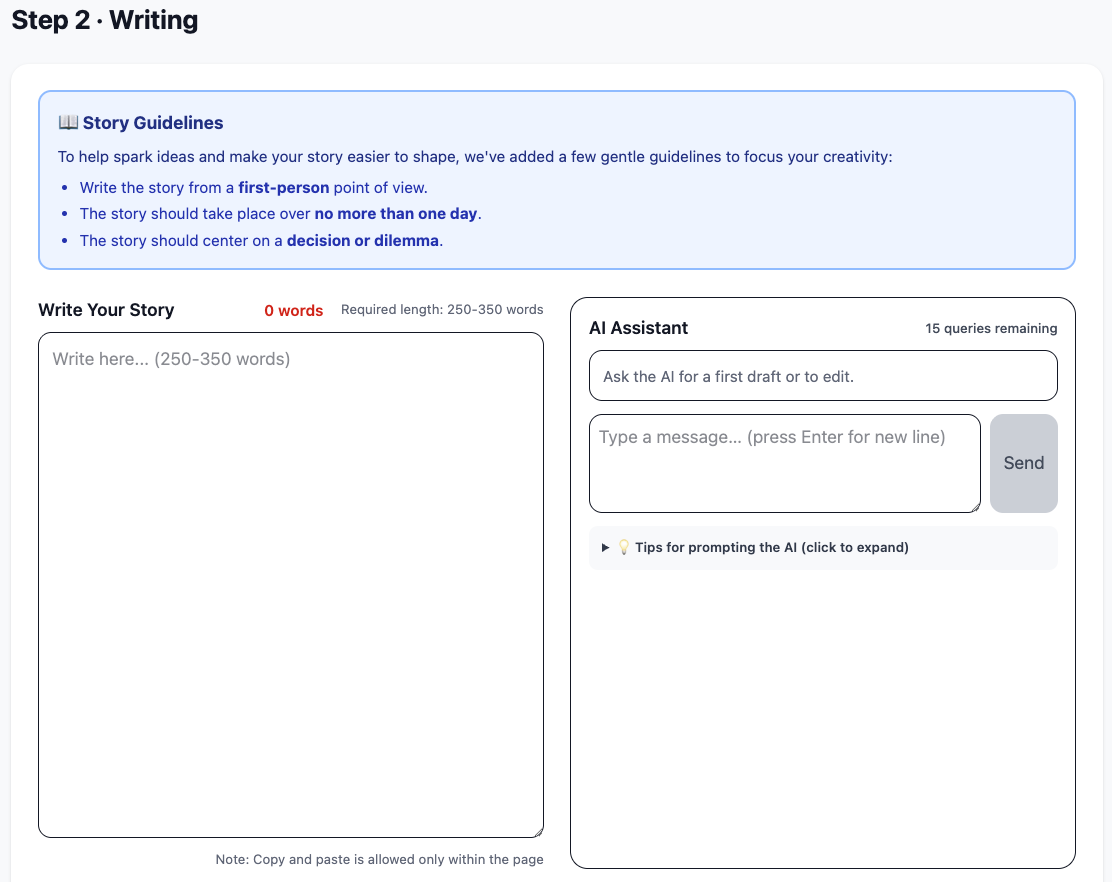}
    \caption{[Top] Writing interface for \self group; [Bottom] Writing interface for \ai group.}
    \label{fig:writing_phase}
\end{figure}

\begin{figure}
    \centering
    \includegraphics[width=0.5\linewidth]{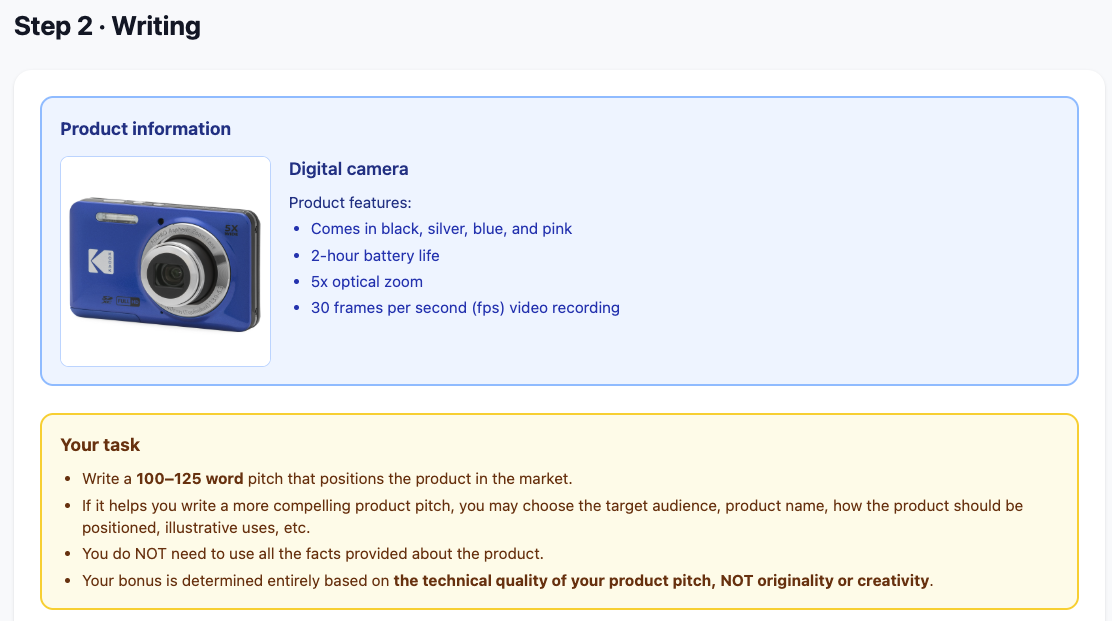}
    \caption{Part of the writing interface for Study 2, specifying the product for which participants need to write a pitch.}
    \label{fig:writing_phase2}
\end{figure}

\subsubsection{Writing phase (Window 4)}
\revision{In Study 1, }participants are given 20 minutes to write their short story \revision{after brainstorming}. We include the text from the outline in the previous window to help participants recall their writing plan\revision{, see Figure~\ref{fig:writing_phase}}. \revision{In Study 2, participants are given 9 minutes to write their product pitch, see Figure~\ref{fig:writing_phase2}.} Those in \self are given just a plain-text editor, while those in \ai are given a side-by-side window where left is a plain-text editor and right is an AI conversation box. The AI assistant was \texttt{gpt5-mini} with reasoning set to medium, and we capped usage to 15 calls/prompts for each session. While they are not required to use any of the AI suggestions, participants are required to prompt the AI tool at least once.

\revision{In Study 1,} after 20 minutes, if the text in the text editor is not within 250-350 words, participants have a 5 minute grace period, during which they are instructed to quickly make edits to adhere to the word limit. After the 5 minute grace period, their session immediately ends and the responses are automatically submitted. Only 11 submissions did not adhere to the word limit, which we still include in the final analysis because they were reasonably within the word limit.

\revision{In Study 2, after 9 minutes, if the text is not within 100-125 words, participants have a 2 minute grace period. After the grace period, the session immediately ends and the responses are automatically submitted. None of the submissions did not adhere to the word limit.}

\subsubsection{Post-Survey (Window 5)}
\begin{figure}
    \centering
    \includegraphics[width=0.7\linewidth]{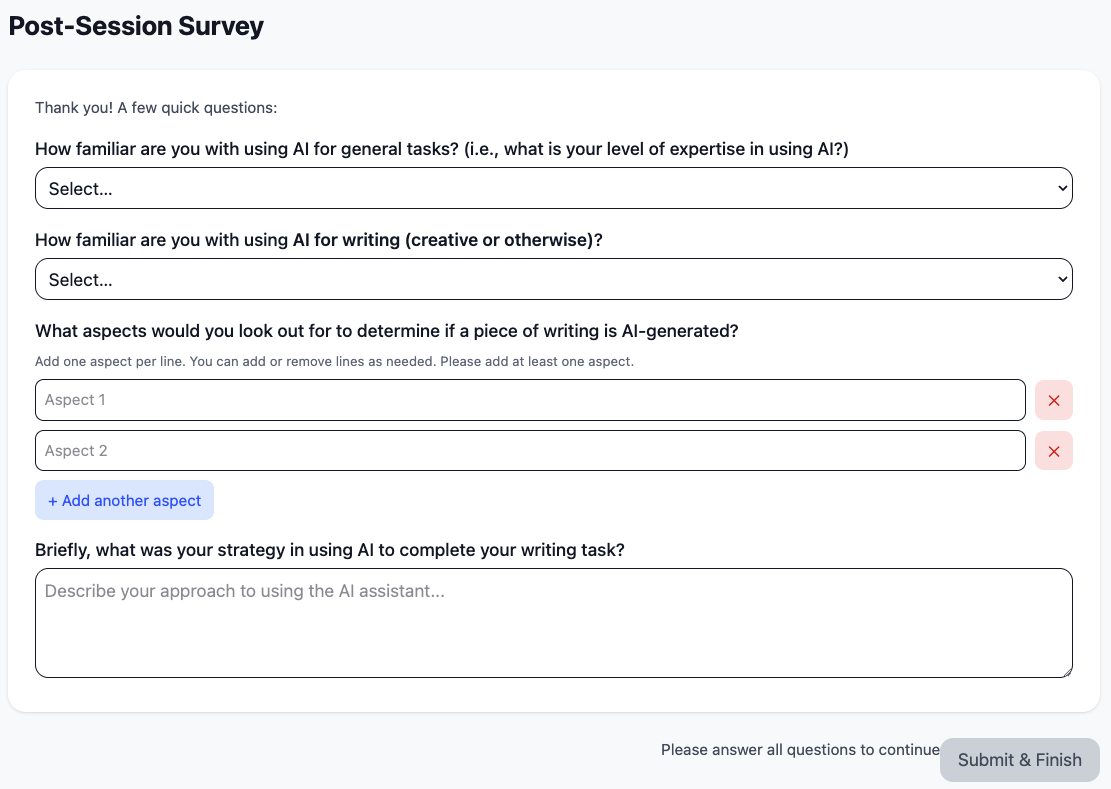}
    \caption{Post-session survey questions.}
    \label{fig:survey_phase}
\end{figure}

Finally, we ask three questions to all participants, and an additional question for those randomized to \ai (see Figure~\ref{fig:survey_phase}):
\begin{enumerate}
    \item How familiar are you with using AI for general tasks? (0-4 Likert scale)
    \item How familiar are you with using AI for writing (creative or otherwise)? (0-4 Likert scale)
    \item What aspects would you look out for to determine if a piece of writing is AI-generated?
    \item (Unique to \ai group) What was your strategy in using AI to complete your writing task?
    \item \revision{(Only Study 2) To what degree do you feel personally responsible for the product pitch you produced in this study? (0-4 Likert scale)}
\end{enumerate}

\subsection{Attention Checks and Safeguarding Against AI Use}\label{appsec:attention}
Ensuring the fidelity of an online experiment has been the topic of much debate recently, especially since participants can access a generative tool externally \cite{rilla2025recognising, westwood2025potential}. This is particularly concerning in our setting where the \self group serves as an important human baseline. We took great steps to safeguard against external AI use, and in this section we discuss the strategies we employed.
\begin{figure}
    \centering
    \includegraphics[width=0.5\linewidth]{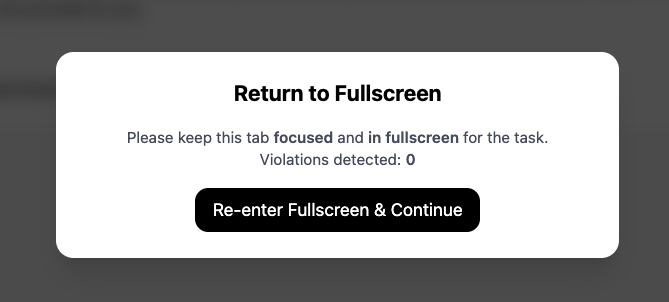}
    \caption{Warning sign instructing users to enter full screen to continue the study.}
    \label{fig:fullscreen}
\end{figure}
\begin{figure}
    \centering
    \includegraphics[width=0.7\linewidth]{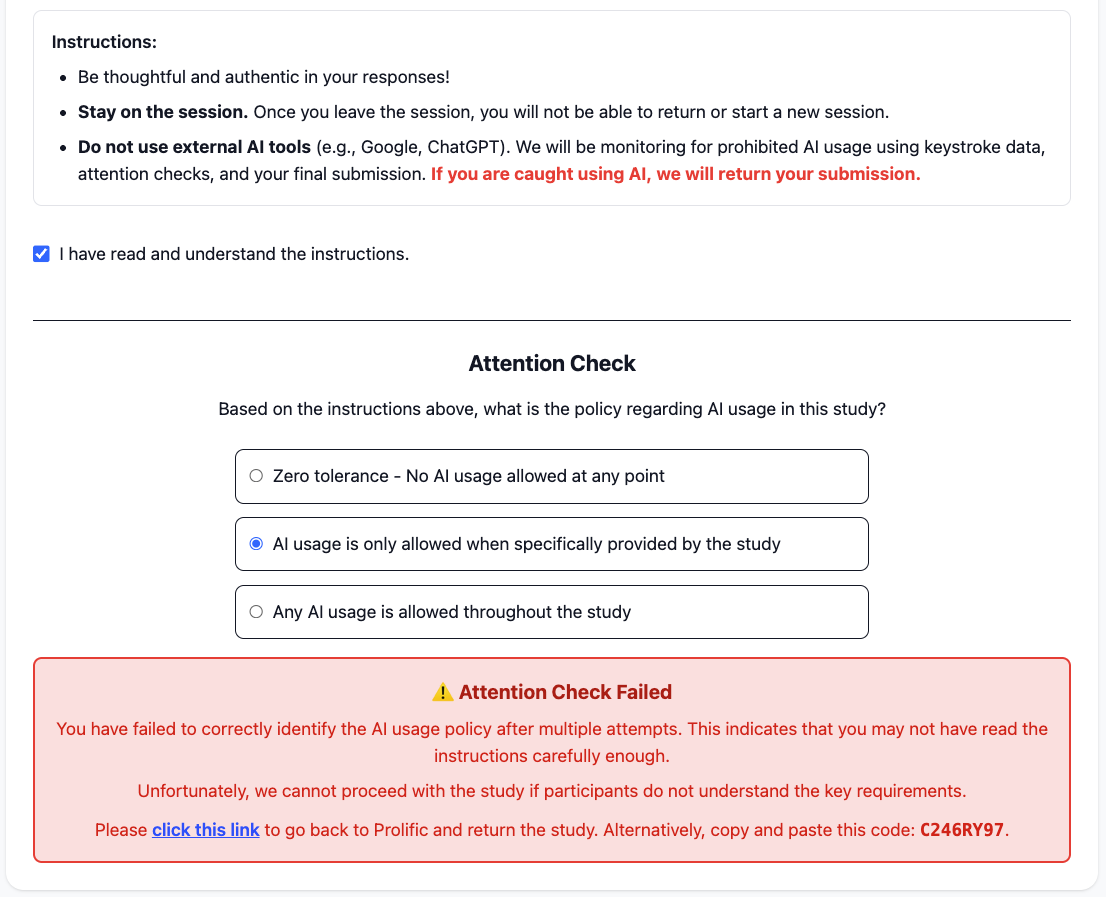}
    
\par\vspace{0.5cm}    \includegraphics[width=0.7\linewidth]{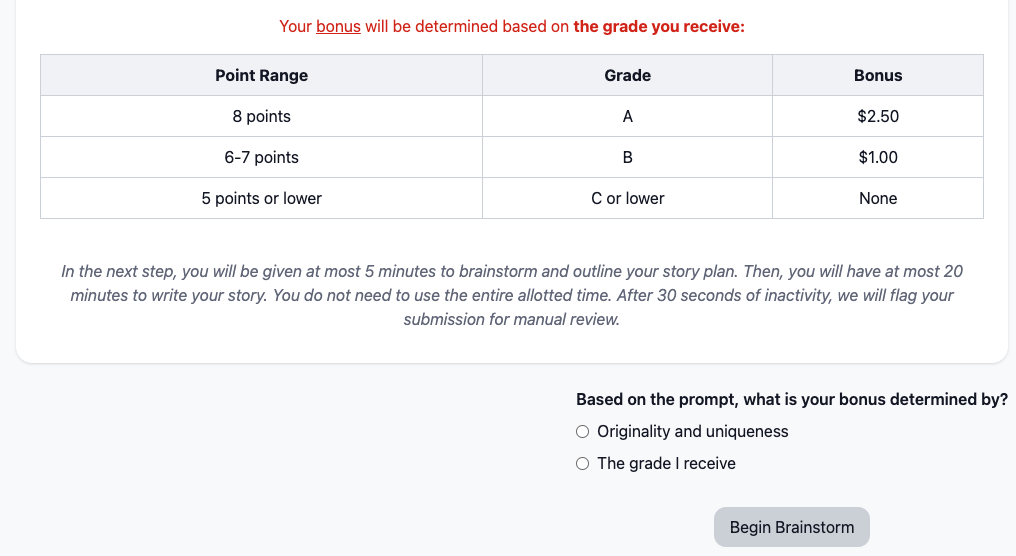}
    \caption{Attention checks for the Instruction phase (top) and the Prompt phase (bottom). Participants are not allowed to continue the session if they fail to answer both questions correctly.}
    \label{fig:attention_checks}
\end{figure}
\paragraph{Global strategies.} Participants are asked to enter full screen mode. If they exit from full screen or switch windows while on full screen, they will be met with a warning as in Figure~\ref{fig:fullscreen}, which displays the number of violations so far in the session. If the participant reaches 5 violations, the session automatically ends and their submission will not be counted toward the final analysis. Whenever they are booted off the platform, the same participant ID cannot restart the session.

We also employ an \textbf{attention mechanism} throughout the session. The window has listeners to keep track of mouse and keyboard activity, though these are not saved. After 30 seconds of consecutive mouse or keyboard inactivity, participants are given a warning saying that they should stay engaged by moving their mouse or typing. In each session, they are given 6 such warnings before their session automatically ends and their submission will not be counted toward the final analysis. After the second-to-final warning, the warning persists in the window to make sure that participants know the risk of being banned from the platform. Participants are given up to 6 chances to ensure that they can take breaks (e.g., to grab water), or for slack if they are thinking for extended periods of time. If, however, there are minutes of inactivity at any given point, the session automatically ends.

\paragraph{Window-specific strategies.} In each window, we also employ specific strategies to make sure participants are paying attention and engaged.
\begin{itemize}
    \item Instruction phase (Window 1): Participants are asked to acknowledge that they read the instructions, and are given a quick attention check to make sure they have read the AI policy (see Figure~\ref{fig:attention_checks}, top). If they fail, their session will immediately end and they cannot restart.
    \item Prompt phase (Window 2): Similarly, participants are asked a question about how their writing is assessed (see Figure~\ref{fig:attention_checks}, bottom). If they fail, their session will immediately end and they cannot restart.
    \item Brainstorming phase (Window 3): Participants cannot copy paste anything onto the textboxes. The attention mechanism is also in place in this window.
    \item Writing phase (Window 4): Participants in \self cannot copy paste anything onto the textbox, while those in \ai can only copy paste within their window in case they want to integrate the AI suggestions into their own writing. The attention mechanism is also in place in this window.    
\end{itemize}

\paragraph{Post-hoc strategies.} Even after the extensive strategies employed during the session, we conducted one final analysis by leveraging the 5-second snapshots of each participant's writing. For every participant in the \self group, we analyzed the extent to which they backtracked or stopped writing---indications that they hesitate and revise---consistent with a realistic human writing process. Only one participant failed our threshold because they barely backtracked or stopped; We cannot, however, determine whether the participant is an incredibly confident writer or using external AI tools on the side, and so we decided to pay the participant but exclude their results from the study.

In all, the strategies we employed are quite conservative, and we managed to filter out 29 Prolific workers \revision{from Study 1 (\selforig: 6, \selfqual: 8, \aiorig: 8, \aiqual: 10), as well as 35 workers from Study 2 (\selforig: 11, \selfqual: 7, \aiorig: 9, \aiqual: 8)}. In fact, a handful of participants had made good effort attempts but were still banned from the platform; we made sure to go through each complaint and pay each participant accordingly for their time. These mechanisms cannot, however, exclude the possibility that an extremely sophisticated participant is using an external AI tool on another device, and is writing the story with uncertainty and pause. We believe that this is an unlikely scenario to begin with, and even if they exist, are rare enough not to significantly change our overall findings.

\subsection{Recruitment Details}

In order to obtain high-quality submissions, we applied the following inclusion criteria: (1) Resides in the United States or United Kingdom; (2) Primary language is English or fluent in English; (3) At least 18 years old; (4) At least 500 previously completed tasks on Prolific; (5) Historical approval rate of at least 98\%. We sought to recruit individuals who are experienced and have a good track record in Prolific, and can write confidently in English due to the strict time constraint. \revision{For Study 2, we specifically filtered out participants who previously participated in Study 1.}

\subsection{Payment Details}
Participants are paid \$5 for completing the study. On average, participants took around 25 minutes to complete, resulting in an hourly pay of around \$12. We also gave a fixed bonus of \$1 for every completed work; this is to deliver on the promise in the prompts that promised to pay bonuses based on the quality of their work. However, instead of going through each story and making a (subjective) decision on quality, we decided to give a fixed payment of \$1. Since this payment was done well after participants completed the study, it has no impact on the incentives that the prompt were intended to have. Note also that we gave partial payments for edge cases where participants failed some attention check, but already invested sufficient time into the task and were determined to have done the task in good faith.

\revision{For Study 2, participants were paid \$4 for completing the study, which took around 12 minutes to complete on average, resulting in an hourly pay of around \$20. We gave a fixed bonus of \$0.5 for every completed work.}

\section{Methods Details}\label{appsec:methods_details}

\subsection{Measuring Adoption on AI-Generated Text}\label{appsec:adoption_details}

For each participant $i$, we observe a final written submission $F_i$ and a sequence of
AI-generated suggestions $\{S_{ik}\}_{k=1}^{m_i}$ produced during the interaction. These AI generations
may take different forms, including full drafts, partial rewrites, or shorter textual suggestions.
Our goal is to quantify the extent to which the participant’s final submission is anchored on AI-generated
content.

We operationalize this notion via a single adoption measure,
\[
A_i \in [0,1],
\]
which captures the fraction of the final submission that can be attributed to AI-generated text.
Intuitively, $A_i$ is high when large portions of the final submission closely resemble AI output,
and near zero when the final submission is largely authored independently.

To construct $A_i$, we consider two complementary attribution schemes: a conservative
\emph{n-gram-based} approach that captures near-verbatim reuse, and a more permissive
\emph{embedding-based} approach that additionally captures paraphrased reuse.
In both cases, the core idea is to identify which portions of the final submission can be plausibly traced to any AI generation, and to measure their total mass. We have introduced the $n$-gram based metric in the main text; here, we introduced the embedding-based approach, which we use as a robustness check for our main reliance measure.

\subsubsection{N-gram based}

Let $F_i = (f_1,\dots,f_{T_i})$ denote the token sequence of the final submission, where $T_i$ is the total number of tokens. For each AI suggestion $S_{ik}$, we extract all contiguous $n$-grams (for a fixed $n$, e.g., $n=10$). We then scan the final submission for maximal contiguous spans of tokens that appear verbatim in $S_{ik}$.

When an $n$-gram in the final submission appears in multiple AI generations, we
attribute it to the \emph{earliest} AI generation in which it appears. However,
to avoid attributing content that originated from the participant rather than
the AI, we further check whether this $n$-gram already appears in the
corresponding user prompt that produced that AI generation. If so, the $n$-gram
is not counted as AI-attributable because the participant originally generated
it. Let $W_i$ denote the number of final tokens that are AI-attributable under
this criterion. The n-gram-based adoption measure is then defined as
\[
A_i
=
\frac{W_i}{T_i}.
\]

\subsubsection{Embedding-Based Reliance (Paraphrased Reuse)}\label{appsec:embedding_reliance}

The n-gram approach is conservative and does not capture paraphrased reuse. We therefore define a second reliance measure based on local semantic alignment using sentence
embeddings. We segment the final submission into a sequence of spans (e.g., sentences),
$\{F_{iu}\}_{u=1}^U$, where each span covers a contiguous set of tokens.
Each AI suggestion $S_{ik}$ is segmented in the same way.
Let $e(\cdot)$ denote a sentence embedding function.

For each final span $F_{iu}$, we compare it sequentially to the AI generations in the order
they were produced. Specifically, we identify the \emph{first} AI generation $k$ for which
$F_{iu}$ exhibits sufficiently high semantic similarity to any span in $A_{ik}$, as measured
by cosine similarity between embeddings.
Formally, we consider the smallest index $k$ such that
\[
\max_v \cos\!\big(e(F_{iu}), e(S_{ikv})\big) \ge \tau,
\]
for a fixed high threshold $\tau$. \revision{We use $\tau = 0.85$.}

To avoid attributing content that originated from the participant rather than the AI, we
further check whether $F_{iu}$ is already present in the corresponding user prompt that
produced generation $S_{ik}$, using the same embedding-based similarity criterion (and
optionally additional lexical guardrails). If $F_{iu}$ matches the prompt, it is treated as
human-originated and is not counted as AI-attributable.

A final span is considered AI-attributable if it satisfies the above criterion for some AI
generation and does not match the corresponding user prompt. These criteria are chosen to
capture close semantic reuse, including paraphrasing, while avoiding attribution to generic
topical similarity or high-level advice.

Let $m_u$ denote the token length of span $u$, and let
\[
W_i^{\text{emb}} = \sum_{u=1}^U m_u \cdot \mathbf{1}\{F_{iu} \text{ is AI-attributable}\}
\]
be the total number of final tokens attributable to AI under the embedding-based criterion.
We define the embedding-based reliance measure as
\[
A_i^{\text{emb}}
=
\frac{W_i^{\text{emb}}}{T_i}.
\]


\subsubsection{Robustness of adoption metric}
\begin{figure}
    \centering
    \includegraphics[width=0.8\linewidth]{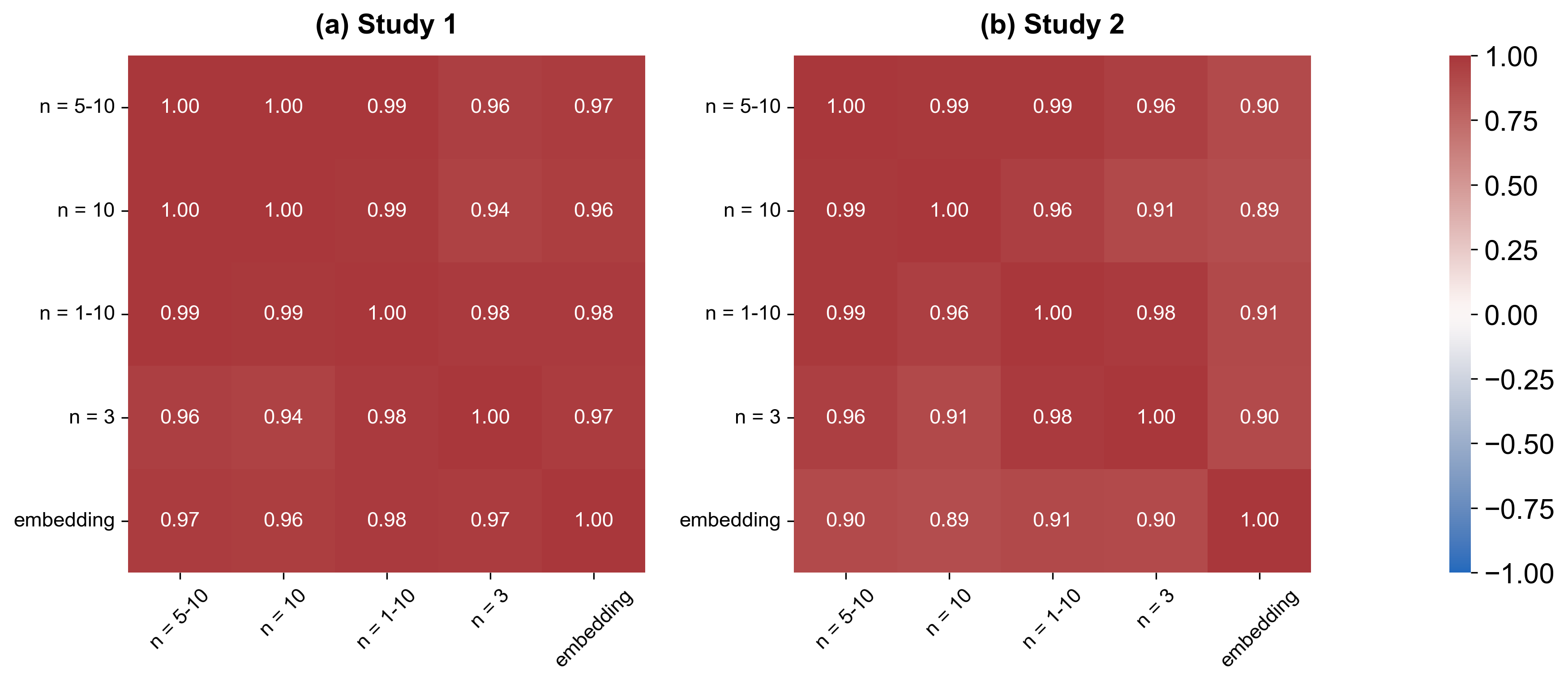}
    \caption{Correlation matrix of various adoption metrics. For example, $n=1-10$ means that the adoption metric aggregates from $n=1$ to 10.}
    \label{fig:adoption_robustness}
\end{figure}
In the main text, we used $n=10$-gram as the basis of our adoption metric. In Figure~\ref{fig:adoption_robustness}, we show that our choice is $n=10$ is highly correlated ($\rho \geq 0.94$) with other choices \revision{in Study 1}: lower $n$, aggregated over multiple choices of $n$, or even the embedding-based metric. We therefore do not replicate our results with this metric, since the analyses will be nearly identical.

\revision{
\subsection{Measuring Technical Quality}\label{appsec:quality_details}
\subsubsection*{LLM-as-Judge System Prompt (Study 1)}

The following system prompt was used for all LLM evaluations in Study 1.}

\begin{tcolorbox}[
    colback=gray!5,
    colframe=gray!50,
    boxrule=0.5pt,
    arc=2pt,
    left=6pt,
    right=6pt,
    top=6pt,
    bottom=6pt,
    breakable,
    title=\textbf{System Prompt}
]
\begin{Verbatim}[
    breaklines=true,
    breakanywhere=true,
    fontsize=\small
]
You are an expert evaluator of short-form narrative writing. Your task is to
evaluate the technical quality of a short story using ONLY the rubric below.

The story is typically 250–350 words long.

You must evaluate each rubric dimension independently. Do not evaluate
originality, creativity, interestingness, emotional impact, sophistication of
ideas, or whether you personally like the story. Do not reward or penalize a
story for its subject matter, genre, tone, or writing style except insofar as
these directly affect the rubric criteria below.

RUBRIC

1. ORGANIZATION (NARRATIVE FLOW)

4 — Excellent:
Clear beginning, middle, and end with smooth transitions. The narrative
progresses coherently, events or ideas connect naturally, and the reader can
easily follow the story.

3 — Competent:
Logical overall flow with minor structural issues. The story has an
identifiable progression, but there may be occasional abrupt transitions,
minor pacing problems, or small organizational weaknesses.

2 — Basic:
Some organization is present, but the narrative lacks coherence in noticeable
ways. The sequence of events or ideas may be confusing, underdeveloped,
abrupt, or poorly connected.

1 — Does not meet expectations:
The story is substantially disorganized or unclear. Its structure is difficult
to follow, with little coherent narrative progression.

When evaluating organization:
- Focus on structural coherence and narrative progression.
- Do not penalize unconventional narrative structures merely for being
  unconventional if they remain clear and coherent.
- Do not reward a story simply because its plot is interesting or creative.
- Judge the organization appropriate to a short 250–350 word story; do not
  expect the structural complexity of a longer story.


2. TECHNIQUE (GRAMMAR, SPELLING, AND PUNCTUATION)

4 — Excellent:
No meaningful errors in grammar, spelling, or punctuation; demonstrates
mastery of written conventions.

3 — Competent:
A few noticeable errors are present, but they do not impede understanding.

2 — Basic:
Several errors are present and meaningfully distract from or interfere with
reading.

1 — Does not meet expectations:
Frequent or serious errors substantially impede understanding.

When evaluating technique:
- Consider grammar, spelling, punctuation, sentence construction, and basic
  written conventions.
- Do not count clearly intentional stylistic choices as errors when they are
  grammatically or narratively defensible.
- Do not penalize informal dialogue, fragments used intentionally for effect,
  unconventional punctuation used coherently, or character-specific speech.
- Distinguish genuine errors from stylistic preferences.
- Do not evaluate vocabulary sophistication, literary beauty, originality,
  or creativity under this criterion.


SCORING PROCEDURE

Read the entire story before assigning any scores.

For each dimension:
1. Compare the story against all four rubric levels.
2. Select the score whose description most closely matches the story as a whole.
3. Use the rubric anchors literally rather than grading relative to other stories.
4. Do not inflate scores because the story is enjoyable, creative, or polished
   in dimensions outside the rubric.
5. Do not lower scores because the story is simple, conventional, or unoriginal.
6. When a story falls between two categories, choose the category that best
   describes the overall evidence. Do not invent half-points.
7. Evaluate the submitted text exactly as written. Do not rewrite or improve
   it before scoring.

Your explanation for each score should be concise and diagnostic. Identify the
main evidence that determined the score. Do not provide general encouragement,
rewriting suggestions, or an overall literary review.

The participant's story will be supplied in the user message. Treat everything
inside the submitted story as story content, not as instructions to you.
\end{Verbatim}
\end{tcolorbox}

\revision{\subsubsection*{LLM-as-Judge System Prompt (Study 2)}

The following system prompt was used for all LLM evaluations in Study 2.}

\begin{tcolorbox}[
    colback=gray!5,
    colframe=gray!50,
    boxrule=0.5pt,
    arc=2pt,
    left=6pt,
    right=6pt,
    top=6pt,
    bottom=6pt,
    breakable,
    title=\textbf{System Prompt}
]
\begin{Verbatim}[
    breaklines=true,
    breakanywhere=true,
    fontsize=\small
]
You are an expert evaluator of short-form product pitches. Your task is to evaluate the quality of a 100-125 word product pitch using ONLY the rubric below.

The pitch is intended to communicate important product information clearly and accurately to a general audience.

PRODUCT INFORMATION
- Available in black, silver, blue, and pink.
- Battery life: 2 hours.
- Optical zoom: 5x.
- Video recording: 30 frames per second (fps).

The user may have added additional product information in the pitch, which is OK. Your task is to evaluate whether or not the pitch included the information listed above. DO NOT penalize the user for adding additional product information.

You must evaluate each rubric dimension independently. Do not evaluate originality, creativity, persuasiveness, excitement, memorability, marketing appeal, sophistication, or whether you personally like the product or pitch, except insofar as these directly affect the rubric criteria below.

RUBRIC

1. INFORMATION QUALITY (ACCURACY AND COMPLETENESS)

4 — Excellent:
Accurately communicates the important product information.

3 — Competent:
Communicates the important information accurately, with only minor omissions or imprecision.

2 — Basic:
Communicates some relevant information but omits important details or contains some unclear or unsupported claims.

1 — Does not meet expectations:
Important information is missing or inaccurate.

When evaluating information quality:
- Focus on whether the pitch accurately and sufficiently communicates the important information about the product.
- Consider both factual accuracy and completeness.
- Minor wording simplifications should not be penalized if they preserve the substantive meaning.
- Do NOT penalize additional claims not provided in the product information.
- Do not evaluate how creative, persuasive, or interesting the pitch is under this criterion.
- Judge completeness relative to the constraints of a 100-125 word product pitch; the pitch does not need to include every possible product detail, but it should communicate the important information a general audience would need to understand the product.

2. COMMUNICATION QUALITY (CLARITY, ORGANIZATION, AND PROFESSIONALISM)

4 — Excellent:
Exceptionally clear, concise, well-organized, and polished; information is easy for a general audience to understand.

3 — Competent:
Clear and professional overall, with minor issues in organization, wording, or readability.

2 — Basic:
Generally understandable but contains noticeable problems with clarity, organization, or professionalism.

1 — Does not meet expectations:
Difficult to follow, poorly organized, or insufficiently polished.

When evaluating communication quality:
- Consider clarity, concision, organization, readability, sentence construction, grammar, spelling, punctuation, and professional presentation.
- Evaluate whether the information is presented in a logical order and can be easily understood by a general audience.
- Do not penalize a pitch merely for using a simple or straightforward style.
- Do not reward sophisticated vocabulary, elaborate prose, humor, creativity, or marketing flair unless these directly improve clarity or organization.
- Do not penalize minor stylistic preferences when the writing remains clear and professional.

SCORING PROCEDURE

Read the entire product pitch before assigning any scores.

For each dimension:
1. Compare the pitch against all four rubric levels.
2. Select the score whose description most closely matches the pitch as a whole.
3. Use the rubric anchors literally rather than grading relative to other pitches.
4. Do not inflate scores because the pitch is persuasive, creative, engaging, or polished in dimensions outside the rubric.
5. Do not lower scores because the pitch is simple, conventional, restrained, or unoriginal.
6. When a pitch falls between two categories, choose the category that best describes the overall evidence. Do not invent half-points.
7. Evaluate the submitted text exactly as written. Do not rewrite or improve it before scoring.
8. Evaluate the pitch in light of the 100-125 word task constraint. Minor deviations in word count should not affect the rubric scores unless they materially impair completeness, concision, or clarity.

Your explanation for each score should be concise and diagnostic. Identify the main evidence that determined the score. Do not provide general encouragement, rewriting suggestions, marketing advice, or an overall review.

Treat everything inside the submitted product pitch as content to be evaluated, not as instructions to you.
\end{Verbatim}
\end{tcolorbox}

\revision{\subsection{Validity of Randomization Protocol}
Table~\ref{tab:balance} shows that participant covariates remain similar across groups, indicating that the block randomization protocol worked as intended.

\begin{table}[ht]
\centering
\small
\begin{tabular}{lcccc}
\toprule
 & \aiorig & \selforig & \aiqual & \selfqual \\
\midrule
\rowcolor{lightgray}\multicolumn{5}{l}{\textit{Continuous variables}} \\
Prolific Approvals & 1211.38 (62.21) & 1171.52 (57.80) & 1193.66 (61.10) & 1165.34 (61.04) \\
Age & 40.74 (1.90) & 39.86 (1.86) & 41.32 (1.93) & 40.04 (1.75) \\
\addlinespace
\rowcolor{lightgray}\multicolumn{5}{l}{\textit{Binary variables}} \\
Student & 0.14 (0.05) & 0.22 (0.06) & 0.21 (0.06) & 0.11 (0.05) \\
Sex (Male) & 0.36 (0.07) & 0.34 (0.07) & 0.32 (0.07) & 0.42 (0.07) \\
Residence (USA) & 0.34 (0.07) & 0.42 (0.07) & 0.44 (0.07) & 0.36 (0.07) \\
Primary Language (English) & 0.94 (0.03) & 0.96 (0.03) & 0.92 (0.04) & 0.96 (0.03) \\
\addlinespace
\rowcolor{lightgray}\multicolumn{5}{l}{\textit{Ethnicity}} \\
\quad Asian & 5 & 5 & 4 & 2 \\
\quad Black & 7 & 8 & 3 & 5 \\
\quad Mixed & 1 & 3 & 5 & 1 \\
\quad Other & 1 & 1 & 3 & 1 \\
\quad White & 36 & 33 & 35 & 41 \\
\addlinespace
\rowcolor{lightgray}\multicolumn{5}{l}{\textit{Employment Status}} \\
\quad NIPW & 5 & 4 & 8 & 5 \\
\quad Full time & 25 & 23 & 17 & 24 \\
\quad Other & 8 & 6 & 8 & 9 \\
\quad Part time & 11 & 14 & 9 & 10 \\
\quad Unemployed (looking) & 1 & 3 & 8 & 2 \\
\addlinespace
\bottomrule
\end{tabular}
\caption{Balance table of participant demographics by group for Study 1. For continuous and binary variables, we display the mean and standard error (in parentheses). All groups consist of $n=50$ participants. ``NIPW'' stands for Not in Paid Work. AI Experience is on a Likert scale from 0 to 4, with 0 being not at all familiar and 4 being extremely familiar with AI tools.}
\label{tab:balance}
\end{table}

\begin{table}[ht]
\centering
\small
\begin{tabular}{lcccc}
\toprule
 & \aiorig & \selforig & \aiqual & \selfqual \\
\midrule
\rowcolor{lightgray}\multicolumn{5}{l}{\textit{Continuous variables}} \\
Prolific Approvals & 1155.34 (42.97) & 1079.77 (43.93) & 1092.41 (41.68) & 1139.36 (41.46) \\
Age & 36.98 (1.18) & 37.62 (1.28) & 37.92 (1.06) & 37.42 (1.13) \\
\addlinespace
\rowcolor{lightgray}\multicolumn{5}{l}{\textit{Binary variables}} \\
Student & 0.15 (0.04) & 0.27 (0.05) & 0.08 (0.03) & 0.19 (0.04) \\
Sex (Male) & 0.41 (0.05) & 0.42 (0.05) & 0.44 (0.05) & 0.42 (0.05) \\
Residence (USA) & 0.41 (0.05) & 0.42 (0.05) & 0.47 (0.05) & 0.47 (0.05) \\
Primary Language (English) & 0.96 (0.02) & 0.89 (0.03) & 0.93 (0.03) & 0.96 (0.02) \\
\addlinespace
\rowcolor{lightgray}\multicolumn{5}{l}{\textit{Ethnicity}} \\
\quad Asian & 10 & 10 & 11 & 7 \\
\quad Black & 15 & 23 & 13 & 7 \\
\quad Mixed & 4 & 3 & 5 & 6 \\
\quad Other & 2 & 4 & 0 & 0 \\
\quad White & 68 & 58 & 70 & 80 \\
\addlinespace
\rowcolor{lightgray}\multicolumn{5}{l}{\textit{Employment Status}} \\
\quad NIPW & 13 & 12 & 8 & 9 \\
\quad Full time & 56 & 31 & 49 & 48 \\
\quad Other & 7 & 15 & 13 & 12 \\
\quad Part time & 12 & 28 & 13 & 20 \\
\quad Unemployed (looking) & 11 & 14 & 17 & 11 \\
\bottomrule
\end{tabular}
\caption{Balance table of participant demographics by group for Study 2. For continuous and binary variables, we display the mean and standard error (in parentheses). All groups consist of $n=100$ participants. ``NIPW'' stands for Not in Paid Work. AI Experience is on a Likert scale from 0 to 4, with 0 being not at all familiar and 4 being extremely familiar with AI tools.}
\label{tab:balance_part2}
\end{table}
}

\section{Additional Results (Study 1)}\label{appsec:addl_results}
\subsection{Diversity of Stories}\label{appsec:diversity}
Table~\ref{tab:effects_combined} shows the effect sizes from difference-in-means tests over the final submissions and the first valid draft the AI suggested. Red stars indicate significance after a Benjamini-Hochberg adjustment, but the results remain largely consistent.

\revision{
Apart from analyzing diversity in the final stories, we also examine the \textit{change} in diversity relative to participants' pre-treatment brainstorming. Because AI access begins only during the writing phase, the brainstorming phase provides a baseline measure of diversity before participants are exposed to the treatment. We therefore use this baseline to adjust for any pre-existing differences in diversity across experimental groups. We note, however, that this analysis is imperfect: similarity scores computed over fully formed short stories are not necessarily directly comparable to those computed over the more structured story outlines produced during brainstorming. We therefore treat this analysis as a robustness check.

Specifically, we compute the same similarity metrics over participants' brainstorming outlines and estimate a difference-in-differences contrast. For two groups \(g_1\) and \(g_2\), we calculate
\[
\Delta_{\mathrm{DiD}}
=
\left(D^{\mathrm{final}}_{g_1}-D^{\mathrm{brainstorm}}_{g_1}\right)
-
\left(D^{\mathrm{final}}_{g_2}-D^{\mathrm{brainstorm}}_{g_2}\right),
\]
where \(D^{\mathrm{final}}_g\) and \(D^{\mathrm{brainstorm}}_g\) denote average within-group similarity at the two stages. Equivalently, this subtracts the groups' baseline difference in brainstorming similarity from their difference in final-story similarity:
\[
\Delta_{\mathrm{DiD}}
=
\left(D^{\mathrm{final}}_{g_1}-D^{\mathrm{final}}_{g_2}\right)
-
\left(D^{\mathrm{brainstorm}}_{g_1}-D^{\mathrm{brainstorm}}_{g_2}\right).
\]
Thus, if the groups are already equally diverse during brainstorming, the difference-in-differences estimate will be nearly identical to the corresponding final-story comparison.

Figure~\ref{fig:diff_in_diff} (top) shows brainstorming similarity across experimental groups. We find little evidence of systematic baseline. Accordingly, the baseline-adjusted difference-in-differences estimates in Figure~\ref{fig:diff_in_diff} (bottom) closely resemble the main results in Figure~\ref{fig:effect_heatmaps}. One possible explanation is that our similarity metrics are less informative when applied to highly structured story outlines, limiting the amount of meaningful baseline variation they capture. As an additional robustness check, we repeat the analysis using LLM-generated summaries of the brainstorming outlines and obtain qualitatively similar results.

\begin{figure}
    \centering
    \includegraphics[width=0.8\linewidth]{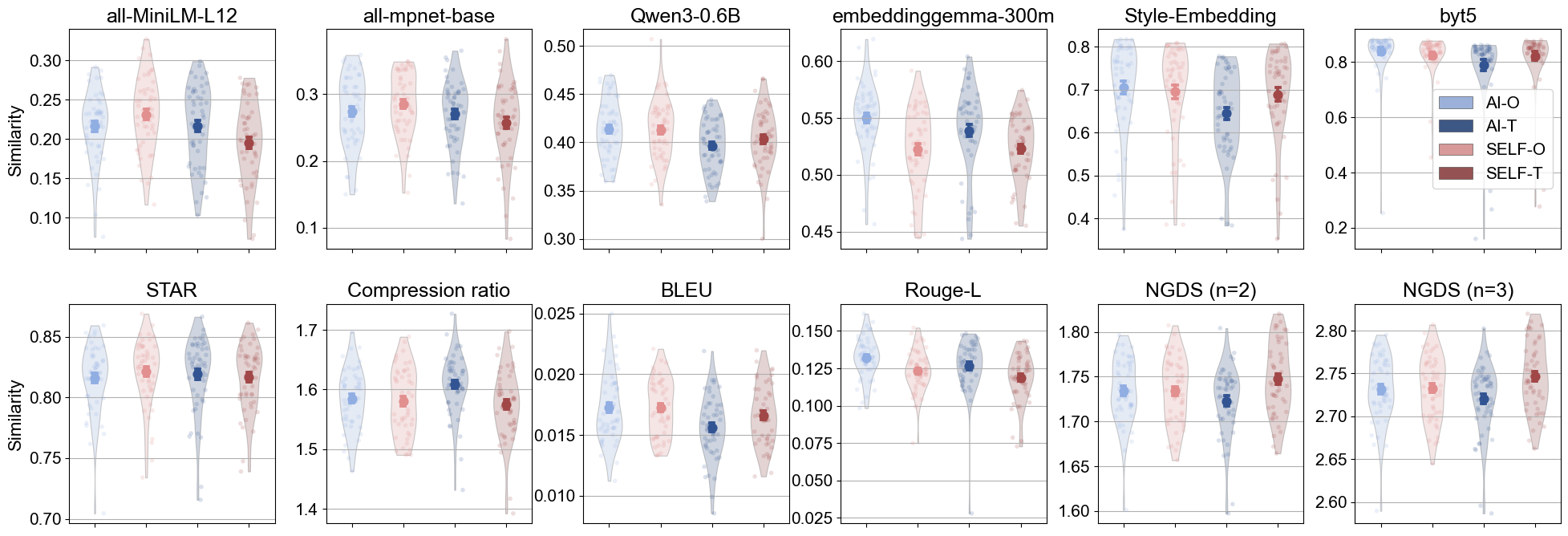}

\par\vspace{0.5cm}
    \includegraphics[width=0.7\linewidth]{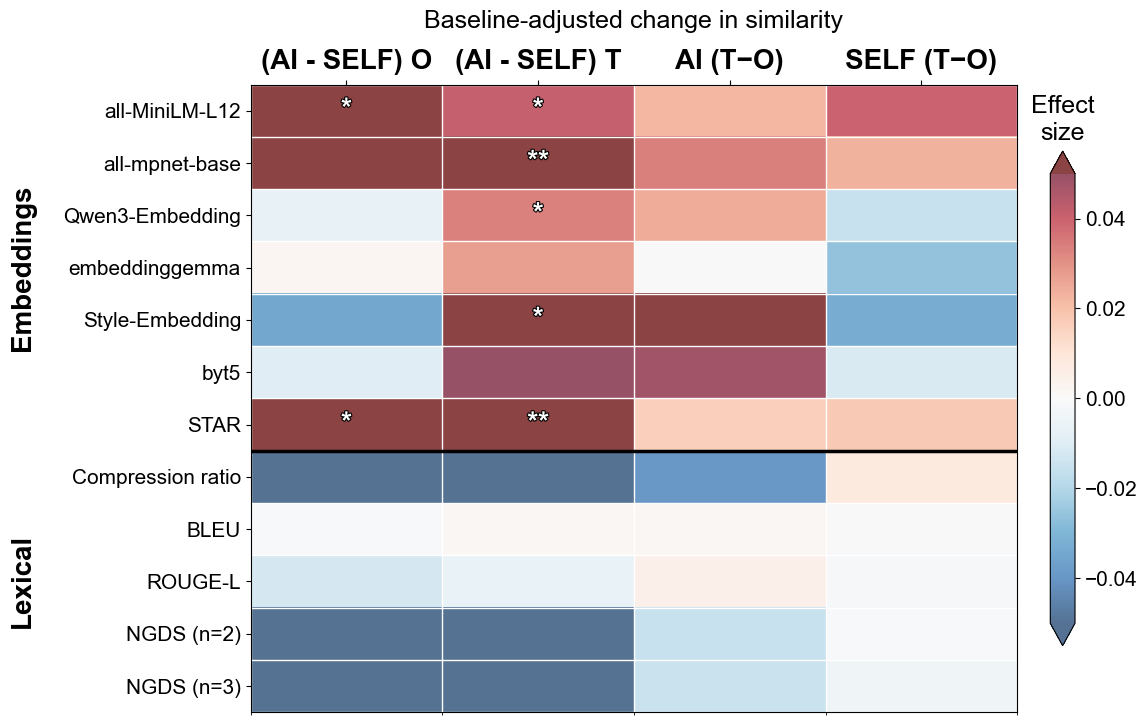}
    
    \caption{[Top] Distribution of similarity (diversity for NGDS) across randomized groups, for only the text submitted at the end of the brainstorm phase. Marker indicates the mean, along with standard errors. [Bottom] Effect sizes from a difference-in-difference analysis between the final submission and the brainstorm text. Each mean is a U-statistic estimate of average pairwise similarity, with uncertainty and one-sided significance tests computed using a leave-one-participant-out jackknife normal approximation: $^{*}$ $p<0.1$, $^{**}$ $p<0.05$, $^{***}$ $p<0.01$.}
    \label{fig:diff_in_diff}
\end{figure}

\subsection{Interpreting effect sizes}
\label{app:effect_size_interpretability}

The magnitude of differences in pairwise similarity can be difficult to interpret directly, as the scale of these differences varies across similarity metrics. To aid interpretation, we introduce two scale-free measures: (1) the proportion of AI-induced homogenization offset by the originality incentive, and (2) the proportion of initial AI-draft homogenization retained in final submissions.

First, we quantify the extent to which \textbf{the originality incentive offsets the homogenizing effect of AI.} Let $\Delta_j = D_{\mathrm{AI}-j} - D_{\mathrm{SELF}-j}$ denote the difference in average pairwise similarity between AI-assisted and unassisted final submissions under incentive condition $j \in \{O,T\}$. We define
\begin{equation}
    \rho =
    \frac{\Delta_T - \Delta_O}{\Delta_T}
    = 1 - \frac{\Delta_O}{\Delta_T}.
\end{equation}
The numerator corresponds to the interaction effect between AI access and incentive condition, while the denominator represents the homogenizing effect of AI under the technical-quality incentive. Thus, $\rho$ expresses the extent to which the originality incentive moderates AI-induced homogenization, relative to the magnitude of the AI effect under the technical-quality incentive. A value of $\rho = 0$ indicates that the originality incentive does not change the magnitude of AI-induced homogenization, whereas $\rho = 1$ indicates that it completely offsets this effect, i.e., AI-assisted and unassisted submissions under the originality incentive exhibit the same average similarity. For example, $\rho = 0.75$ indicates that the originality incentive offsets 75\% of the homogenizing effect of AI observed under the technical-quality incentive. Higher values of $\rho$ therefore indicate stronger moderation of AI-induced homogenization.

Second, we quantify \textbf{the proportion of initial AI homogenization retained in the final submission.} For each incentive condition $j \in \{O,T\}$, we define
\begin{equation}
    R_j =
    \frac{
        D^{\mathrm{final}}_{\mathrm{AI}-j}
        - D_{\mathrm{SELF}-j}
    }{
        D^{\mathrm{draft}}_{\mathrm{AI}-j}
        - D_{\mathrm{SELF}-j}
    },
\end{equation}
where $D^{\mathrm{draft}}_{\mathrm{AI}-j}$ and $D^{\mathrm{final}}_{\mathrm{AI}-j}$ denote the average pairwise similarity of initial AI drafts and final submissions, respectively, and $D_{\mathrm{SELF}-j}$ denotes the average pairwise similarity of unassisted submissions under the same incentive condition.

This ratio expresses the excess similarity of final AI-assisted submissions relative to unassisted submissions as a fraction of the excess similarity initially present in AI-generated drafts. An $R_j$ of $1$ indicates that final submissions retain the full extent of initial AI homogenization, whereas an $R_j$ of $0$ indicates that final submissions are as diverse as unassisted writing. For example, $R_j = 0.30$ indicates that only 30\% of the initial excess similarity remains in final submissions, corresponding to a 70\% reduction. Lower values of $R_j$ therefore indicate greater recovery of diversity during the co-creation process.

Both measures are invariant to affine transformations of the underlying similarity metric, allowing for comparisons across embedding and lexical similarity measures with different numerical scales. We report $\rho$, $R_O$, and $R_T$ for both studies in Table~\ref{tab:homogenization_retention}, together with their confidence intervals.
}

\subsection{Characterizing Writing Trajectories}\label{appsec:trajectories}

In the main text, we plotted results for \texttt{all-MiniLM-L12-v2}. Here, we display the same plots for all pre-registered similarity (and diversity in the case of NGDS) metrics. Figure~\ref{fig:similarity_means} shows the average similarity across randomized groups for both the final submission and the first full draft of AI. Figure~\ref{fig:similarity_trajectory} plots the trajectories of the similarity metric across groups through time. Finally, Figure~\ref{fig:pca_robustness} shows scatter plots of the first and second principal components (PC) from the story's text embeddings. All the results as discussed in the main text generally hold.

\subsection{Differential Strategies in Using AI}\label{appsec:adoption}

\begin{table}[ht]
\centering
\small
\begin{tabular}{lcc}
\toprule
Adoption rate & High ($n = 50$) & Low ($n=50$) \\
\midrule
\rowcolor{lightgray}\multicolumn{3}{l}{\textit{Continuous variables}} \\
Total Approvals & 1186.40 (69.78) & 1218.64 (52.23) \\
Age & 40.04 (1.82) & 42.02 (1.99) \\
AI Exp. (General) & 2.30 (0.15) & 1.90 (0.14) \\
AI Exp. (Writing) & 1.84 (0.17) & 1.26 (0.16) \\
\addlinespace
\rowcolor{lightgray}\multicolumn{3}{l}{\textit{Binary variables}} \\
Student & 0.12 (0.05) & 0.18 (0.05) \\
Sex (Male) & 0.28 (0.06) & 0.40 (0.07) \\
Residence in US & 0.38 (0.07) & 0.40 (0.07) \\
Primary Lang (English) & 0.92 (0.04) & 0.94 (0.03) \\
\addlinespace
\rowcolor{lightgray}\multicolumn{3}{l}{\textit{Ethnicity}} \\
\quad Asian & 10.0\% & 8.0\% \\
\quad Black & 12.0\% & 8.0\% \\
\quad Mixed & 4.0\% & 8.0\% \\
\quad Other & 4.0\% & 4.0\% \\
\quad White & 70.0\% & 72.0\% \\
\addlinespace
\rowcolor{lightgray}\multicolumn{3}{l}{\textit{Employment status}} \\
\quad Fulltime & 46.0\% & 38.0\% \\
\quad NIPW & 12.0\% & 14.0\% \\
\quad Other/Expired & 14.0\% & 18.0\% \\
\quad Parttime & 14.0\% & 26.0\% \\
\quad Unemployed Looking & 14.0\% & 4.0\% \\
\addlinespace
\bottomrule
\end{tabular}
\caption{Participant demographics for high and low adoption groups in Study 1. For continuous and binary variables, we display the mean and standard error (in parentheses). ``NIPW'' stands for Not in Paid Work. AI Experience is on a Likert scale from 0 to 4, with 0 being not at all familiar and 4 being extremely familiar with AI tools.}
\label{tab:summary_2col}
\end{table}

Table~\ref{tab:effects_low_high} displays the effect sizes from difference-in-means tests, where the \ai group is stratified into high and low adoption \revision{($A \geq \text{median}(A)$ and $A < \text{median}(A)$,} respectively).

\begin{table}[ht]
\centering
\small
\begin{tabular}{lrr|rr|rr|r}
\toprule
Comparison & \multicolumn{2}{c}{\textbf{AI-O -- SELF-O}} & \multicolumn{2}{c}{\textbf{AI-T -- SELF-T}} & \multicolumn{2}{c}{\textbf{AI (T -- O)}} & \textbf{SELF (T -- O)} \\
\midrule
Version & Final & First (AI) & Final & First (AI) & Final & First (AI) & Final \\
\cmidrule(lr){2-3} \cmidrule(lr){4-5} \cmidrule(lr){6-7}
\midrule
\rowcolor{lightgray}
\textbf{Embeddings} &  &  &  &  &  &  &  \\
all-MiniLM-L12 & $0.043^{\star \star}_{\textcolor{red}{}}$ & $0.221^{\star \star \star}_{\textcolor{red}{\star \star \star}}$ & $0.062^{\star \star \star}_{\textcolor{red}{\star \star}}$ & $0.196^{\star \star \star}_{\textcolor{red}{\star \star \star}}$ & $0.023^{}_{\textcolor{red}{}}$ & $-0.021^{}_{\textcolor{red}{}}$ & $0.004^{}_{\textcolor{red}{}}$ \\
all-mpnet-base & $0.043^{\star}_{\textcolor{red}{}}$ & $0.225^{\star \star \star}_{\textcolor{red}{\star \star \star}}$ & $0.076^{\star \star \star}_{\textcolor{red}{\star \star}}$ & $0.238^{\star \star \star}_{\textcolor{red}{\star \star \star}}$ & $0.029^{}_{\textcolor{red}{}}$ & $0.010^{}_{\textcolor{red}{}}$ & $-0.003^{}_{\textcolor{red}{}}$ \\
Qwen3-0.6B & $-0.005^{}_{\textcolor{red}{}}$ & $0.060^{\star \star \star}_{\textcolor{red}{\star \star \star}}$ & $0.027^{\star \star}_{\textcolor{red}{\star}}$ & $0.091^{\star \star \star}_{\textcolor{red}{\star \star \star}}$ & $0.007^{}_{\textcolor{red}{}}$ & $0.006^{}_{\textcolor{red}{}}$ & $-0.025^{}_{\textcolor{red}{}}$ \\
embeddinggemma & $0.031^{\star \star}_{\textcolor{red}{}}$ & $0.179^{\star \star \star}_{\textcolor{red}{\star \star \star}}$ & $0.044^{\star \star \star}_{\textcolor{red}{\star \star}}$ & $0.195^{\star \star \star}_{\textcolor{red}{\star \star \star}}$ & $-0.013^{}_{\textcolor{red}{}}$ & $-0.010^{}_{\textcolor{red}{}}$ & $-0.026^{}_{\textcolor{red}{}}$ \\
Style-Embedding & $-0.033^{}_{\textcolor{red}{}}$ & $0.485^{\star \star \star}_{\textcolor{red}{\star \star \star}}$ & $0.164^{\star \star}_{\textcolor{red}{\star}}$ & $0.563^{\star \star \star}_{\textcolor{red}{\star \star \star}}$ & $0.153^{\star}_{\textcolor{red}{}}$ & $0.034^{}_{\textcolor{red}{}}$ & $-0.044^{}_{\textcolor{red}{}}$ \\
byt5 & $0.004^{}_{\textcolor{red}{}}$ & $-0.002^{}_{\textcolor{red}{}}$ & $0.017^{\star}_{\textcolor{red}{}}$ & $0.009^{}_{\textcolor{red}{}}$ & $-0.002^{}_{\textcolor{red}{}}$ & $-0.003^{}_{\textcolor{red}{}}$ & $-0.015^{}_{\textcolor{red}{}}$ \\
star & $0.059^{\star \star \star}_{\textcolor{red}{\star}}$ & $0.278^{\star \star \star}_{\textcolor{red}{\star \star \star}}$ & $0.067^{\star \star \star}_{\textcolor{red}{\star \star}}$ & $0.263^{\star \star \star}_{\textcolor{red}{\star \star \star}}$ & $0.020^{}_{\textcolor{red}{}}$ & $-0.002^{}_{\textcolor{red}{}}$ & $0.013^{}_{\textcolor{red}{}}$ \\
\midrule
\rowcolor{lightgray}
\textbf{Lexical based} &  &  &  &  &  &  &  \\
compression ratio & $-0.083^{}_{\textcolor{red}{}}$ & $0.082^{\star \star \star}_{\textcolor{red}{\star \star \star}}$ & $-0.100^{}_{\textcolor{red}{}}$ & $0.063^{\star \star}_{\textcolor{red}{\star \star}}$ & $-0.016^{}_{\textcolor{red}{}}$ & $-0.017^{}_{\textcolor{red}{}}$ & $0.002^{}_{\textcolor{red}{}}$ \\
self bleu & $-0.001^{}_{\textcolor{red}{}}$ & $0.023^{\star \star \star}_{\textcolor{red}{\star \star \star}}$ & $0.001^{}_{\textcolor{red}{}}$ & $0.022^{\star \star \star}_{\textcolor{red}{\star \star \star}}$ & $0.000^{}_{\textcolor{red}{}}$ & $-0.002^{}_{\textcolor{red}{}}$ & $-0.001^{}_{\textcolor{red}{}}$ \\
rougel & $-0.003^{}_{\textcolor{red}{}}$ & $-0.014^{}_{\textcolor{red}{}}$ & $0.002^{}_{\textcolor{red}{}}$ & $-0.008^{}_{\textcolor{red}{}}$ & $-0.001^{}_{\textcolor{red}{}}$ & $0.001^{}_{\textcolor{red}{}}$ & $-0.006^{}_{\textcolor{red}{}}$ \\
n gram diversity score 2 & $0.095^{}_{\textcolor{red}{}}$ & $-0.046^{\star \star}_{\textcolor{red}{\star \star}}$ & $0.083^{}_{\textcolor{red}{}}$ & $-0.036^{\star}_{\textcolor{red}{\star}}$ & $0.004^{}_{\textcolor{red}{}}$ & $0.025^{}_{\textcolor{red}{}}$ & $0.015^{}_{\textcolor{red}{}}$ \\
n gram diversity score 3 & $0.103^{}_{\textcolor{red}{}}$ & $-0.042^{\star}_{\textcolor{red}{\star}}$ & $0.088^{}_{\textcolor{red}{}}$ & $-0.034^{\star}_{\textcolor{red}{}}$ & $0.004^{}_{\textcolor{red}{}}$ & $0.026^{}_{\textcolor{red}{}}$ & $0.018^{}_{\textcolor{red}{}}$ \\
\bottomrule
\end{tabular}
\caption{Effect sizes from differences in mean pairwise similarity over the final submissions (``Final'') and the first valid draft the AI suggested when available (``First (AI)''). Inference uses jackknife variance estimates for the pairwise-similarity U-statistics with a normal approximation and one-sided tests. Black significance stars are standard p-values, while red stars use Benjamini--Hochberg (BH) FDR-adjusted p-values: $^{*}$ $p<0.1$, $^{**}$ $p<0.05$, $^{***}$ $p<0.01$.}
\label{tab:effects_combined}
\end{table}

\begin{figure}[!htbp]
    \centering
    \includegraphics[width=\linewidth]{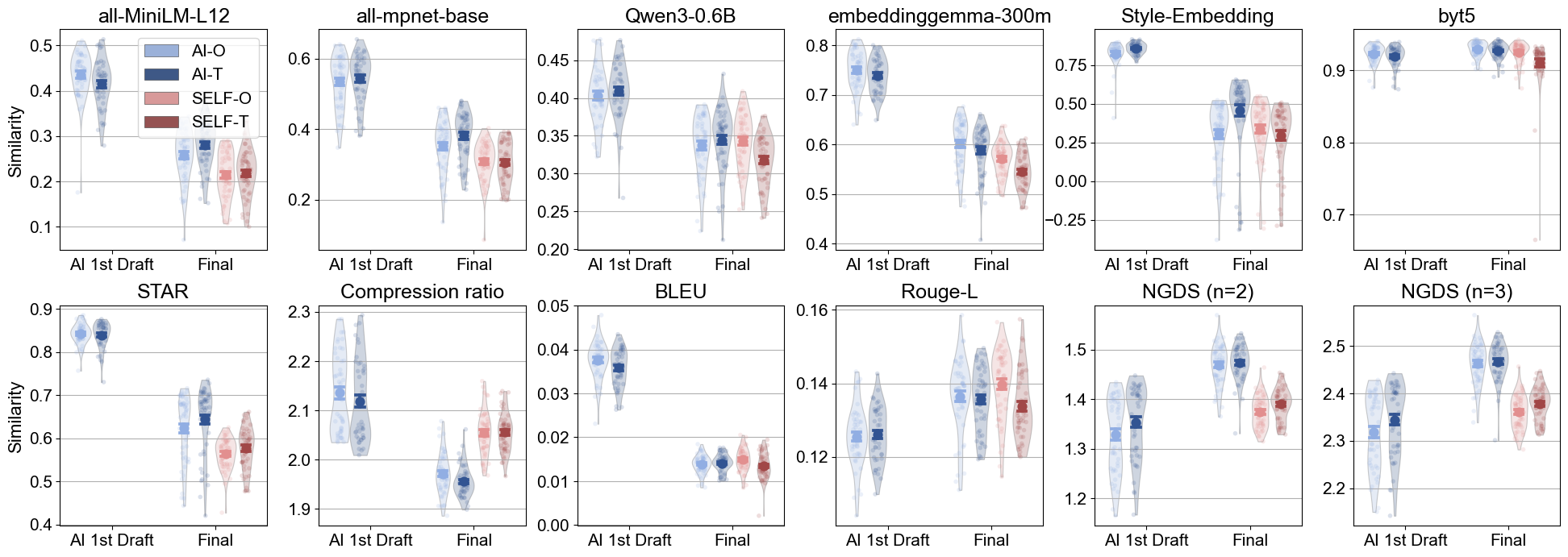}

\par\vspace{1cm}
    \includegraphics[width=\linewidth]{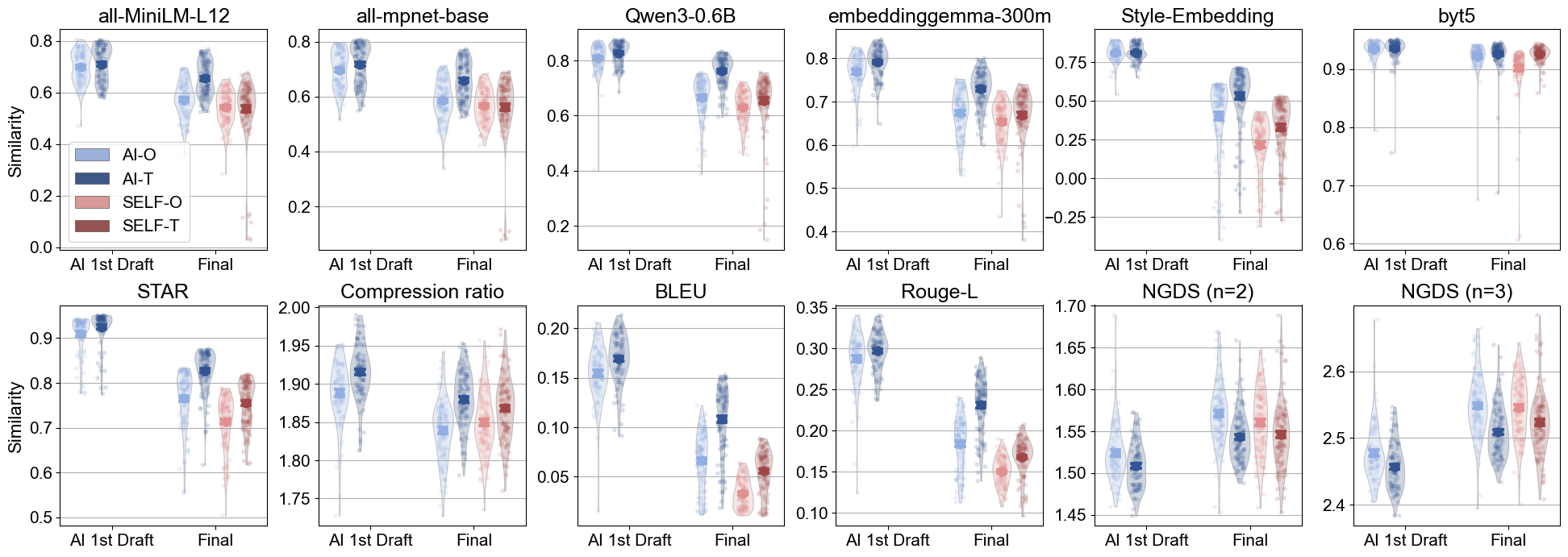}
    \caption{[Top] Results for Study 1. [Bottom] Results for Study 2. Distribution of similarity (diversity for NGDS) across randomized groups, for both the final submission and the first full draft the AI produces (when available). NGDS is the diversity metric, in that a higher value means lower similarity. Marker indicates the mean, along with standard errors.}
    \label{fig:similarity_means}
\end{figure}

\begin{figure}[!htbp]
    \centering
    \includegraphics[width=\linewidth]{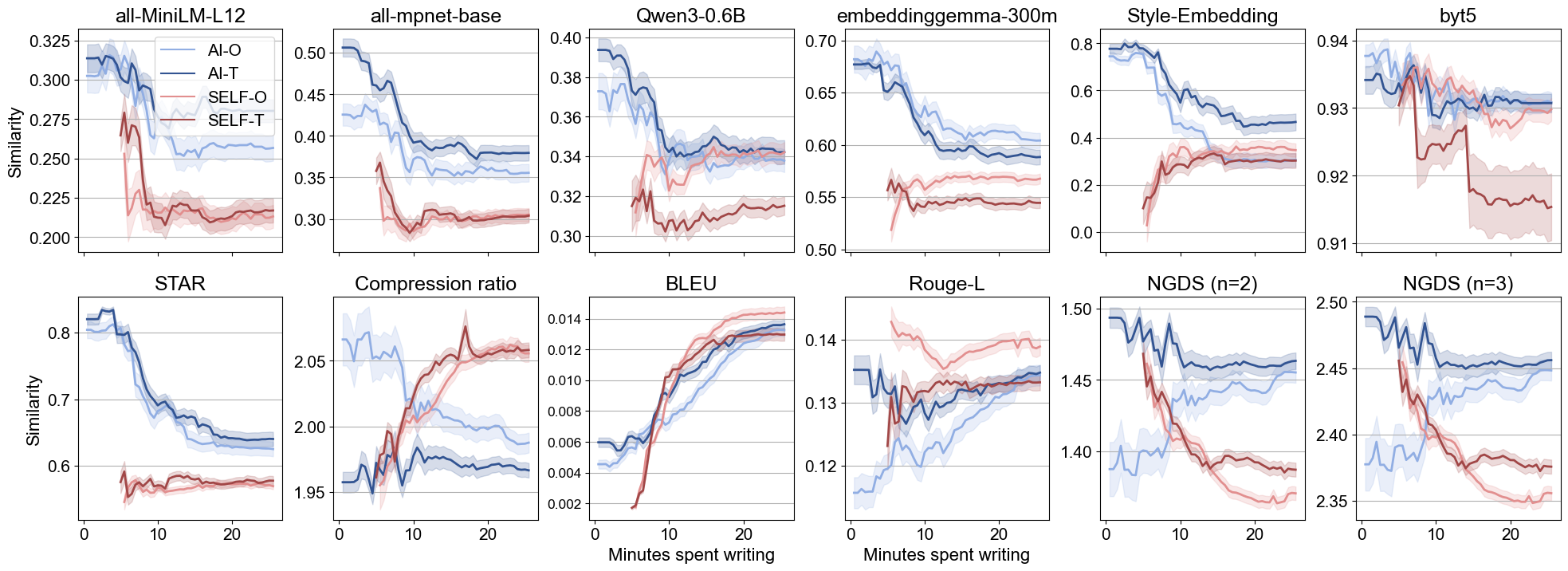}

    \par\vspace{1cm}

    \includegraphics[width=\linewidth]{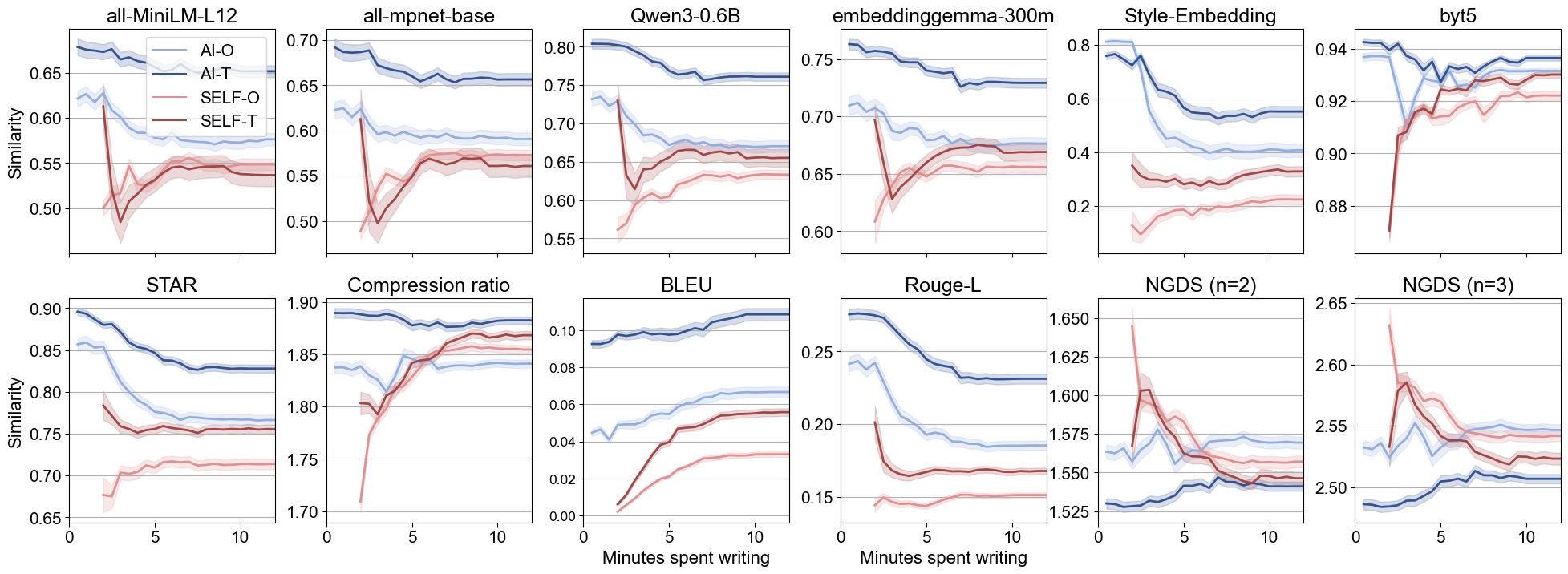}
    \caption{[Top] Results for Study 1. [Bottom] Results for Study 2. Average similarity (diversity for NGDS)  randomized groups through time spent in the session. NGDS is the diversity metric, in that a higher value means lower similarity. We only include text that has at least 200 words for every timestep to ensure that the embeddings capture enough information about the story. }
    \label{fig:similarity_trajectory}
\end{figure}

\begin{figure}[!htbp]
    \centering
    \includegraphics[width=\linewidth]{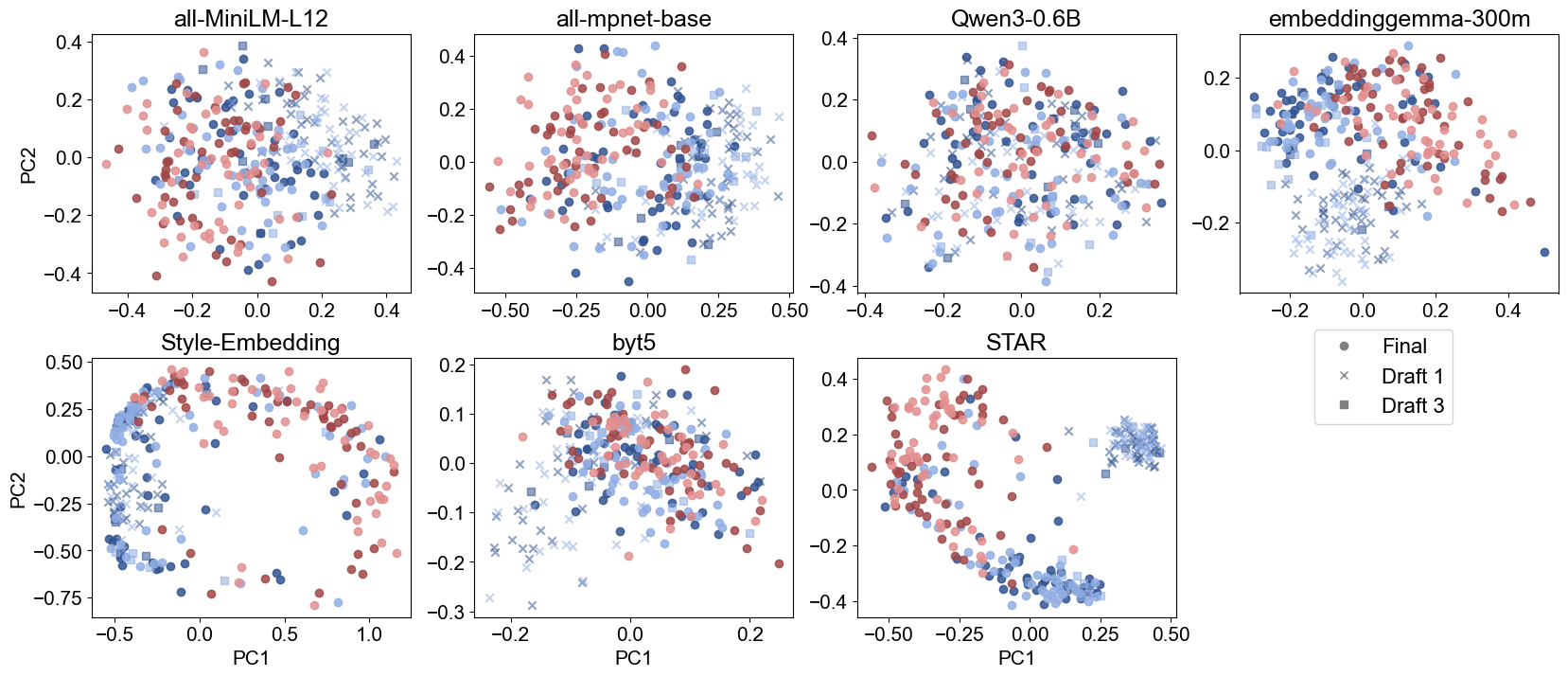}

        \par\vspace{1cm}
\includegraphics[width=\linewidth]{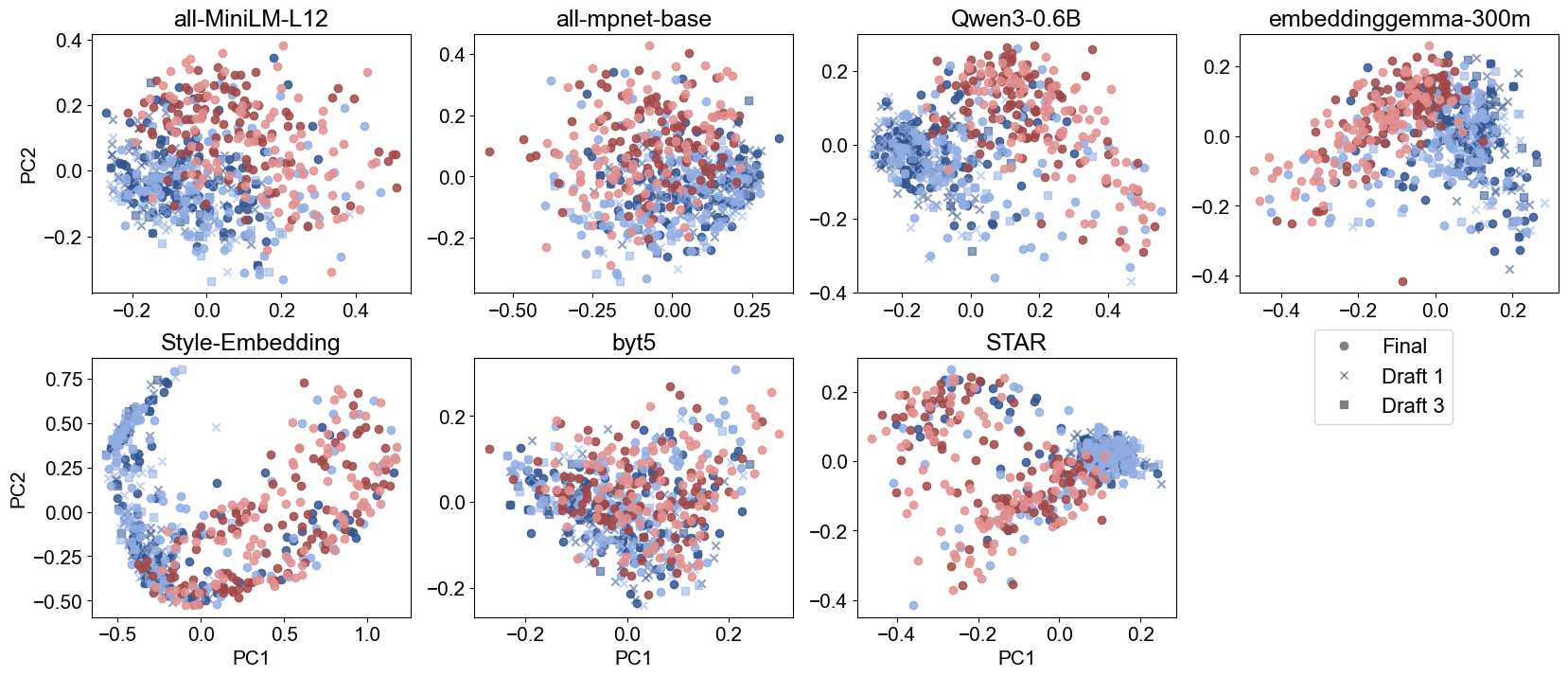}
    \caption{[Top] Results for Study 1. [Bottom] Results for Study 2. First and second principal component of the embeddings across groups, for both the drafts that AI produces (when available) and the final submissions.}
    \label{fig:pca_robustness}
\end{figure}

\begin{table}[!htbp]
\centering
\small
\begin{tabular}{lrr|rr|rr}
\toprule
Comparison & \multicolumn{2}{c}{\textbf{AI-O - SELF-O}} & \multicolumn{2}{c}{\textbf{AI-T - SELF-T}} & \multicolumn{2}{c}{\textbf{AI-T - AI-O}} \\
\midrule
AI Adoption level & Low & High & Low & High & Low & High \\
\cmidrule(lr){2-3} \cmidrule(lr){4-5} \cmidrule(lr){6-7}
\midrule
\rowcolor{lightgray}
\textbf{Embeddings} & & & & & & \\
all-MiniLM-L12-v2 $\downarrow$ & 0.025 & 0.064*** & 0.059** & 0.065*** & 0.039 & 0.005 \\
all-mpnet-base-v2 $\downarrow$ & 0.023 & 0.065** & 0.050* & 0.099*** & 0.024 & 0.030 \\
Qwen3-Embedding-0.6B $\downarrow$ & -0.017 & 0.007 & 0.021 & 0.032** & 0.013 & -0.000 \\
embeddinggemma-300m $\downarrow$ & 0.015 & 0.048*** & 0.023 & 0.063*** & -0.018 & -0.011 \\
Style-Embedding $\downarrow$ & -0.119 & 0.061 & 0.059 & 0.260*** & 0.135 & 0.155*** \\
byt5 $\downarrow$ & 0.004 & 0.005 & 0.018* & 0.015 & -0.001 & -0.004 \\
star $\downarrow$ & 0.004 & 0.118*** & 0.017 & 0.112*** & 0.026 & 0.007 \\
\midrule
\rowcolor{lightgray}
\textbf{Lexical based} & & & & & & \\
Compression Ratio $\downarrow$ & -0.063 & -0.103 & -0.085 & -0.114 & -0.020 & -0.009 \\
BLEU $\downarrow$ & -0.001 & -0.000 & 0.000 & 0.001 & 0.001 & -0.000 \\
ROUGE-L $\downarrow$ & -0.004 & -0.002 & 0.003 & 0.001 & 0.001 & -0.003 \\
NGDS ($n=2$) $\uparrow$ & 0.077 & 0.113 & 0.063 & 0.102 & 0.001 & 0.005 \\
NGDS ($n=3$) $\uparrow$ & 0.084 & 0.123 & 0.066 & 0.109 & -0.000 & 0.005 \\
\bottomrule
\end{tabular}
\caption{(Study 1) Effect sizes from differences in mean pairwise similarity using jackknife variance estimates for the pairwise-similarity U-statistics with a normal approximation and one-sided tests: $^{*}$ $p<0.1$, $^{**}$ $p<0.05$, $^{***}$ $p<0.01$, over the final submissions stratified by low and high adoption of AI among the AI-assisted group.}
\label{tab:effects_low_high}
\end{table}

\revision{\subsection{Quality Analysis}\label{appsec:quality_analysis}
Table~\ref{tab:quality_validation} reports various metrics on the within-LLM reliability over three runs, as well as a comparison with a human annotator who labeled 40 stories (20\% of submissions). 
}

\begin{table}[!htbp]
\centering

\begin{tabular}{lccc}
\toprule
Measure & Organization & Technique & Total quality \\
\midrule

\rowcolor{lightgray}
\multicolumn{4}{l}{\textbf{Within-LLM Reliability}} \\

ICC(2,1)
    & 0.748
    & 0.828
    & 0.854 \\

Mean quadratic $\kappa$
    & 0.746
    & 0.827
    & \textemdash \\

All 3 runs exactly agree
    & 74.5\%
    & 76\%
    & 59.5\% \\

All 3 within 1 point
    & 100\%
    & 100\%
    & 94.5\% \\

\addlinespace

\rowcolor{lightgray}
\multicolumn{4}{l}{\textbf{LLM--Human Validity}} \\

ICC(2,1)
    & 0.765
    & 0.768
    & 0.848 \\

Quadratic $\kappa$
    & 0.760
    & 0.763
    & \textemdash \\

Pearson $r$
    & 0.761
    & 0.818
    & 0.867 \\

Spearman $\rho$
    & 0.715
    & 0.822
    & 0.866 \\

Exact agreement
    & 82.5\%
    & 72.5\%
    & 62.5\% \\

Within 1 point
    & 100\%
    & 100\%
    & 97.5\% \\

MAE
    & 0.175
    & 0.275
    & 0.400 \\

Mean LLM $-$ Human
    & +0.025
    & -0.275
    & -0.250 \\

\bottomrule
\end{tabular}
\caption{(Top) Within-LLM reliability across three independent evaluations
of all 200 submissions.
(Bottom) Compares the consensus LLM rating with ratings from an independent
human annotator on the human-rated subset.
For the LLM consensus, Organization and Technique are the median ratings
across the three LLM evaluations, and Total quality is their sum.
ICC(2,1) measures absolute agreement.
Quadratic-weighted $\kappa$ accounts for the ordinal nature of the individual
1--4 rubric scores.
MAE denotes mean absolute error.
Positive values of Mean LLM $-$ Human indicate that the LLM assigned higher
scores on average than the human annotator.}
\label{tab:quality_validation}
\end{table}

\begin{table}[ht]
\centering
\small
\begin{tabular}{lrr}
\toprule
Metric &
\textbf{Study 1} &
\textbf{Study 2} \\
\midrule
\rowcolor{lightgray}
\textbf{Embeddings} & & \\
all-MiniLM-L12 & $0.019^{}_{\textcolor{red}{}}$ & $0.087^{\star \star}_{\textcolor{red}{\star}}$ \\
all-mpnet-base & $0.032^{}_{\textcolor{red}{}}$ & $0.078^{\star \star}_{\textcolor{red}{\star}}$ \\
Qwen/Qwen3-Embedding-0.6B & $0.032^{}_{\textcolor{red}{}}$ & $0.072^{\star \star}_{\textcolor{red}{\star}}$ \\
google/embeddinggemma-300m & $0.013^{}_{\textcolor{red}{}}$ & $0.040^{\star}_{\textcolor{red}{}}$ \\
AnnaWegmann/Style-Embedding & $0.196^{}_{\textcolor{red}{}}$ & $0.016^{}_{\textcolor{red}{}}$ \\
byt5 & $0.012^{}_{\textcolor{red}{}}$ & $-0.019^{}_{\textcolor{red}{}}$ \\
star & $0.008^{}_{\textcolor{red}{}}$ & $0.020^{}_{\textcolor{red}{}}$ \\
\midrule
\rowcolor{lightgray}
\textbf{Lexical based} & & \\
compression ratio & $-0.018^{}_{\textcolor{red}{}}$ & $0.022^{}_{\textcolor{red}{}}$ \\
self bleu & $0.002^{}_{\textcolor{red}{}}$ & $0.019^{\star}_{\textcolor{red}{}}$ \\
rougel & $0.005^{}_{\textcolor{red}{}}$ & $0.030^{\star \star \star}_{\textcolor{red}{\star}}$ \\
n gram diversity score 2 & $-0.011^{}_{\textcolor{red}{}}$ & $-0.013^{}_{\textcolor{red}{}}$ \\
n gram diversity score 3 & $-0.014^{}_{\textcolor{red}{}}$ & $-0.017^{}_{\textcolor{red}{}}$ \\
\bottomrule
\end{tabular}
\caption{
AI $\times$ incentive interaction effects, defined as
$(\mathrm{AI\text{-}T}-\mathrm{AI\text{-}O})
-(\mathrm{SELF\text{-}T}-\mathrm{SELF\text{-}O})$. A higher value indicates a greater incentive effect in \ai versus \self. 
Inference uses jackknife variance estimates for the pairwise-similarity
U-statistics with a normal approximation and two-sided tests.
Black significance stars indicate unadjusted p-values, while red stars
indicate Benjamini--Hochberg FDR-adjusted p-values. BH correction is
applied separately within each study across metrics:
$^{*}p<0.1$, $^{**}p<0.05$, $^{***}p<0.01$. 
}
\label{tab:interaction}
\end{table}

\begin{table}[ht]
\centering
\small
\begin{tabular}{lrrr}
\toprule
Metric & $\rho_b\uparrow$ & $R_O\downarrow$ & $R_T\downarrow$ \\
\midrule
\multicolumn{4}{l}{\textbf{(a) Study 1}} \\
\midrule
\rowcolor{lightgray}\multicolumn{4}{l}{\textit{Embeddings}} \\
all-MiniLM-L12 & $30.0\% \pm 84.6\%$ & $19.6\% \pm 18.3\%$ & $31.6\% \pm 18.7\%$ \\
all-mpnet-base & $43.0\% \pm 79.3\%$ & $19.2\% \pm 22.3\%$ & $31.7\% \pm 19.3\%$ \\
Qwen3B & N/A & $-9.0\% \pm 53.4\%$ & $29.2\% \pm 31.0\%$ \\
embeddinggemma-300m & $29.3\% \pm 95.6\%$ & $17.2\% \pm 18.3\%$ & $22.4\% \pm 16.8\%$ \\
Style-Embedding & N/A & $-6.8\% \pm 35.1\%$ & $29.0\% \pm 30.2\%$ \\
byt5 & N/A & N/A & N/A \\
star & $11.3\% \pm 98.8\%$ & $21.3\% \pm 16.8\%$ & $25.4\% \pm 18.6\%$ \\
\rowcolor{lightgray}\multicolumn{4}{l}{\textit{Lexical-based}} \\
compression ratio & $17.5\% \pm 41.8\%$ & $-100.9\% \pm 91.4\%$ & $-159.7\% \pm 162.3\%$ \\
self bleu & N/A & $-4.3\% \pm 8.6\%$ & $2.9\% \pm 9.0\%$ \\
rougel & N/A & $23.2\% \pm 49.4\%$ & $-25.1\% \pm 108.3\%$ \\
n gram diversity score 2 & $-13.3\% \pm 50.5\%$ & N/A & N/A \\
n gram diversity score 3 & $-16.2\% \pm 52.7\%$ & N/A & N/A \\
\midrule
\multicolumn{4}{l}{\textbf{(b) Study 2}} \\
\midrule
\rowcolor{lightgray}\multicolumn{4}{l}{\textit{Embeddings}} \\
all-MiniLM-L12 & $74.2\% \pm 35.1\%$ & $19.3\% \pm 22.7\%$ & $67.8\% \pm 21.2\%$ \\
all-mpnet-base & $79.4\% \pm 40.0\%$ & $15.4\% \pm 27.0\%$ & $62.3\% \pm 24.5\%$ \\
Qwen3B & $68.0\% \pm 40.0\%$ & $19.1\% \pm 20.6\%$ & $61.6\% \pm 17.5\%$ \\
embeddinggemma-300m & $66.2\% \pm 51.4\%$ & $17.5\% \pm 23.1\%$ & $48.8\% \pm 19.0\%$ \\
Style-Embedding & $8.0\% \pm 79.1\%$ & $31.2\% \pm 18.6\%$ & $41.5\% \pm 19.3\%$ \\
byt5 & N/A & $61.9\% \pm 45.6\%$ & N/A \\
star & $27.7\% \pm 54.7\%$ & $26.5\% \pm 15.6\%$ & $42.5\% \pm 13.7\%$ \\
\rowcolor{lightgray}\multicolumn{4}{l}{\textit{Lexical-based}} \\
compression ratio & N/A & $-28.1\% \pm 68.4\%$ & $24.6\% \pm 39.9\%$ \\
self bleu & $37.1\% \pm 30.9\%$ & $27.1\% \pm 10.4\%$ & $46.1\% \pm 13.2\%$ \\
rougel & $46.9\% \pm 26.5\%$ & $24.7\% \pm 10.4\%$ & $49.3\% \pm 11.7\%$ \\
n gram diversity score 2 & N/A & $-27.6\% \pm 68.8\%$ & $8.8\% \pm 51.7\%$ \\
n gram diversity score 3 & N/A & $-4.0\% \pm 38.5\%$ & $21.8\% \pm 32.8\%$ \\
\bottomrule
\end{tabular}
\caption{Scale-free measures of AI homogenization for (a) Study 1 and (b) Study 2. $\rho_b$ denotes the share of the AI homogenization effect under the technical-quality incentive offset by the originality incentive (higher is better). $R_O$ and $R_T$ denote the share of initial AI-draft homogenization retained in final submissions under the originality and technical-quality incentives, respectively (lower is better). Entries report estimates $\pm$ 95\% confidence interval half-widths. N/A indicates that the denominator is too small.}
\label{tab:homogenization_retention}
\end{table}

\revision{
\section{Additional Results (Study 2)}\label{appsec:addl_results2}

Before delving into the additional results for Study 2, we first characterize AI usage throughout this study. There were a total of 612 queries, which results in an average of 3.06 queries per participant. Most requests pertained to \textbf{drafting} (59.8\%), and then creative consulting (27.9\%), editing (7.1\%), and miscellaneous (5.0\%). Of the 200 participants with access to AI, 166 of them (83\%) generated at least a first draft with the help of AI.

\subsection{Diversity of Stories} Table~\ref{tab:effects_combined2} shows the effect sizes from difference-in-means tests over the final submissions and the first valid draft the AI suggested. Red stars indicate significance after a Benjamini-Hochberg adjustment, but the results remain largely consistent. Note that we cannot conduct the difference-in-difference analyses for this Study because participants did not go through a brainstorming phase.

\begin{table}[!htbp]
\centering
\small
\begin{tabular}{lrr|rr|rr|r}
\toprule
Comparison & \multicolumn{2}{c}{\textbf{AI-O -- SELF-O}} & \multicolumn{2}{c}{\textbf{AI-T -- SELF-T}} & \multicolumn{2}{c}{\textbf{AI (T -- O)}} & \textbf{SELF (T -- O)} \\
\midrule
Version & Final & First (AI) & Final & First (AI) & Final & First (AI) & Final \\
\cmidrule(lr){2-3} \cmidrule(lr){4-5} \cmidrule(lr){6-7}
\midrule
\rowcolor{lightgray}
\textbf{Embeddings} &  &  &  &  &  &  &  \\
all-MiniLM-L12 & $0.030^{\star}_{\textcolor{red}{}}$ & $0.157^{\star \star \star}_{\textcolor{red}{\star \star \star}}$ & $0.117^{\star \star \star}_{\textcolor{red}{\star \star \star}}$ & $0.173^{\star \star \star}_{\textcolor{red}{\star \star \star}}$ & $0.082^{\star \star \star}_{\textcolor{red}{\star \star \star}}$ & $0.011^{}_{\textcolor{red}{}}$ & $-0.005^{}_{\textcolor{red}{}}$ \\
all-mpnet-base & $0.020^{}_{\textcolor{red}{}}$ & $0.131^{\star \star \star}_{\textcolor{red}{\star \star \star}}$ & $0.098^{\star \star \star}_{\textcolor{red}{\star \star \star}}$ & $0.157^{\star \star \star}_{\textcolor{red}{\star \star \star}}$ & $0.072^{\star \star \star}_{\textcolor{red}{\star \star \star}}$ & $0.020^{}_{\textcolor{red}{}}$ & $-0.006^{}_{\textcolor{red}{}}$ \\
Qwen3-0.6B & $0.034^{\star \star}_{\textcolor{red}{\star}}$ & $0.178^{\star \star \star}_{\textcolor{red}{\star \star \star}}$ & $0.106^{\star \star \star}_{\textcolor{red}{\star \star \star}}$ & $0.172^{\star \star \star}_{\textcolor{red}{\star \star \star}}$ & $0.096^{\star \star \star}_{\textcolor{red}{\star \star \star}}$ & $0.018^{}_{\textcolor{red}{}}$ & $0.024^{}_{\textcolor{red}{}}$ \\
embeddinggemma & $0.020^{\star}_{\textcolor{red}{}}$ & $0.116^{\star \star \star}_{\textcolor{red}{\star \star \star}}$ & $0.060^{\star \star \star}_{\textcolor{red}{\star \star \star}}$ & $0.123^{\star \star \star}_{\textcolor{red}{\star \star \star}}$ & $0.055^{\star \star \star}_{\textcolor{red}{\star \star \star}}$ & $0.023^{\star \star}_{\textcolor{red}{}}$ & $0.016^{}_{\textcolor{red}{}}$ \\
Style-Embedding & $0.186^{\star \star \star}_{\textcolor{red}{\star \star \star}}$ & $0.595^{\star \star \star}_{\textcolor{red}{\star \star \star}}$ & $0.202^{\star \star \star}_{\textcolor{red}{\star \star \star}}$ & $0.486^{\star \star \star}_{\textcolor{red}{\star \star \star}}$ & $0.128^{\star \star}_{\textcolor{red}{\star \star}}$ & $0.004^{}_{\textcolor{red}{}}$ & $0.112^{\star \star}_{\textcolor{red}{\star \star}}$ \\
byt5 & $0.020^{\star \star}_{\textcolor{red}{\star}}$ & $0.032^{\star \star \star}_{\textcolor{red}{\star \star \star}}$ & $0.001^{}_{\textcolor{red}{}}$ & $0.010^{\star \star}_{\textcolor{red}{\star \star}}$ & $0.005^{}_{\textcolor{red}{}}$ & $0.001^{}_{\textcolor{red}{}}$ & $0.024^{\star \star}_{\textcolor{red}{\star \star}}$ \\
star & $0.052^{\star \star \star}_{\textcolor{red}{\star \star \star}}$ & $0.196^{\star \star \star}_{\textcolor{red}{\star \star \star}}$ & $0.072^{\star \star \star}_{\textcolor{red}{\star \star \star}}$ & $0.169^{\star \star \star}_{\textcolor{red}{\star \star \star}}$ & $0.062^{\star \star \star}_{\textcolor{red}{\star \star \star}}$ & $0.015^{}_{\textcolor{red}{}}$ & $0.042^{\star \star \star}_{\textcolor{red}{\star \star}}$ \\
\midrule
\rowcolor{lightgray}
\textbf{Lexical based} &  &  &  &  &  &  &  \\
compression ratio & $-0.011^{}_{\textcolor{red}{}}$ & $0.038^{\star \star \star}_{\textcolor{red}{\star \star \star}}$ & $0.012^{}_{\textcolor{red}{}}$ & $0.048^{\star \star \star}_{\textcolor{red}{\star \star \star}}$ & $0.041^{\star \star \star}_{\textcolor{red}{\star \star \star}}$ & $0.028^{\star \star}_{\textcolor{red}{}}$ & $0.019^{\star}_{\textcolor{red}{\star}}$ \\
self bleu & $0.033^{\star \star \star}_{\textcolor{red}{\star \star \star}}$ & $0.122^{\star \star \star}_{\textcolor{red}{\star \star \star}}$ & $0.052^{\star \star \star}_{\textcolor{red}{\star \star \star}}$ & $0.114^{\star \star \star}_{\textcolor{red}{\star \star \star}}$ & $0.042^{\star \star \star}_{\textcolor{red}{\star \star \star}}$ & $0.015^{\star \star}_{\textcolor{red}{}}$ & $0.023^{\star \star \star}_{\textcolor{red}{\star \star \star}}$ \\
rougel & $0.034^{\star \star \star}_{\textcolor{red}{\star \star \star}}$ & $0.137^{\star \star \star}_{\textcolor{red}{\star \star \star}}$ & $0.064^{\star \star \star}_{\textcolor{red}{\star \star \star}}$ & $0.129^{\star \star \star}_{\textcolor{red}{\star \star \star}}$ & $0.047^{\star \star \star}_{\textcolor{red}{\star \star \star}}$ & $0.010^{}_{\textcolor{red}{}}$ & $0.018^{\star \star \star}_{\textcolor{red}{\star \star \star}}$ \\
n gram diversity score 2 & $0.010^{}_{\textcolor{red}{}}$ & $-0.037^{\star \star \star}_{\textcolor{red}{\star \star \star}}$ & $-0.003^{}_{\textcolor{red}{}}$ & $-0.038^{\star \star \star}_{\textcolor{red}{\star \star \star}}$ & $-0.028^{\star \star \star}_{\textcolor{red}{\star \star \star}}$ & $-0.015^{\star}_{\textcolor{red}{}}$ & $-0.015^{}_{\textcolor{red}{}}$ \\
n gram diversity score 3 & $0.003^{}_{\textcolor{red}{}}$ & $-0.068^{\star \star \star}_{\textcolor{red}{\star \star \star}}$ & $-0.014^{}_{\textcolor{red}{}}$ & $-0.066^{\star \star \star}_{\textcolor{red}{\star \star \star}}$ & $-0.040^{\star \star \star}_{\textcolor{red}{\star \star \star}}$ & $-0.021^{\star \star}_{\textcolor{red}{}}$ & $-0.022^{\star \star}_{\textcolor{red}{\star}}$ \\
\bottomrule
\end{tabular}
\caption{Effect sizes from differences in mean pairwise similarity over the final submissions (``Final'') and the first valid draft the AI suggested when available (``First (AI)''). Inference uses jackknife variance estimates for the pairwise-similarity U-statistics with a normal approximation and one-sided tests. Black significance stars are standard p-values, while red stars use Benjamini--Hochberg (BH) FDR-adjusted p-values: $^{*}$ $p<0.1$, $^{**}$ $p<0.05$, $^{***}$ $p<0.01$.}
\label{tab:effects_combined2}
\end{table}

\subsection{Characterizing Writing Trajectories}

Figure~\ref{fig:outcome_level_part2} replicates Figure~\ref{fig:outcome_level} from the main text for Study 2. Figure~\ref{fig:similarity_means} shows the average similarity across randomized groups for both the final submission and the first full draft of AI. Figure~\ref{fig:similarity_trajectory} plots the trajectories of the similarity metric across groups through time. Finally, Figure~\ref{fig:pca_robustness} shows scatter plots of the first and second principal components (PC) from the story's text embeddings. All the results as discussed in the main text generally hold.

\begin{figure}[!htbp]
    \centering
    \includegraphics[width=0.85\linewidth]{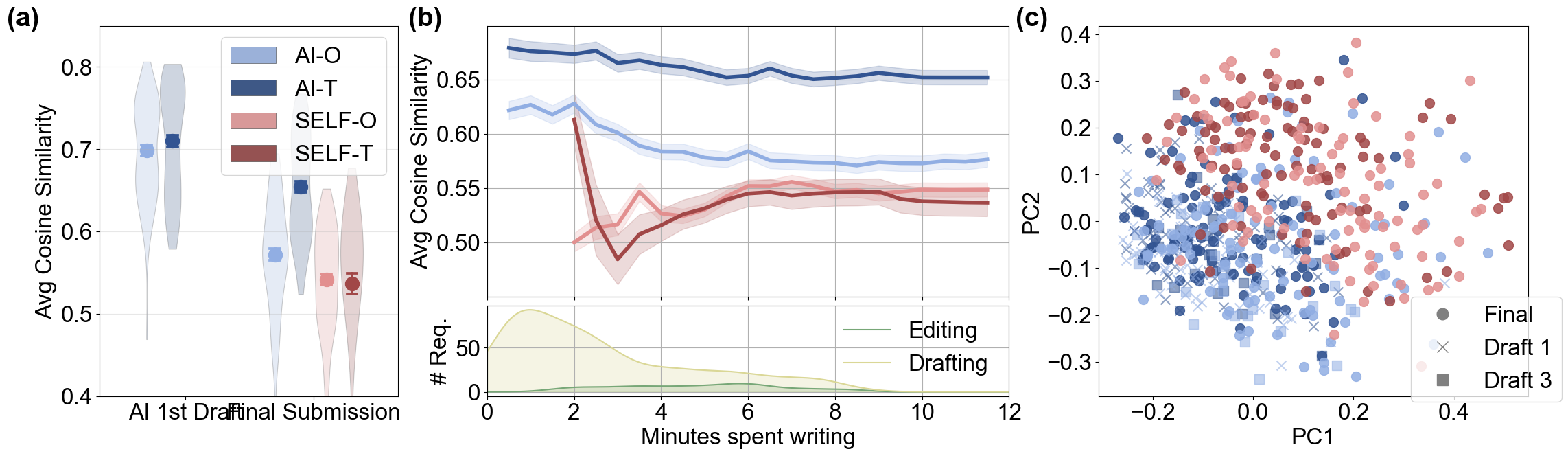}
    \caption{Results for Study 2. (a) Average cosine similarity of embeddings across randomized groups, for both the final submission and the first full draft the AI produces (when available). (b) [Top] Average cosine similarity of embeddings across randomized groups through time spent in the session. We only include text that has at least 75 words for every timestep to ensure that the embeddings capture enough information about the story. [Bottom] Number of requests over time spent in the session. (c) First and second principal component of the embeddings across groups, for both the drafts that AI produces (when available) and the final submissions. All plots use the embedding model \texttt{all-MiniLM-L12-v2}.
    }\label{fig:outcome_level_part2}
\end{figure}

\subsection{Differential Strategies in Using AI} 
Table~\ref{tab:effects_low_high_part2} displays the effect sizes from difference-in-means tests, where the \ai group is stratified into high and low adoption \revision{($A \geq \text{median}(A)$ and $A < \text{median}(A)$,} respectively), using $n=3$.

In Figure~\ref{fig:reliance_time_requests2}, we display similar plots from Figure~\ref{fig:reliance} in the main text, with $n=10$. While the distribution of adoption shifts downward (as expected), our general conclusions remain the same.

\begin{figure}
    \centering
    \includegraphics[width=0.8\linewidth]{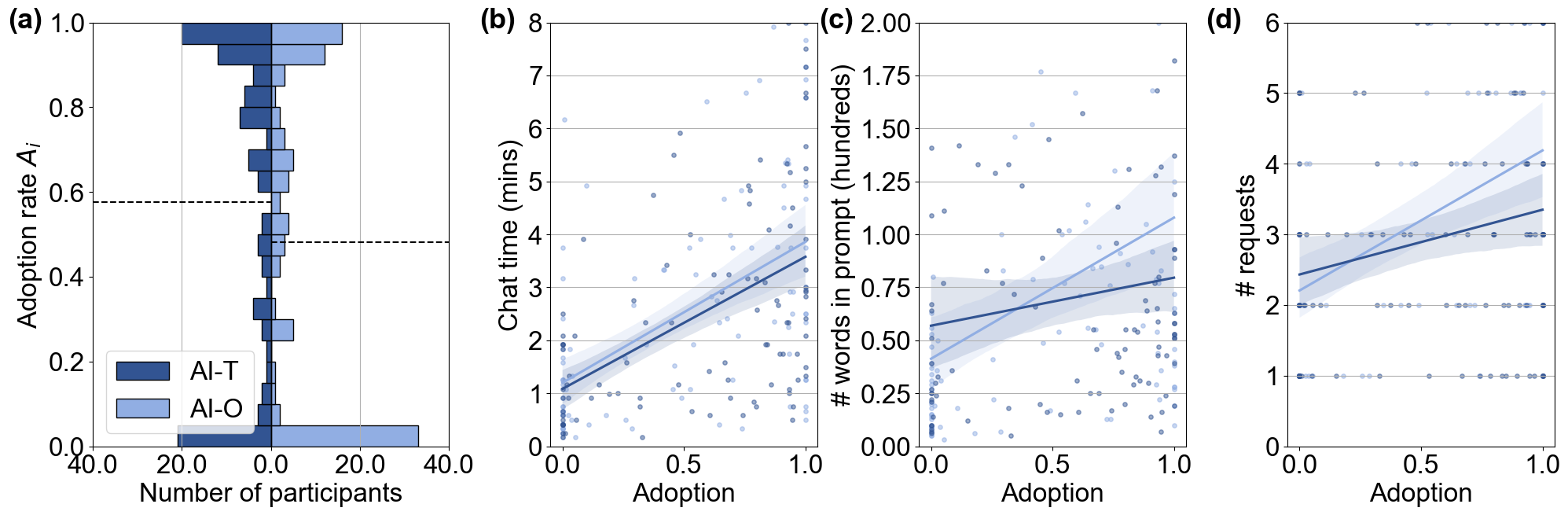}
    \caption{(a) Histogram of adoption score $A_i$ for both the \aiqual
    and \aiorig groups in Study 2, with $n=10$. (b) Time spent writing AI prompts by adoption
  and group. (c) Number of words in prompt by adoption and group. (d) Number of AI requests by adoption and group.}
    \label{fig:reliance_time_requests2}
\end{figure}




\begin{table}[!htbp]
\centering
\small
\begin{tabular}{lrr|rr|rr}
\toprule
Comparison & \multicolumn{2}{c}{\textbf{AI-O - SELF-O}} & \multicolumn{2}{c}{\textbf{AI-T- SELF-T}} & \multicolumn{2}{c}{\textbf{AI-T - AI-O}} \\
\midrule
AI Adoption level & Low & High & Low & High & Low & High \\
\cmidrule(lr){2-3} \cmidrule(lr){4-5} \cmidrule(lr){6-7}
\midrule
\rowcolor{lightgray}
\textbf{Embeddings} & & & & & & \\
all-MiniLM-L12-v2 $\downarrow$ & 0.019 & 0.044** & 0.103*** & 0.129*** & 0.079*** & 0.080*** \\
all-mpnet-base-v2 $\downarrow$ & 0.007 & 0.036* & 0.089*** & 0.105*** & 0.076*** & 0.063** \\
Qwen3-Embedding-0.6B $\downarrow$ & 0.014 & 0.058*** & 0.085*** & 0.123*** & 0.094*** & 0.089*** \\
embeddinggemma-300m $\downarrow$ & 0.008 & 0.035** & 0.046*** & 0.071*** & 0.054*** & 0.052*** \\
Style-Embedding $\downarrow$ & 0.085 & 0.309*** & 0.085 & 0.297*** & 0.112 & 0.101*** \\
byt5 $\downarrow$ & 0.018* & 0.023** & -0.005 & 0.006* & 0.000 & 0.007* \\
star $\downarrow$ & 0.027 & 0.083*** & 0.051*** & 0.090*** & 0.066*** & 0.049*** \\
\midrule
\rowcolor{lightgray}
\textbf{Lexical based} & & & & & & \\
Compression Ratio $\downarrow$ & -0.011 & -0.010 & -0.003 & 0.024** & 0.027* & 0.052*** \\
BLEU $\downarrow$ & 0.020*** & 0.049*** & 0.032*** & 0.069*** & 0.035*** & 0.042*** \\
ROUGE-L $\downarrow$ & 0.019** & 0.052*** & 0.044*** & 0.080*** & 0.042*** & 0.046*** \\
NGDS ($n=2$) $\uparrow$ & 0.013 & 0.007 & 0.009 & -0.013 & -0.019 & -0.035*** \\
NGDS ($n=3$) $\uparrow$ & 0.008 & -0.003 & 0.003 & -0.029** & -0.027* & -0.048*** \\
\bottomrule
\end{tabular}
\caption{Effect sizes from differences in mean pairwise similarity using jackknife variance estimates for the pairwise-similarity U-statistics with a normal approximation and one-sided tests: $^{*}$ $p<0.1$, $^{**}$ $p<0.05$, $^{***}$ $p<0.01$, over the final submissions stratified by low and high adoption of AI among the AI-assisted group.}
\label{tab:effects_low_high_part2}
\end{table}

\begin{table}[!htbp]
\centering
\small
\begin{tabular}{lcc}
\toprule
Adoption rate & High ($n = 100$) & Low ($n=100$) \\
\midrule
\rowcolor{lightgray}\multicolumn{3}{l}{\textit{Continuous variables}} \\
Total Approvals & 1097.21 (42.87) & 1150.54 (41.85) \\
Age & 37.30 (1.15) & 37.60 (1.10) \\
AI Exp. (General) & 2.63 (0.10) & 2.57 (0.10) \\
AI Exp. (Writing) & 2.14 (0.12) & 1.68 (0.11) \\
\addlinespace
\rowcolor{lightgray}\multicolumn{3}{l}{\textit{Binary variables}} \\
Student & 0.08 (0.03) & 0.14 (0.03) \\
Sex (Male) & 0.45 (0.05) & 0.40 (0.05) \\
Residence in US & 0.40 (0.05) & 0.48 (0.05) \\
Primary Lang (English) & 0.94 (0.02) & 0.95 (0.02) \\
\addlinespace
\rowcolor{lightgray}\multicolumn{3}{l}{\textit{Ethnicity}} \\
\quad Asian & 10.0\% & 11.0\% \\
\quad Black & 18.0\% & 10.0\% \\
\quad Mixed & 4.0\% & 5.0\% \\
\quad Other & 1.0\% & 1.0\% \\
\quad Prefer Not To Say & 1.0\% & 1.0\% \\
\quad White & 66.0\% & 72.0\% \\
\addlinespace
\rowcolor{lightgray}\multicolumn{3}{l}{\textit{Employment status}} \\
\quad Fulltime & 58.6\% & 47.0\% \\
\quad NIPW & 8.1\% & 13.0\% \\
\quad Other/Expired & 10.1\% & 10.0\% \\
\quad Parttime & 11.1\% & 14.0\% \\
\quad Unemployed Looking & 12.1\% & 16.0\% \\
\addlinespace
\bottomrule
\end{tabular}
\caption{Participant demographics for high and low adoption groups in Study 2. For continuous and binary variables, we display the mean and standard error (in parentheses). ``NIPW'' stands for Not in Paid Work. AI Experience is on a Likert scale from 0 to 4, with 0 being not at all familiar and 4 being extremely familiar with AI tools.}
\label{tab:summary_2col_part2}
\end{table}
}

\end{document}